\documentclass[12pt]{article}
\usepackage[reqno]{amsmath2000}
\usepackage{amsthm2000}
\usepackage[dvips]{color,graphicx}
\usepackage{amssymb} 
\def\href#1#2{#2}

\def\Inv{\text{\rm Inv}} 

\def\i{i}
\def\IP{\relax{\rm I\kern-.18em P}}

\newcommand{\nc}{\newcommand}
\newcommand{\beq}{\begin{equation}}
\newcommand{\be}{\begin{equation}}
\newcommand{\eeq}{\end{equation}}
\newcommand{\ee}{\end{equation}}
\newcommand{\beqa}{\begin{eqnarray}}
\newcommand{\ba}{\begin{eqnarray}}
\newcommand{\eeqa}{\end{eqnarray}}
\newcommand{\ea}{\end{eqnarray}}
\def\sgn{{\rm sgn}} 
\def\tq{\tilde{q}} 
\newcommand{\pa}{\partial}

\def\Mat{{\mbox{\rm Mat}}}

\def\ik{{\sf k}}%\def\ik{{\rm k}}
\def\min{{\mbox{\rm min\/}}}
\def\cH{{\cal H}}
\def\Im{{\mbox{\rm Im}}}

\def\id{{\rm id}}
\def\tr{{\rm tr}}
\def\cS{{\cal S}}
\def\tF{{\rm F}}
\def\tCS{{\rm CS}}
\def\tf{{f}}
\def\tL{{\rm L}}
\def\cC{{\cal C}}

\def\tA{{\rm A}} 
\def\rmA{{\rm A}}
\def\rma{{\rm a}}
\def\astk{\, , \, } 
\def\bc{\bar\gamma} 
\def\c{\gamma} 
\def\pl{\partial} 
\def\bpl{\bar \partial} 
\def\H3p{H_3^+}
\def\QR{\mathbb{R}} 
\def\QC{\mathbb{C}} 
\def\a{\alpha}
\def\b{\beta} 
\def\c{\gamma} 
\def\d{\delta} 
\def\e{\epsilon} 
\def\ga{\gamma} 

\def\nof{{\stackrel{\scriptscriptstyle{\circ}}
{\scriptscriptstyle{\circ}}}} 
\def\no{\raisebox{-2pt}{$\nof\:\!$}}
 
\def\raa{{r}} 

\def\mH{\Sigma} 
\def\tF{{\sf F}}
\def\hf{\hat f}
\def\astkn{\stackrel{\ast}{,}}
\def\bw{\bar w} 
\def\bW{{\overset{\underline{\ \ }}{W}}\,\!} 

\def\cW{{\cal W}}  
\def\bcW{{\overset{\underline{\ \ }}{\cal W}}\,\!}
\def\cV{{\cal V}} 
\def\cJ{{\cal J}} 
 
\def\ap{{\alpha '}}

\def\bT{{\overset{\underline{\ \ }}{T}}\,\!}
\def\sg{{\cal G}} 
\def\asg{{\widehat{\sg}}}
\def\bJ{{\bar J}}
\def\bz{{\bar z}}
\def\pl{\partial}
\def\bpl{\bar \partial} 
\def\Ad{{\rm Ad}}
\def\ad{{\rm ad}} 

\def\dim{{\rm dim}}
\def\rank{{\rm rank}}
\def\id{{\rm id}}
\def\c{\gamma} 
 
\def\bj{{\bar \jmath}} 
\def\cA{{\cal A}}
\def\cH{{\cal H}}
\def\H{{\cal H}}
\def\cJ{{\cal J}}

\nc{\nn}{\nonumber}
\def\e{\epsilon}
\def\o{\otimes}

\def\a{\alpha}
\def\b{\beta} 
 
\def\IP{\relax{\rm I\kern-.18em P}}
\def\tr{{\rm tr\ }}

\def\QC{\mathbb{C}}

\def\QR{\mathbb{R}}

\def\R{\QR}

\def\bi{{\bar \imath}}
\def\bj{{\bar \jmath}} 
\def\bn{{\bar n}} 
\def\cU{{\cal U}}
\newcommand{\SLC}{{\rm SL(2,\QC \rm)}}
\newcommand{\SLR}{{\rm SL(2,\QR \rm)}}
\newcommand{\SU}{{\rm SU(2)}}

\newcommand{\Ga}{\Gamma}

\newcommand{\CH}{{\mathcal H}}

\newcommand{\SH}{{\mathsf H}}

\newcommand{\SL}{{\mathsf L}}
\newcommand{\bL}{\bar{\SL}}

\def\U{{U}}

\def\vac{|0\rangle} 
\def\ppe{\hspace*{-2.5mm}}
\def\ew{\hspace*{-1mm}}
\newcommand{\Fus}[6]{F_{{\scriptstyle #1}{\scriptstyle #2}}
  \hspace*{.3mm}\displaystyle{[} \ew \begin{array}{ll} {\scriptstyle #3 }
  \ppe & {\scriptstyle #4} \ppe \\[-2mm] {\scriptstyle #5}\ppe &
  {\scriptstyle #6}\ew \end{array}\displaystyle{]}}

\newcommand{\CG}[6]{\displaystyle{[} \,\ew \begin{array}{lll} 
  {\scriptstyle #1} \ppe
  & {\scriptstyle #2} \ppe & {\scriptstyle #3} \ew \\[-2mm] {\scriptstyle
  #4} \ppe & {\scriptstyle #5}\ppe & {\scriptstyle #6} \ew\end{array}
  \displaystyle{]}}
\newcommand{\SJS}[6]{ \displaystyle{\{ }  \ew \begin{array}{lll} 
  {\scriptstyle #1} \ppe  & 
  {\scriptstyle #2} \ppe & {\scriptstyle #3}
  \ppe \\[-2mm]{\scriptstyle #4}  \ppe & {\scriptstyle #5} \ppe &
 {\scriptstyle #6} \ew \end{array} \displaystyle{\} } }

\def\orb{{\mbox{\rm \scriptsize orb}}}
\def\nn{\nonumber}

\def\cH{{\cal H}}

\def\S{{\cal S}}

\def\a{\alpha}

\def\ra{\rightarrow} 
\def\hx{\hat x}

\newcommand{\vvert}[3]{\displaystyle{(} \ew \begin{array}{ll} 
  \ & \hspace*{-4mm} {\scriptstyle #1} \\[-2mm] {\scriptstyle #2} 
  \ppe & {\scriptstyle #3} \ew \end{array} \displaystyle{)}}

\input epsf
\newcount\figno
\def\fig#1#2#3{
\par\begingroup\parindent=0pt\leftskip=1cm\rightskip=1cm\parindent=0pt
\baselineskip=15pt
\global\advance\figno by 1
\epsfxsize=#3
\centerline{\epsfbox{#2}}
\vskip 12pt
{\bf \small Figure \the\figno:} {\small #1}\par
\endgroup\par
}
\def\figlabel#1{\xdef#1{\the\figno 
\mbox{ }}}
\def\encadremath#1{\vbox{\hrule\hbox{\vrule\kern8pt\vbox{\kern8pt
\hbox{$\displaystyle #1$}\kern8pt}
\kern8pt\vrule}\hrule}}

\begin{document}
\baselineskip=19pt
\title{\bf Lectures on Branes in Curved Backgrounds\\[1cm]}
\author {{\sc Volker Schomerus} \\[2mm]
                  Albert-Einstein-Institut
      \\ Am M{\"u}hlenberg 1, D--14476 Golm, Germany \\[1mm]
  {\tt email:vschomer@aei.mpg.de}}
\vskip.8cm
\date{January, 2002}
%
%%%%%%%%%%%%%%%%%%%%%%  DOCUMENT %%%%%%%%%%%%%%%%%%%%%%%%%%%%%%%%%%%%%%
%\begin{document}
\begin{titlepage}      \maketitle       \thispagestyle{empty}

\vskip1cm
\begin{abstract}
\noindent     
These lectures provide an introduction to the microscopic 
description of branes in curved backgrounds. After a brief
reminder of the flat space theory, the basic principles and 
techniques of (rational) boundary conformal field theory 
are presented in the second lecture. The general formalism 
is then illustrated through a detailed discussion of 
branes on compact group manifolds. In the final lecture, 
many more recent developments are reviewed, including some 
results for non-compact target spaces.  
\end{abstract}
\vspace*{-14.9cm}
{\tt {AEI-2002-076 \hfill hep-th/0209241 }}
\bigskip\vfill
\noindent
{Lectures presented at the third RTN school on 
`The Quantum Structure of Spacetime and the Geometric Nature 
of Fundamental Interactions', Utrecht, January 17-22, 2002.} 
\end{titlepage}

\tableofcontents
\newpage

\section{Introduction}

During the last years the study of branes has been a powerful
tool to gain new insights into various string-dualities and 
thereby into the non-perturbative aspects of string theory. As 
long as the string length is much smaller than any other scale 
in a given problem, one can replace genuine string theory 
computations by a supergravity analysis. But the current 
problems in field and string theory challenge us more and 
more to go beyond  this supergravity approximation. For 
example, {\em string effects} are very relevant for all 
realistic string compactifications with N=1 supersymmetry 
on the brane. When the compactification scale gets small, 
even simple information such as the spectrum of BPS-branes 
is not protected by supersymmetry and can deviate drastically 
 (see e.g.\ \cite{Doug3,Doug4}) from the gravity `predictions'. 
Through the AdS/CFT correspondence string effects also have direct 
implications on non-perturbative aspects of gauge theories 
with finite (small) `t Hooft coupling (see e.g.\ \cite{AGMOO}) 
since the latter is tied up with the curvature radius of the 
AdS space.  
\smallskip

String corrections to supergravity can be studied with 
methods of 2-dimensional conformal field theory which 
provide an exact construction of the string perturbation 
expansion. When D-branes are present, the world-surface
of strings can end on them \cite{Pollec}. Hence, the strings' 
parametrization fields $X$ live on a 2D surface $\Sigma$ with 
boundaries. The choice of the boundary condition 
encodes the geometry of the brane. If we consider e.g.\ branes 
in a D-dimensional background with metric $g_{\mu\nu }(X)$, 
B-field $B_{\mu\nu}(X)$ and constant dilaton, the 
associated 2D field theory is the non-linear 
$\sigma$-model 
$$ S[X] \ = \ \frac{1}{4 \pi \ap} \int_\Sigma d^2z 
\left( g_{\mu \nu} (X) + B_{\mu \nu}(X) \right) 
\pl X^\mu \bpl X^\nu + \dots $$ 
where the dots stand for contributions from 
world-sheet fermions. This action has to be supplemented 
with appropriate boundary conditions on $X = X(z,\bz)$. 
\smallskip
   
The simplest situation occurs when both $g$ and $B$ are 
constant. Then the action $S$ is quadratic and hence we
are dealing with a 2D free field theory which can be 
solved easily. But when the background is curved, the 
metric cannot be constant and we are suddenly facing 
the problem of constructing a 2D interacting model. This 
is where the powerful techniques of (boundary) conformal 
field theory step in. 
\smallskip 

For closed strings, much of the relevant technology had 
been developed more than 10 years ago (see e.g. \cite{Ginslec,
Cardlec,FrMaSeBook}) following the seminal paper by Belavin, 
Polyakov and Zamolodchikov \cite{BPZ}. World-sheet theories with 
boundaries, however, received only very limited attention, 
with a few notable exceptions \cite{Card84,Card86,Card89,
Lewe1,BiaSag1,BiPrSa1,BiaSag2,PrSaSt1,PrSaSt2}. The 
situation changed in '96 when Polchinski demonstrated that 
understanding branes requires to study open strings with 
non-trivial boundary conditions \cite{Pollec}. This 
discovery prompted rapid and beautiful new developments 
in boundary conformal field theory and its applications to 
string theory. While most of the initial work focused on 
branes in toroidal compactifications and orbifolds thereof, 
it soon became clear that dealing with less trivial Calabi-Yau 
compactifications would require much more sophisticated 
methods (see \cite{RecSch1,FucSch2}). 
\smallskip 

These lectures intend to provide a self-contained introduction 
to some of the most important ideas and results in boundary 
conformal field theory. We shall begin with an extensive discussion 
of branes in flat backgrounds, including a derivation of brane 
non-commutativity. This part is technically quite transparent
and it will serve as a guide line during our ascend through 
the general structure of boundary conformal field theory. Once 
we have left flat space, we shall explain notions related 
to the {\em bulk theory}, analyze the importance of one-point 
functions and show how they can be encoded in {\em boundary 
states}. Furthermore, we derive the famous {\em Cardy constraint} 
and two important {\em sewing relations} for the bulk one-point 
functions and the boundary operator product expansions. The second
part concludes by presenting a generic family of exact solutions 
which not only applies to a very large class of backgrounds but 
also turns out to be fundamental for many of the more recent 
generalizations. Along the way we fill in some material that 
helps to bridge back to flat space. In the third lecture, the 
whole technology is then applied to the construction of branes 
on group manifolds. In this example, geometric ideas match nicely 
with the algebraic approach of boundary conformal field theory. 
We will close with a short guide to further developments and 
to some of the existing literature.  
\smallskip

Assuming that the reader is familiar with some basic concepts 
from open string theory, we will switch rather freely between 
string and conformal field theory terminology. Particular 
aspects of world-sheet or space-time supersymmetry are mostly 
ignored. For much of what we are about to explain, they are 
either irrelevant or can easily be incorporated using only 
standard ingredients (see e.g.\ \cite{GrScWi1Book,GrScWi2Book,
Pol1Book,Pol2Book}). Let us also mention that there exist 
several nice and rather complementary lecture notes on boundary 
conformal field theory \cite{PetZublec} including applications 
to condensed matter theory \cite{Salelec} and open strings 
\cite{Gablec,AngSaglec}. 
\medskip 

Before concluding this introduction, I would like to thank the 
the organizers of the winter school in Utrecht for a very enjoyable 
meeting and the participants for their encouraging feedback and  
interesting questions. These lectures grew out of several other 
courses, e.g.\ at the Erwin Schr\"odinger Institute in Vienna 
and at the $2^{nd}$ Lisbon School on Superstrings. I am grateful 
for many useful comments during those earlier events that helped 
to improve the presentation. Finally, I also thank Stefan 
Fredenhagen, Thomas Quella and Andreas Recknagel for reading these 
notes carefully and for all their constructive remarks. 

\newpage

\section{Flat Backgrounds and Free Field Theory} 
\setcounter{equation}{0}

Our aim in the first lecture is to review the microscopic theory 
of branes in flat backgrounds. The corresponding world-sheet theory 
can be solved with elementary methods (see Section 2.1). We shall 
use it to compute the coupling of closed strings to the brane
(Section 2.2) and the scattering amplitudes of open string modes
(Section 2.3). In a certain limit, the latter give rise to 
non-commutative Yang-Mills theory. The presentation is tailored 
to prepare for the construction of branes in curved backgrounds. 
   
\subsection{Solution of the world-sheet theory} 

\paragraph{The problem.} 

\def\cX{\QR^D} 
We want to study the motion of open strings in a D-dimensional 
Euclidean background $\QR^D$ that is equipped with a 
metric $g_{\mu \nu}$ and an anti-symmetric field $B_{\mu \nu}$. 
For the moment we shall assume that both background fields are 
constant. The world-surface of an open string is parametrized 
by a field $X: \Sigma \rightarrow \cX$. Here, $\Sigma$ is the 
world-sheet of the string, i.e.\ the strip $[0,\pi] \times \QR$ 
or, equivalently, the upper half plane 
$$\Sigma \ = \ \{\  z\, \in\, \QC\ | \ \Im z \, \geq\,  0\ \}
\ \ . $$  
These two realizations of the world-sheet are related by the 
exponential map. The motion of open strings in the given 
background geometry $(g,B)$ is controlled by the following 
quadratic action functional, 
\be S(X) \  = \ \frac{1}{4\pi \ap} \, \int_\Sigma 
  d^2 z \, \left( g_{\mu \nu} + B_{\mu \nu} \right) 
      \pl X^\mu \bpl X^\nu \ \ .  \label{flatact} \ee
It is important to notice that for constant $B$ the 
world-sheet action can be re-written in the form, 
$$ S(X) \ = \ \   \ \frac{1}{4\pi \ap} \, \int_\Sigma 
  d^2 z \, g_{\mu \nu} \pl X^\mu \bpl X^\nu \  + \ 
 \frac{1}{2\pi \ap} \, \int_\QR du \, B_{\mu\nu} 
  X^\mu \pl_u X^\nu \ \  $$
where the second term involving the B-field is a pure boundary 
term and we have used the decomposition $z = u + i v$, i.e.\ the 
coordinate $u$ parametrizes the boundary of $\Sigma$. Hence, the 
B-field does not affect the dynamics in the interior of $\Sigma$, 
but it provides a linear background $A_\mu (X)= B_{\mu \nu} X^\nu$ 
to which the end-points of the open strings couple as if they were 
charged particles in a magnetic background.  The complete 
description of the system requires to specify boundary conditions 
for the parametrization field $X$. To this end, we single out some 
$d$-dimensional hyper-plane $V$ in $\cX$ and demand 
\ba 
\left( \pl_u X^\mu (z,\bz) \right)_{z = \bz} ^\perp & = & 0 
\label{BC1}\\[2mm] 
\left( g_{\mu \nu} \pl_v X^\nu (z,\bz) \right)^{\|}_{z = \bz} & = & 
\left( i B_{\mu\nu } \pl_u X^\nu (z,\bz) \right)^{\|}_{z = \bz} \ \ . 
\label{BC2} \ea       
The symbols $\perp (\|) $ refer to directions perpendicular 
(parallel) to the hyper-surface $V$. With the first relation, 
we impose Dirichlet boundary conditions in the directions 
perpendicular to $V$, thereby restricting the endpoints of open 
strings to move along $V$. In other words, with our boundary 
conditions we have placed a D-brane along $V$. For $B = 0$, the 
second condition reduces to usual Neumann boundary condition 
along $V$. Hence, a non-vanishing B-fields gives rise to a 
deformation of Neumann boundary conditions. This has some 
interesting effects which we shall address below. 
\smallskip 

Without restriction we can assume that the $d$-dimensional plane 
$V$ is stretched out along the plane defined by the equations
$x^a = x_0^a,\ a = d+1, \dots, D,$ where the parameters $x_0^a$ 
describe the brane's transverse location. Dirichlet boundary 
conditions are then imposed for $X^a$ with $a = d+1, \dots, D$.   

\paragraph{Solution transverse to the brane.} 

For the transverse directions $a = d+1, \dots, D$, the fields
$X^a$ are constructed through the following formula for the 
general solution of the 2-dimensional Laplace equation
$ \pl \bpl X^a (z,\bz) =0$  with Dirichlet boundary conditions,  
\be X^a(z,\bz) \ = \ x^a_0 + i \sqrt{\frac{\ap}{2}} \, 
 \sum_{n \neq 0} \, \frac{\a^a_n}{n} \left( z^{-n} - 
 \bz^{-n}\right)  \ \  . \label{DX} 
\ee
In the quantum theory, the objects $\a^a_n$ become operators 
obeying the following relations
\be [\, \a^a_n , \a^b_m\, ] \ = \ n\, g^{ab}\, 
   \delta_{n,-m} \ \ \ , \ \ \ 
   (\a^a_n)^* \ = \ \a^a_{-n}\ \ .  \label{flatcomm1}
\ee
The commutation relations for $\a_n^a$ ensure that the field 
$X^a$ and its time derivative possess the usual canonical 
commutator. Reality of the bosonic field $X^a$ is encoded in 
the behavior of $\a^a_n$ under conjugation. 
\smallskip 

The operators $\a^a_n, n \neq 0,$ act as creation and annihilation 
operators on the Fock space $\cH^a_0 = \cV_0$ which is generated 
by $\a^a_n, n < 0,$ from a unique ground state $\vac$ subject
to the conditions 
$$ \a^a_n \, \vac  \ = \ 0 \ \ \mbox{ for } \ \ \ n > 0
\ \ . $$ 
This construction of the state space $\cH^a_0$ along with the 
formula (\ref{DX}) provides the complete solution for any direction 
transverse to our brane $V$.  
\medskip 

Before we turn to the directions along the brane, let us briefly 
remark that the operators $\a^a_n$ can be obtained as the 
Fourier modes 
\be \a^a_n \ = \ \ {\frac{1}{2\pi i}} \int_C z^{n}\, J^a(z)\, dz
                  - {\frac{1}{2\pi i}} \int_C \bar z^{n}\,
                    \bJ^a(\bar z) \,d\bar z\ \ . 
\label{a1} 
\ee
Here, $C$ is a semi-circle in $\mH$ centered around the point $z=0$  
and $J^a, \bJ^a$ denote the usual chiral currents
\ba J^a(z) & = & \i \pl X^a(z,\bz) \ = \ \sqrt {\frac{\ap}{2}} 
 \sum_n \, \a^a_n\, z^{-n-1}\ \ , \nn \\[2mm] 
  \bJ^a(\bz) & = & \i \bpl X^a(z,\bz) \ = \ - \sqrt {\frac{\ap}{2}} 
 \sum_n \, \a^a_n\, \bz^{-n-1}\ \ . \nn 
\ea
From these explicit formulas we read off that the currents obey 
\be J^a(z) = - \bJ^a(\bz) 
   \label{DBC} \ee 
all along the real line $z = \bz$. This relation is equivalent 
to the Dirichlet boundary condition and it tells us that the two 
sets of conserved currents in our theory are identified along the 
real line. Hence, there is only a single set of currents living 
on the boundary, while there are two sets throughout the bulk of 
the world-sheet.

\paragraph{Solution along the brane.} 
Let us now repeat the above free field theory analysis 
for the directions along the branes which are subject to 
the boundary condition (\ref{BC2}). The fields $X^i, i = 1, \dots, 
d,$ are once more constructed using the  general solution  of the 
wave equation  
\ba X^i(z,\bz) & = & \hx^i  - i \sqrt{\frac{\ap}{2}}\, \a_0^i\, \ln z\bz 
     - i \sqrt{\frac{\ap}{2}}\, {B^i}_j \, \a_0^j \ln \frac{z}{\bz} + 
  \nn \\[2mm] & &    
       + i \sqrt{\frac{\ap}{2}} \, 
 \sum_{n \neq 0} \, \frac{\a^i_n}{n} \left( z^{-n} + \bz^{-n}\right) 
 + i \sqrt{\frac{\ap}{2}} \, 
 \sum_{n \neq 0} \, \frac{{B^i}_j \a^j_n}{n} 
  \left( z^{-n} - \bz^{-n}\right) 
\ \  \label{NX} \ea   
where summation over $j = 1, \dots, d,$ is understood. In passing 
to the quantum theory, $\hx^i, \a^i_n$ become operators satisfying 
\ba [ \, \a_n^i , \a_m^j \, ] \ = \ n \, G^{ij}\, \delta_{n,-m} 
   \ \ \ \ & , & \ \ \ \ 
   [ \, \hx^i, \a_n^j\, ] \ = \ i \sqrt{\ap} \, G^{ij}\,  \delta_{0,n} 
  \ \ , \label{flatcomm2a} \\[2mm]  
   [\, \hx^i , \hx^j\, ] & = & i  \, \Theta^{ij} \ \ \label{flatcomm2b}
\ea
Furthermore, they obey the reality properties $(\hx^i)^* = \hx^i$ and $
(a^i_n)^* = a^i_{-n}$. The commutation relations involve new structure 
constants $G^{ij}$ and $\Theta^{ij}$ which are obtained from the background 
fields through 
\be \label{BtoT} 
G^{ij} \ = \ \left( \frac{1}{g+B} \right)_{\rm S}^{ij} 
   \ \ \ \ , \ \ \ \ 
\Theta^{ij} \ = \ \left( \frac{\ap}{g+B} \right)_{\rm A}^{ij} 
\ \ . 
\ee
Here, ${\rm S}$ or ${\rm A}$ mean that the expression in brackets
gets symmetrized or anti-symmetrized, respectively. Note that the 
matrix $\Theta$ vanishes if and only if the B-field vanishes. A 
non-zero $\Theta$ causes the center of mass coordinates $\hx^i$ 
of the open string to be quantized. We shall see below that this 
has very interesting consequences.   
\smallskip

Once more, the operators $\a^i_n$ act as creation and annihilation 
operators but now there exists a $d$-parameter family of ground states 
$|k\rangle$ which are parametrized by a momentum $k = (k_i)_{i = 
1, \dots, d}$, 
$$ \a^i_0 \, |k\rangle   \ = \ \sqrt{\ap}\, G^{ij}\, k_j \, |k\rangle 
\ \ . $$ 
If we denote the associated Fock spaces by $\cV_k$, the state 
space $\cH^B$ for the directions along the brane can be written as 
a direct integral $\cH^B  = \int_k 
d^dk \cV_k$. On this state space we can also represent the position 
operators $\hx^i$ as simple shifts of the momentum,
$$ \exp(i k'_i \, \hx^i) \, |k \rangle \ = \ e^{\frac{i}{2} k \times k'}\, 
   |k+k'\rangle \ \  $$ 
where the vector product $\times$ is defined through $k \times k' 
= k_i \Theta^{ij} k'_j$. The fields we shall consider below involve 
only exponentials of $\hx^i$ and not $\hx^i$ itself.  
\medskip

From $X^i$ we obtain the chiral currents $J^i$ and $\bJ^i$ 
in the same way as above
\ba J^i(z) & = & \i \pl X^i(z,\bz) \ = \ \sqrt {\frac{\ap}{2}} 
 \sum_n \, {\left(1 + B\right)^i}_j \a^j_n z^{-n-1}\ \ , 
\label{JB} \\[2mm] 
  \bJ^i(\bz) & = & \i \bpl X^i(z,\bz) \ = \ \sqrt {\frac{\ap}{2}} 
 \sum_n \,  {\left(1 - B\right)^i}_j  \a^j_n \bz^{-n-1}\ \ .  
\label{bJB} \ea
These currents obey a linear boundary condition all along the real 
line $z = \bz$ which is equivalent to the condition (\ref{BC2}),   
\be \label{glueN} 
 J^i(z) \ = \ 
   \left( \frac{1 + B}{1 - B} \right)^{i}_{\ j} \bJ^j(\bz) 
  \ =: \ \left( \Omega^B \bJ \right)^i(\bz)  \ . 
\ee   
In this case, there appears a non-trivial map $\Omega^B$ that 
rotates the anti-holomorphic fields before they are identified 
with their holomorphic counterparts. It replaces the simple sign 
that we found in eq.\ (\ref{DBC}) for the directions transverse 
to the brane. The map $\Omega^B$ also shows up in the formula
\be \a^i_n \ = \ \ {\frac{1}{2\pi i}} \int_C z^{n}\, J^i(z)\, dz
                  + {\frac{1}{2\pi i}} \int_C \bar z^{n}\,
      \left(\Omega^B\bJ\right)^i (\bar z) \,d\bar z\ \ 
\label{a2} \ee
which is used to obtain the oscillators $\a^i_n$ from the 
local fields $J^i$ and $\bJ^i$. These remarks complete our 
solution of the world-sheet theory. 

\subsection{The closed string sector}

We are now prepared to discuss some of the bulk fields and their 
properties. After a few brief remarks on  the Virasoro fields, we 
shall explain how to obtain the vertex operators for closed string 
tachyons and compute their couplings to the brane along $V$. 
These couplings encode all information about the density 
distribution of the brane. 

\paragraph{The Virasoro field.} Along with the chiral currents
$J^\mu $ and $\bJ_\mu$, there exists another very important pair 
of chiral fields, namely the Virasoro fields $T, \bT$. The 
holomorphic field $T$ is obtained from the chiral currents 
$J^\mu$ (the index $\mu$ is taken from $\mu = 1, \dots, D$) 
through the prescription   
$$ T(z) \ := \ \frac{1}{\ap} \,\sum g_{\mu\nu} \, :\! J^\mu 
   J^\nu\!: (z) \ := \ 
  \frac{1}{\ap} \, \lim_{w \ra z} \left(g_{\mu\nu} J^\mu(w) 
   J^\nu(z) - \frac{\ap}{2} \frac{D}{(w-z)^2} \right) \ \ . 
$$ 
Here, we use the limiting procedure on the right hand side to 
define the conformal normal ordering $: \cdot :$. For the 
anti-holomorphic partner $\bT$ we employ the same construction 
with currents $\bJ^\mu$ instead of $J^\mu$. The boundary conditions 
for the chiral currents (\ref{DBC}),(\ref{glueN}) imply that 
the two Virasoro fields coincide along the boundary $z = \bz$,  
\be  T(z) \ = \ \bT(\bz) \ \ . \label{glueTflat} \ee
Such a relation can be seen to prevent world-sheet momentum 
from leaking out across the boundary of $\Sigma$. Technically, 
it allows us to construct the following modes  
\be 
\label{Virmod} 
  L_n  \ := \ {\frac{1}{2\pi i}} \int_C z^{n+1}\, T(z)\, dz
                  - {\frac{1}{2\pi i}} \int_C \bar z^{n+1}\,
                    \bT(\bar z) \,d\bar z \ \ . 
\ee
The elements $L_n$ generate one copy of the Virasoro algebra 
with central charge $c=D$. They can also be expressed through 
the oscillators $\a^\mu_n$, 
\be  
L_n \ = \ \frac12 \sum_{n = -\infty}^\infty 
            G_{i j} \ \no \a^i_{m-n} \a^j_n \no    
      +  \frac12 \sum_{n = -\infty}^\infty 
            g_{a b} \ \no \a^a_{m-n} \a^b_n \no   
\label{Ln} \ee
where $ G$ with lower indices denotes the inverse of $(G^{ij})$
and it appears in the expression for $L_n$ because of the factors 
$(1\pm B)$ in eqs.\ (\ref{JB}),(\ref{bJB}). The symbol $\no \cdot 
\no$ stands for operator normal ordering, i.e.\ it instructs us 
to move all the annihilation operators $\a^\mu_n, n \geq 0,$ to 
the right of the creation operators $\a^\mu_n, n <0$.        
 
\paragraph{One-point functions.} 
Let us now turn to the family of bulk fields that are associated
with closed string tachyons. These are defined by the expression
\be \phi_{k,k} (z,\bz) \ : = \ :\! e^{ikX(z,\bz)} \! : 
 \ = \ \sum_{n=1}^\infty \frac{(ik)^n}{n!} 
   \, :\! X^n\! : (z,\bz) \ \ . \label{Bvert} 
\ee 
Here $k = (k_\mu)$ and we have suppressed all indices on the fields
$X$ and the momenta $k$. Furthermore, we extended the prescription 
$: \cdot :$ for the conformal normal ordering to arbitrary powers of 
the bosonic field $X$. The $n^{th}$-order normal ordered product is 
defined recursively 
by 
$$ :\! X^{\mu_1}(z_1,\bz_1) \cdots X^{\mu_n}(z_n,\bz_n)\!: \ = \ 
 X^{\mu_1}(z_1,\bz_1) \cdots X^{\mu_n}(z_n,\bz_n) + 
\sum \mbox{subtractions} 
$$ 
where the sum runs over all ways of choosing pairs of fields
from the product and replacing them by $(-1)$ times the free 
propagator, i.e.\    
$$  X^\mu(z_1,\bz_1) X^\nu(z_2,\bz_2) \ \longrightarrow \ 
-\,  \langle X^\mu(z_1,\bz_1) X^\nu(z_2,\bz_2) \rangle  \ = \ 
- \ap \, g^{\mu\nu} \, \ln |z_1-z_2| \ \ . $$ 
For more details see e.g.\ \cite{Pol1Book}. Note that the normal 
ordering prescription for bulk fields is the same as on the full 
complex plane because it uses the propagator of the free bosonic 
field on the full complex plane. This differs from the propagator 
on the upper half plane
\ba \langle  X^\mu(z_1,\bz_1) X^\nu(z_2,\bz_2) \rangle_B 
   & = & 
 - \ap \, g^{\mu\nu} \ln |z_1-z_2| + \ap\, g^{\mu\nu} 
  \ln |z_1-\bz_2| \nn \\[2mm] 
& &  -  \ap\, G^{\mu \nu} \ln |z_1 - \bz_2|^2 - 
  \frac{\Theta^{\mu\nu}}{2\pi} \ln 
  \frac{z_1 - \bz_2}{\bz_1 - z_2}        
\label{Dprop}
\ea 
by  terms which are regular in the upper half plane $\Im\/ z > 0$ 
and become singular only along the boundary. In writing down 
eq.\ (\ref{Dprop}), we promoted $G$ and $\Theta$ to $D \times 
D$-matrices such that all new elements vanish. 
\smallskip
         
After these remarks it is easy to rewrite the bulk field 
$\phi_{k,k}(z,\bz)$ in terms of operator normal ordering, i.e.\  
such that annihilation operators stand to the right of the 
creation operators, 
\be  \phi_{k,k} (z,\bz) \ : = \ \frac{1}
   {|z-\bz|^{\ap (k^2 - 2 k\cdot k)}} 
    \  \no e^{ik X (z,\bz)} \no  
    \label{BvertD} \ \ . 
\ee 
In the exponent we use $k \cdot k = G^{i j}k_i k_j$ and $k^2 = g^{\mu\nu}
k_\nu k_\mu$. The 
singularity is related to the second and third term in the propagator
(\ref{Dprop}). In the operator normal ordering we agree to treat 
$\hat x^\mu$ as a creation operator, i.e.\ we move it to the left. 
Now that the tachyon vertex operators are well defined and 
conveniently expressed through the operator normal ordering, we 
can compute all their correlation functions and in particular the 
one-point function on the upper half plane. With our brane localized 
along $V = \{ x^a = x_0^a\}$ these one-point functions are given by 
\be \langle \phi_{k,k}(z,\bz)\rangle_{x_0} \ = \ 
 \delta^{(d)} (k_i) \ \frac{e^{ik_ax^a_0}} 
{|z-\bz|^{\ap k^2}} \ \ .
\label{D1pt} 
\ee 
From the set of all these one-point functions we can recover the 
parameters $x^a_0$, i.e.\ the brane's transverse position is \
completely specified by the one-point function of the bulk tachyon 
vertex operators. We shall see that a similar statement remains 
true for branes in curved backgrounds. 
\medskip 

It will be useful to understand how a density distribution of the 
brane can be read off from the one-point functions. In string theory, 
one-point functions of bulk fields describe how closed string 
modes couple to the D-brane. Our formula (\ref{D1pt}) implies that 
the coupling is completely delocalized in the directions of momentum 
space that are normal to the brane. Hence, after Fourier transformation,  
closed strings are seen to couple to some object that it localized along 
$x^a = x^a_0$ in position space. This is just the location of our brane 
in the background. In formulas we find   
\be \lim_{\ap \rightarrow 0} \langle \phi_{k,k}(z,\bz)
 \rangle_{x_0} \ =  \ \delta^{(d)} (k_i)\, e^{i k_a x^a_0} \ 
     \sim \  \int d^D x\, 
       \delta^{(D-d)}(x^a - x^a_0) \, \phi_{k,k}(x) 
\label{flatgeom} 
\ee
where $\phi_{k,k}(x) = \exp(i k_\mu x^\mu)$ is the wave function 
of a scalar particle moving in $\QR^D$ with momentum $k$. The 
first factor $\delta^{(D-d)}(x^a-x^a_0)$ in the integrand is 
interpreted as the density distribution of the brane.    

\subsection{The open string sector}  

After our discussion of bulk fields we now turn to a new set of 
fields which can be inserted at points on the boundary of the 
world-sheet. Such boundary fields are associated to the modes of open 
strings on the brane. We will briefly talk about their spectrum before 
we compute correlators of the tachyon vertex operators. The results of 
these computations can be expressed with the help of the non-commutative 
Moyal-Weyl product. We review the latter to make this presentation 
self-contained. Finally, we shall argue that - in a certain decoupling 
limit - the scattering amplitudes of massless open string modes can be 
reproduced by the so-called non-commutative Yang-Mills theory. 

\paragraph{Spectrum of boundary fields.} Boundary fields are in 
one-to-one correspondence with states of the boundary theory. The
space of these states was constructed explicitly when we solved the 
model in the first subsection. We remind the reader that it is given by 
$$ \H_{(B,d)} \ = \int d^d k \, \cV_k  \otimes 
    \cV^{\otimes_{D-d}}_0\ \ . $$
Here the integral over momenta came with the directions along 
the brane while the $D-d$ factors $\cV_0$ are associated with the 
transverse space. 

$\H_{(B,d)}$ is the space on which all our bulk and boundary fields 
act. In particular, through eq.\ (\ref{Virmod}), it carries an action 
of the Virasoro algebra. Among the Virasoro modes, $L_0$ is distinguished 
because it agrees with the Hamiltonian of the world-sheet theory up to a 
simple shift, i.e.\ $H = L_0 - D/24$. Using the explicit formula 
(\ref{Ln}) for $L_0$ it is rather easy to calculate the partition function 
of the theory, 
\be \label{flatPF}  
  Z_{(B,d)}(q) \ := \ {\mbox{\rm tr}}_\cH \left( q^H \right) \ = \ 
  \frac{1}{\eta^D(q)} \int d^d k \ q^{\frac{\ap k \cdot k}{2}}
\ee  
where $\eta(q) = q^{1/24} \prod (1-q^n)$ is Dedekind's 
$\eta$-function. The factor $1/\eta^D(q)$ is associated with the 
oscillations of the bosonic string in the D-dimensional flat backgound. 
In addition, open strings can move along the brane and this motion gives 
rise to the integral over the $d$-dimensional center of mass momentum 
$k^{\|}$. The term $\ap k\cdot k/2$ in the exponent is the kinetic 
energy of a particle moving in $d$-dimensional flat space with metric 
$G$. Our discussion here shows that the partition function $Z$ is an 
important quantity containing quite detailed information about the 
boundary condition.

\paragraph{The Weyl product.}  
\def\tF{{\rm F}}
Before we start discussing open string scattering amplitudes, 
we want to recall some elementary mathematical results on the
quantization of a very simple classical system. It consists
of a $d$-dimensional linear space $V$ along with a constant 
anti-symmetric $d\times d$ matrix $\Theta^{i j}$. The
latter defines a Poisson bracket for functions on $V$. When 
evaluated on the coordinate functions $x^j,j = 1,\dots,d,$ 
the Poisson structure reads, 
\be \{\, x^i \, , \, x^j \, \} \ = \ \Theta^{i j } 
\ \ . \label{PB} \ee
Quantization means to associate a self-adjoint operator 
$\hx^j:\H \rightarrow \H$ on some state space $\H$ to 
each coordinate function such that
\be [\, \hx^i \, , \, \hx^j \, ] \ = \ i \, 
\Theta^{i j} \ \ . \label{QPB} \ee
More generally, one would like to associate a self-adjoint 
operator $\tF = Q(f)$ to any real valued function $f$ on $V$ 
such the commutator $[\tF_1,\tF_2]$ is approximated by the 
Poisson bracket $\{f_1,f_2\}$ in a sense that we shall make
more precise below. An appropriate mapping $f \rightarrow 
\tF = Q(f)$ was suggested by Weyl \cite{Weyl},   
$$ \tF \ = \ Q(f) = \ \int\,  d^dk \, \hat f (k) \ 
\exp(i k_j \hx^j ) \ \  $$ 
where $\hf(k)$ denotes the Fourier transform of $f$. The 
operator $\tF$ is trace class, if $f$ is smooth and 
decreases, together with all its derivatives, faster 
than the reciprocal of any polynomial at infinity. A 
detailed discussion of appropriate spaces of functions
can be found e.g.\ in \cite{KonSch2}.   
\smallskip

We want to compute the product of any two operators 
$Q(f)$ and $Q(g)$ and compare this to the operator 
which Weyl's formula assigns to the Poisson bracket of 
the two functions $f$ and $g$. Using the famous 
Baker-Campbell-Hausdorff formula one finds the following 
auxiliary result for the product of two exponentials
\be \exp(i k_i \hat x^i )\, \exp (i k'_j \hat x^j) 
\ = \ \exp( - \frac{i}{2} k_i \Theta^{i j} k'_j) 
 \, \exp(i (k+k')_i \hx^i ) \ \ . 
\label{Qmult} \ee
As one can show by a short computation, this formula implies
that the product of two operators $Q(f)$ and $Q(g)$ is given 
by 
\ba 
Q(f) \, Q(g) & = &  Q(f \ast g)  \ \ \ \ \ \  
\mbox{ where } \nn \\[2mm] 
f \ast g \, (x) & = & \exp ( -\frac {i}{2} \Theta^{\mu \nu} 
\pl_\mu \tilde \pl_\nu ) \  f (x) g(\tilde x) |_{\tilde x = x } 
\ \ . \label{MWprod} \ea
The multiplication $\ast$ defined in the second row is known 
as the {\em Moyal product} \cite{Moyal} associated with the 
constant anti-symmetric matrix $\Theta$. It is an associative 
and non-commutative product for functions on the $d$-dimensional 
space $V$. Moreover, to leading order in the number of 
derivatives, one finds
$$  [\, f \astkn g \, ]  \ := \  f \ast g - g \ast f 
%   f g - \frac{i}{2} \Theta^{\mu \nu} \pl_mu f \pl_\nu g 
% - g f + \frac{i}{2} \Theta^{\mu \nu} \pl_mu g \pl_\nu f + \ 
%   \dots \nn \\[2mm] 
  \ = \  - i \Theta^{\mu \nu}\, \pl_\mu f \pl_\nu g \, + \, \dots  
  \ =\ -i \{ f \, , \, g\, \} \ + \ \dots \ \ . 
$$
Hence, the Moyal-commutator of the functions $f$ and $g$ 
is approximated by the Poisson bracket of these functions. 
In the same sense, the commutator of the operators $Q(f)$ 
and $Q(g)$ is approximated by the operator $Q(\{f,g\})$.

\paragraph{Correlation functions.} 

Following the standard wisdom of conformal field theory,
there is a boundary field associated with each state in 
the space $\H_{(B,d)}$. For the ground states $|k\rangle$, 
the corresponding fields are the `open string tachyon 
vertex operators',  
$$  \psi_k(u) \ := \ \no e^{i k_i X^i(u)} \no 
 \ = \ e^{i k_i X^i_<(u)}
      \, e^{i k_i X^i_>(u)}
$$ 
where 
\beqa 
X^\mu_>(u) & = & -  i\, \sqrt{2 \a'} \, \a^\mu_0\,  \ln u 
   + i \, \sqrt{ 2 \a'} \, \sum_{n>0} 
   \frac{\a^\mu_n}{n}  u^{-n} \ \ , \\[2mm]
X^\mu_< (u)& = & \hat x^\mu  + i \, \sqrt{2\a'} 
 \sum_{n<0} \frac{\a^\mu_n}{n}  u^{-n} \ \ . 
\eeqa
From our exact construction of the theory it is rather 
straightforward now to compute all the correlation functions 
of these tachyonic vertex operators. Before we present the 
result of this computation, let us introduce the {\em 
decoupling limit} of a functional $F(\ap;g,B)$. It is 
defined by \cite{SeiWit99}
$$ F^{DL}(g,B) \ = \ \lim_{\e \rightarrow 0} 
   F(\e; g \e^2,  B \e) \ \ . $$
We shall explain the idea behind this limit at the end of the 
section. For the moment, let us return to the correlators we 
were about to compute. They can be evaluated easily with the 
help of the Baker-Campbell-Hausdorff formula, 
\ba 
\langle \psi_{k_1}(u_1) \ \cdots \ \psi_{k_n}(u_n) \rangle 
 & = & \prod_{r < s} e^{-\frac{i}{2} k_{r} \times k_{s}} \ 
\frac{\d(\sum_r k_r)}{|u_r - u_s|^{\ap k_r \cdot k_s}}
\nn \\[2mm] 
&\stackrel{DL}{\rightarrow}  & 
\prod_{r < s} e^{- \frac{i}{2} k_{r} \times k_{s}}
        \  {\d({\textstyle \sum_r} k_r)} \ \ . 
\label{oscorr}\ea 
Here, $r,s = 1,\dots,n,$ and we have used the notation 
$k \times k' = k_i \Theta^{ij} k'_j$, as before. Note that 
the phase factors that appear when we evaluate the correlation 
function of the exponential field are identical to the phase 
factors we encountered in multiplying two exponential functions 
using the Moyal-Weyl product (see eq.\ (\ref{Qmult})). In the 
decoupling limit, $\a' G^{ij}$ vanishes and hence we are 
left with the phase factors and a $\delta$ function that 
enforces momentum conservation. Hence, in this limit of the 
theory, the correlation functions are determined entirely by 
the Moyal-Weyl product, i.e.\ 
$$  
\langle \psi_{k_1}(u_1) \ \cdots \ \psi_{k_n}(u_n) \rangle^{DL}  
\ = \ \int_V d^dx \, e_{k_1} \ast \dots \ast e_{k_n} 
\ \ $$
where $e_k = \exp (i k_j x^j)$ is the exponential function. 
If we introduce the fields $\psi[f](u)$ by 
$$ \psi[f](u) \ = \ \int\,  d^dk \, \hat f(k) \, \psi_k(u) $$
then the result can be restated as follows
\be
\langle \psi[f_1](u_1) \ \cdots \ \psi[f_n](u_n) \rangle^{DL}  
\ = \ \int_V d^dx \, f_1 \ast \dots \ast f_n
\ \ . \label{flattopcorr} \ee
In conclusion, the decoupling limit of all boundary 
correlation `functions' is independent of the world-sheet 
coordinates $u_r$ and its value can be computed from the 
non-commutative Moyal-Weyl product associated with the 
anti-symmetric tensor $\Theta$. A related observation was 
made by several authors \cite{DouHul,CheKro,ArArSh,ChuHo1,Scho}. 
The formulation we have presented here was found in \cite{Scho}.  
\smallskip

Our results have a rather nice physical explanation. Recall that 
the action functional (\ref{flatact}) for open strings has two terms: 
to begin with there is a boundary term which describes the motion of 
the charged open string ends in a magnetic field. It is well known 
that the coordinates of charged particles in a magnetic background 
have a non-vanishing Poisson bracket and hence they do not commute 
after quantization. To reach such a particle limit, we have to 
suppress the string oscillations, i.e\ we have to send $\a'$ to zero. 
The B-field should be scaled down at the same rate so that $B/\a'$ 
remains constant. But even in this limiting regime the resulting 
theory for the string endpoints does not approach the theory for 
charged particles in a magnetic field because of the bulk term in 
the action (\ref{flatact}) for open strings. This term makes the 
open string ends remember that they are attached to a string which 
becomes very stiff as we try to turn off the oscillations. 
Consequently, the open string ends dissipate energy into these 
tails and the strength of this dissipation is given by $g/\a'$
(see also \cite{CallThor}). If we want to suppress this effect, 
the closed string metric $g$ (more precisely its components $g_{ij}$ 
along the brane) has to vanish at a faster rate than $\a'$. All this 
in achieved by the decoupling limit we defined above. It also ensures 
that the open string metric $G$ remains finite.

\paragraph{Non-commutative Yang-Mills theory.}  
In a supersymmetric string vacuum, the scalar tachyon for
which our previous discussion was most relevant does not 
arise but there appears a massless vector field associated 
with the 2D boundary fields 
\be \no J^\mu(u) \exp(i k_i X^i (u) ) \no  \ = \ 
   J^\mu_<(u) \,  \no \exp(i k_i X^i(u)) \no  + 
  \no \exp(i k_i X^i(u)) \no \,  J^\mu_>(u) \ \ . 
\label{MVO} \ee 
Once more, normal ordering for these boundary operator means
to move all the annihilation operators to the right of the 
creation operators. Consequently, $J_>(u)$ is defined by 
$$
J^\mu_>(u) \ = \ \sum_{n> -1}\  \a^\mu_n \, u^{-n-1} \ \ \ \ 
\mbox{ and } \ \ \ \ J^\mu_<  \ = \ J^\mu - J^\mu_> 
\ = \ \sum_{n\leq -1} \ \a^\mu_n \, u^{-n-1} \ \ \ .  
$$
To compute the correlation functions of the above fields in 
the decoupling limit, we proceed in two steps. The first one 
is to remove all the currents from the correlators with the help 
of Ward identities. Once we are left with correlation functions 
of the exponential fields, we can then use the results from the 
discussion above. Since we are only interested in the decoupling 
limit of the correlation function, we can drop sub-leading terms 
whenever they arise in the computation.  
\smallskip

Let us discuss this in some more detail. Using a bit of algebra, 
it is not difficult to derive that 
\ba 
\left( J^\mu(u_1) J^\nu(u_2) \right)_{\rm sing} & := & 
  [\, J^\mu_> (u_1) , J^\nu (u_2) \, ] \ = \ 
%%%%%   [\, J^\mu_> (u_1) , J^\nu_< (u_2) \, ] \ = \ 
  \frac{\a'}{2} \frac{G^{\mu \nu}}{(u_1 - u_2)^2}   
\label{JonJ} \ \ \\[2mm] 
\left( J^\mu(u_1) \psi_k (u_2) \right)_{\rm sing} & := & 
  [\, J^\mu_>(u_1) , \psi_k(u_2)\, ] \ = \ 
   \frac{\a'}{2} \frac{ G^{\mu i} k_i } {u_1 - u_2} \psi_k(u_2)  
\label{Jonpsi} \ \ . 
\ea
The commutators we have listed here compute the singular part of 
the corresponding operator product expansions. This is indicated 
in the notation on the left hand side. The two commutation relations
along with the properties  $J^\mu_>(u) |0\rangle = 0$ and $\langle 0|
J^\mu_<(u) = 0$ of the vacuum can be used to remove all currents from 
an arbitrary correlator. In order to do so, we commute the lowering 
term $J^\mu_>$ of each current insertion to the right until it meets 
the vacuum and similarly the raising terms are all pushed to the left. 
The commutation relations give rise to two different contributions. 
If one currents hits another, both of them disappear from the 
correlator and we obtain a factor of $\a' G^{\mu\nu}$. Commuting a 
current through an exponential, on the other hand, removes only one 
current insertion and furnishes a factor $\a' G^{\mu i} k_i$ 
instead. Since both factors are of the same order in $\a'$, the 
leading contribution to an n-point function is obtained when we 
contract as many pairs of currents as possible, i.e.\ $n/2$ for 
even $n$ and $(n-1)/2$ for odd $n$. In the latter case, the last 
current will necessarily lead to a linear dependence on the 
momentum $k$. 
\smallskip   
     
It follows from these general remarks on the evaluation of n-point 
functions that the three- and four-point functions both contain 
terms which are second order in $\a'$. All higher correlators are 
subleading. While the dominant contributions to the three-point 
function contain a factor linear in the momenta $k$, the corresponding 
factor for the four-point function is independent of $k$. Both these 
factors are finally multiplied with correlators (\ref{oscorr}) of the 
exponential fields. The latter certainly introduce a strong $k$ 
dependence whenever there is a non-vanishing B-field.     
\smallskip 

Our final task now is to interpret the resulting expressions for 
correlators as vertices of some effective low-energy field theory 
on the brane. The structure of the terms we have just outlined 
shares all the essential features with the vertices in Yang-Mills 
theory. For instance, the three gluon vertex is linear in the 
external momenta just as the factor we have talked about in our 
evaluation of the three-point function. The non-vanishing B-field,
however, causes all the vertices to be multiplied by momentum 
dependent phase factors. Hence, after Fourier transformation, 
we expect the fields in the effective field theory to be multiplied 
with the Moyal-product rather than the ordinary point-wise 
multiplication. 
\smallskip 

Though our arguments have been a bit sketchy, all our conclusions can 
be confirmed by an exact computation. Taking the physical state 
conditions into account, the effective action for a stack of $M$ 
branes is indeed given by the Yang-Mills action for fields 
$A_\mu \in \Mat_M({\mbox{\it Fun\/}}(\QR^D))$ on a non-commutative 
$\mathbb{R}^D$ \cite{SeiWit99}, 
\ba
 \cS_N(A) & = & \frac{1}{4} \, \int  d^D x \ \mbox{\rm tr \/} 
   \left( \,  F_{\mu \nu} \ast F^{\mu \nu}\, \right)\nn \\[2mm] 
 \mbox{where} \ \ \  F_{\mu \nu}(A)  & = & 
\partial_\mu A_\nu - \partial_\nu A_\mu + i [A_\mu \astkn A_\nu]\nn 
\ea 
and with $\ast$ being the Moyal product as before. The integration 
extends over the world-volume of the brane. In the formula we suppressed 
fermionic contributions (which appear for superstring theories) and we 
assumed that $D = d$. The action for lower dimensional branes is obtained 
by dimensional reduction. One can easily see that this non-commutative 
Yang-Mills theory is invariant under the following gauge transformations 
$$ A_\mu \ \longrightarrow \ A_\mu + \partial_\mu \lambda + 
  i [ A^\mu \astkn \lambda ] $$ 
for $\lambda \in \Mat_M({\mbox{\it Fun\/}}(\QR^D))$. The relation between 
branes in flat space and non-commutative geometry has been the main 
motivation during the last years to study non-commutative field theories. 
In particular, there has been significant progress in constructing 
their classical solutions (see e.g.\ \cite{NekSch,GoMiSt,GroNek1,Poly,
AgGoMiSt,HaKrLa}). The latter allow for an interpretation as condensates
on branes. We shall come back to related issues in the third lecture when 
we analyse branes on a 3-sphere. For an overview over many of the recent 
developments in this field and other aspects that we have not touched, 
we recommend e.g.\ \cite{SeiWit99,KonSch1,KonSch2,Harvlec,DouNek} and 
references therein.  
\newpage

\section{2D Boundary conformal field theory} 
\setcounter{equation}{0}

We now want to extend the microscopic formalism to branes in 
general backgrounds. The extension relies heavily on methods 
and ideas from conformal field theory. After a brief review
of some basic concepts from bulk conformal field theory, we 
explain how branes can be described through boundary conformal 
field theory. In particular we shall argue that they are  
uniquely characterized by the way in which they couple to closed 
string modes or, in terms of the world-sheet theory, 
by the one-point functions of bulk fields. Using world-sheet 
duality it is then possible to determine the corresponding open 
string spectra. Along the way we shall derive a number of 
algebraic relations for the couplings of closed strings to 
the brane and the interaction of open strings. Universal 
solutions of these relations are the subject of the last 
subsection.  

\subsection{Some background from CFT}   

In an attempt to make these lectures self-contained, we shall 
begin our discussion with some more or less well known material 
on conformal field theory (CFT). Readers who have been exposed 
to CFT may skip most of this subsection where we present some 
notations and the basic data of the bulk theory. These include 
the space of bulk 
fields, the bulk partition function and the operator product expansion. 
In the second subsection we collect some background material on 
chiral algebras which arise as symmetries of 2D bulk and boundary 
CFTs.
 
\subsubsection{The bulk theory.} 

Bulk conformal field theories, i.e.\ 2D CFTs defined on the full 
complex plane, appear in the world-sheet desription of closed strings. 
Their state spaces $\cH^{(P)}$ contain all the closed string modes 
and the coefficients $C = C^{(P)}$ of their operator product expansions 
encode closed string interactions. There exist many non-trivial examples 
of 2D conformal field theories, and hence of exactly solvable string 
backgrounds. These are constructed with the help of certain infinite
dimensional symmetries known as chiral or W-algebras. 

\paragraph{Bulk fields and bulk OPE.} 
\def\bV{{\overset{\underline{\ \, }}{V}}\,\!} 
\def\bcV{{\overset{\underline{\ \, }}{\cV}}\,\!} 
\def\bh{\bar h} 
All constructions of boundary conformal field theories start 
from the data of a usual conformal field theory on the complex 
plane which we shall refer to as {\em bulk theory}. It consists 
of a space $\cH^{(P)}$ of states equipped with the action of a 
Hamiltonian $H^{(P)}$ and of field operators $\varphi(z,\bar z)$. 
According to the famous state-field correspondence, the latter 
can be labeled by elements in the state space $\cH^{(P)}$, 
\be \varphi(z,\bar z) \ = \ \Phi^{(P)}(|\varphi\rangle ;z,\bar z) 
\ \ \ 
   \mbox{ for all } \ \ \ |\varphi\rangle \in \cH^{(P)}\ \ . 
\label{stfldcor}
\ee
The reverse relation is given by $\varphi(0,0) | 0\rangle 
= |\varphi \rangle$ where $|0\rangle$ denotes the unique 
vacuum state in the state space $\cH^{(P)}$ of the bulk 
theory. 
\medskip

Among the fields of a CFT one distinguishes so-called {\em chiral 
fields} which depend on only one of the coordinates $z$ or $\bar z$ 
so that they are either holomorphic, $W = W(z)$, or anti-holomorphic, 
$\bW = \bW(\bar z)$. The most important of these chiral fields, the 
Virasoro fields $T(z)$ and $\bT(z)$, come with the stress tensor and
hence they are present in any CFT. But in most models there exist
further (anti-)\-holomorphic fields whose Laurent modes $W_n$ and 
$\bW_n$ defined through 
\be    W(z) \ = \ \sum \, W_n \ z^{-n-h} \ \ \ , \ \ \ 
  \bW(\bar z) \ = \ \sum \, \bW_n \ \bar z^{-n-\bar h} \ \ , 
\label{modexp}
\ee  
generate two commuting {\em chiral algebras}, $\cW$ and $\bcW$.  
The numbers $h$ and $\bar h$ are the (half-) integer conformal weights 
(scaling dimension) of $W$ and $\bW$. Throughout this text we shall 
assume the two chiral algebras $\cW$ and $\bcW$ to be isomorphic. 
We encountered an example of such a chiral algebra in the first 
lecture. There it was generated by the Laurent modes $\a^\mu_n$ 
and $\bar \a^\mu_n$ of the currents $J(z)$ and $\bJ (\bz)$.  
\smallskip 

In general, the state space $\cH^{(P)}$ of the bulk theory admits a 
decomposition into irreducible representations $\cV_i$ and $\bcV_\bi$ 
of the two commuting chiral algebras, 
\be
 \cH^{(P)} \ = \ {\bigoplus}_{i,\bi} \ n^{(P)}_{i\bi} \ \cV_i \o \bcV_\bi\ \ . 
\label{CSspace} 
\ee
For simplicity we shall assume that the multiplicities satisfy 
$n^{(P)}_{i\bi} \in \{0,1\}$. Higher multiplicities can easily be 
incorporated but they would require additional indices. We reserve 
the label $i=0$ for the vacuum representation $\cV_0$ of the chiral 
algebra. It is mapped to $\cW$ via the state-field correspondence 
$\Phi^{(P)}$. As suggested by our notation, the set of representations 
often turns out to be discrete. This is in contrast to the situation 
we met in our discussion of strings in $\QR^D$ where $i$ ran over the 
continuum of closed string momenta. The discrete sum in eq.\ (\ref{CSspace}) 
signals that our analysis focuses primarily on compact backgrounds, even 
though some of the general ideas apply to non-compact situations as well 
(see last Section).  
\smallskip 

Each irreducible representation $\cV_i$ of $\cW$ acquires an
integer grading under the action of the Virasoro mode $L_0$ and hence 
it may be decomposed as $\cV_i = \bigoplus_{n\geq 0} V_i^n$. The 
subspace $V_i^0$ of ground states in $\cV_i$ carries an irreducible 
action of all the zero modes $W_0$. We will denote the corresponding 
linear maps by $X^i_W$,
\be    X^i_W \ := \ W_0\ |_{V_i^0} : V_i^0 \ \longrightarrow \ 
                 V_i^0 \ \ \ \mbox{ for all chiral fields $W$} 
\ \ . 
\label{Xdef}
\ee       
The whole irreducible representation $\cV^i$ may be recovered 
from the elements of the finite-dimen\-sional subspace $V_i^0$ 
by acting with $W_n,\ n < 0$.   
\smallskip

Using the state-field correspondence $\Phi^{(P)}$, we can assign 
fields to all states in $V_i^0 \o \bV_\bi^0$. We shall assemble 
them into a single object which one can regard as a matrix of 
fields after choosing some basis $|e^a_i\rangle \o |\bar e^b_\bi
\rangle \in V_i^0 \o \bV_\bi^0$, 
\be \varphi_{i,\bi}(z,\bar z) \ : = \ \left( 
   \Phi^{(P)}(|e^a_i\rangle \o |\bar e^b_\bi\rangle ; z,\bar z)
   \ \right) \ \ . 
\label{phiij}
\ee
The matrix elements are labeled by $a = 1,\dots, \dim V_i^0$ and 
$b = 1, \dots, \dim \bV_\bi^0$. We shall refer to these field
multiplets  as closed string vertex operators or {\em primary 
fields}. All other fields in the theory can be obtained by 
multiplying with chiral fields and their derivatives.% 
\medskip%

So far, we have merely talked about the space of bulk fields. But 
more data are needed to characterize a closed string background. 
These are encoded in the short distance singularities of correlation 
functions or, equivalently, in the structure constants of the 
operator product expansions
\be \label{POPE} 
 \varphi_{i,\bi} (z_1,\bz_1) \varphi_{j,\bj}(z_2,\bz_2) \ = \ \sum_{n \bn} 
  \, {C_{i\bi;j\bj}}^{n\bn} \, z_{12}^{h_n-h_i - h_j}  
   \bz_{12}^{\bh_n - \bh_i - \bh_j} 
     \varphi_{n,\bn}(z_2,\bz_2) + \dots \ \ . 
\ee 
Here, $z_{12} = z_1-z_2$ and $h_i, \bh_\bi$ denote the conformal weights 
of the field $\varphi_{i,\bi}$, i.e.\ the values of $L_0$ and $\bL_0$ on 
$V_i^0 \o \bV_i^0$. The numbers $C$ describe the scattering amplitude for 
two closed string modes combining into a single one (``pant diagram'').  
\medskip

Together, the state space (\ref{CSspace}) and the set of couplings 
$C$ in rel.\ (\ref{POPE}) can be shown to specify the bulk theory 
completely. We shall assume that these data are given to us, i.e.\ 
that the closed string background has been solved already.   

\paragraph{Example: The free boson.} Our use of the state field 
correspondence $\Phi$ may seem a bit formal at first, but it can 
clarify some conceptual issues and even simplifies many equations  
later on. Since it certainly takes time to get used to this rather 
abstract formalism, let us pause for a moment and illustrate the 
general concepts in the 
example of a single free boson. In this case, the state space of 
the bulk theory is given by 
\be \label{bulkFB}  
\cH^{(P)} \ = \ \int dk\  \cV_k \otimes \bcV_k \ \ . 
\ee
As long as we do not compactify the theory, there is a continuum of 
sectors parametrized by $i = k = \bi$. In the first lectures we have 
discussed how chiral currents $J(z)$ and the Virasoro field act 
$T(z)$ act on $\cV_k$. The same discussion applies to their 
anti-holomorphic partners. Hence, the formula (\ref{bulkFB}) 
provides a decomposition of the space of bulk fields into irreducible 
representations of the chiral algebra that is generated by the modes
$\a_n$ and $\bar \a_n$. States in $\cV_k \o \bcV_k$ are used to  
describe all the modes of a closed string that moves with center of 
mass momentum $k$ through the flat space. 
\smallskip 

The ground states $|k\rangle \o |k\rangle$ are non-degenerate in this 
case and hence they give rise to a single bulk field $\varphi_{k,k}
(z,\bz)$ for each momentum $k$. These fields are the familiar 
exponential fields, 
$$ \varphi_{k,k}(z,\bz) \ = \ \Phi^{(P)}(|k\rangle \otimes |k\rangle;z,\bz) 
 \ = \ : \exp (i k X(z,\bz)) :\ \ .   $$
Their correlation functions are rather easy to compute (see e.g.\ 
\cite{Pol1Book}). From such expressions one can read off the 
following short distance expansion 
$$ \varphi_{k_1,k_1} (z_1,\bz_1)  \varphi_{k_2,k_2} (z_2,\bz_2) 
  \ \sim \ \int dk \, \delta(k_1+k_2 -k) \, |z_1-z_2|^{\ap 
     (k^2_1 + k^2_2 - k^2)} \varphi_{k,k}(z_2,\bz_2) + \dots 
\ \ . 
$$ 
Comparison with our general form (\ref{POPE}) of the operator 
products shows that the coefficients $C$ are simply given by 
the $\delta$ function which expresses momentum conservation. 
Note that the exponent and the coefficient of the short distance 
singularity are a direct consequence of the equation of motion 
$\Delta X(z,\bz) = 0$ for the free bosonic field. In fact, the 
equation implies that correlators of $X$ itself possess the 
usual logarithmic singularity when two coordinates approach 
each other. After exponentiation, this gives rise to the 
leading term in the operator product expansion of the fields  
$\varphi_{k,k}$. In this sense, the short distance singularity 
encodes the dynamics of the bulk field and hence characterizes 
the background of the model.  
  
\subsubsection{Chiral algebras}  

Chiral algebras can be considered as symmetries of 2D conformal 
field theory. Since they play such a crucial role for all exact
solutions, we shall briefly go through the most important 
notions in the representation theory of chiral algebras. These
include the set $\cJ$ of representations, the modular S-matrix
$S$, the fusion rules $N$ and the fusing matrix $F$. The general 
concepts are illustrated in the case of the U(1)-current algebra. 
    
\paragraph{Representation theory.} Chiral- or W-algebras are 
generated by the modes of a finite set of chiral fields $W^\nu_n$.
These algebras mimic the role played by Lie algebras in 
atomic physics. Recall that transition amplitudes in atomic 
physics can be expressed as products of Clebsch-Gordan 
coefficients and so-called reduced matrix elements. While the 
former are purely representation theoretic data which depend only on 
the symmetry of the theory, the latter contain all the information 
about the physics of the specific system. Similarly, amplitudes in 
conformal field theory are built from representation theoretic 
data of W-algebras along with structure constants of the various
operator product expansions, the latter being the reduced matrix 
elements of conformal field theory. In the (rational) conformal 
bootstrap, the structure constants are determined as solutions of 
certain algebraic equations which arise as factorization constraints 
and we will have to say a lot more about such equations as we proceed. 
Constructing the representation theoretic data, on the other hand, is 
a mathematical problem which is the same for all models that possess
the same W-symmetry. Throughout most of the following text we shall 
not be concerned with this part of the analysis and simply use the 
known results. But we decided to include at least a short general 
review on representation theory of W-algebras.  
\smallskip

We consider a finite number of bosonic chiral fields $W^\nu(z)$ with 
positive integer conformal dimension $h_\nu$ and require that there 
is one distinguished chiral field $T(z)$ of conformal dimension $h 
= 2$ whose modes $L_n$ satisfy the usual Virasoro relations for 
central charge $c$. Their commutation relations with the Laurent 
modes $W^\nu_n$ of $W^\nu(z)$ are assumed to be of the form  
\be  [L_n , W^\nu_m ] \ = \ \bigl( n(h_\nu -1) - m \bigr) \, 
W^\nu_{n+m}\ \ .  \label{LWcomm} \ee
In addition, the modes of the generating chiral fields also
possess commutation relations among each other which need not 
be linear in the fields. The algebra generated by the modes 
$W^\nu_n$ is the chiral or W-algebra $\cW$ (for a precise 
definition and examples see \cite{SchBou} and in particular 
\cite{Bonn}). We shall also demand that $\cW$ comes equipped 
with a $\ast$-operation.  
\smallskip 

Sectors $\cV_i$ of the chiral algebra are irreducible (unitary) 
representations of $\cW$ for which the spectrum of $L_0$ is bounded 
from below. Our requirement on the spectrum of $L_0$ along with 
the commutation relations (\ref{LWcomm}) implies that any $\cV_i$ 
contains a sub-space $V_i^0$ of ground states which are annihilated 
by all modes $W^\nu_n$ such that $n>0$. The spaces $V_i^0$ carry an 
irreducible representation of the zero mode algebra $\cW^0$, i.e.\ 
of the algebra that is generated by the zero modes $W^\nu_0$, and we 
can use the operators $W^\nu_n, n<0,$ to create the whole sector 
$\cV_i$ out of states in $V_i^0$. Unitarity of the sectors means 
that the space $\cV_i$ may be equipped with a non-negative bi-linear 
form which is compatible with the $\ast$-operation on $\cW$. This 
requirement imposes a constraint on the allowed representations of 
the zero mode algebra on ground states. Hence, one can associate a 
representation $V_i^0$ of the zero mode algebra to every sector 
$\cV_i$, but for most chiral algebras the converse is not true.
In other words, the sectors $\cV_i$ of $\cW$ are labeled by elements 
$i$ taken from a subset $\cJ$ within the set of all irreducible 
(unitary) representations of the zero mode algebra. 
\smallskip 

For a given sector $\cV_i$ let us denote by $h_i$ the lowest eigenvalue 
of the Virasoro mode $L_0$. Furthermore, we introduce the {\em character}
$$   \chi_i(q) \ = \ \mbox{\rm tr}_{\cV_i}\, \bigl( q^{L_0 - \frac{c}{24}}
  \bigr)  
 \ \ . $$ 
The full set of these characters $\chi_i, i \in \cJ,$ has the remarkable 
property to close under modular conjugation, i.e.\ there exists a complex 
valued matrix $S = (S_{ij})$ such that 
\be \chi_i(\tq ) \ = \ S_{ij} \, \chi_j(q)  \  \ 
\label{Smatrix}
\ee 
where $\tq = \exp (-2\pi i/\tau)$ for $q = \exp(2\pi i\tau)$, as before. 
This {\em S-matrix} is symmetric and unitary. Once $S$ has been 
constructed for a given chiral algebra $\cW$, one can introduce the 
numbers 
\be \label{Verl} {N_{ij}}^k \ = \ \sum_l 
\frac{S_{il} S_{jl} S^*_{kl}}{S_{0l}} \ \ . \ee 
Quite remarkably, they turn out to be non-negative integers. This 
property, however, possesses a direct explanation in representation 
theory. In fact, there exists a product $\circ$ of sectors -- known 
as the fusion product -- such that ${N_{ij}}^k$ describe analogues of 
the Clebsch-Gordan multiplicities for the decomposition of $i \circ j$ 
into the irreducible sectors $k$ \cite{Verl}. For this reason, we refer 
to $N$ as the {\em fusion rules} of $\cW$. The relation (\ref{Verl}) 
between the fusion rules $N$ and the matrix $S$ is called the {\em 
Verlinde formula}. 
\medskip 

The {\em fusing matrix} $F$ is the last quantity in the representation 
theory of chiral algebras which plays an important role below. 
Unfortunately, it is not as easy to describe. To begin with, let 
us be a bit more explicit about the fusion product. Its definition 
is based on the following family of homomorphisms (see e.g.\ 
\cite{MooSei2}) 
\ba \delta_z ( W^\nu_n ) & := & e^{-z L_{-1}} W^\nu_n e^{zL_{-1}} \o 
   {\bf 1}  + {\bf 1} \o W^\nu_n \nn \\[2mm] & = &  
   \sum_{m=0} \, \left( \begin{array}{c} h_\nu + n - 1 \\ m \end{array} 
   \right)\, z^{n + h_\nu -1-m} \, W^\nu_{1+m-h_\nu} 
   \o {\bf 1} + {\bf 1} \o W^\nu_n 
\ea
which is defined for $n > - h_\nu$. The condition on $n$ guarantees 
that the sum on the right hand side terminates after a finite number 
of terms. Suppose now that we are given two sectors $\cV_j$ and 
$\cV_{i}$. With the help of $\delta_z$ we define an action of 
the modes $W^\nu_n, n > - h_\nu,$ on their product. This action can 
be used to search for ground states and hence for sectors $k$ in 
the fusion product $j \circ i$. To any three such labels $j,i,
k$ there is assigned an intertwiner 
$$V \vvert{\ j}{k}{i}(z): \ \cV_{j} \o \cV_{i} \ \rightarrow 
\ \cV_{k}$$ 
which intertwines between the action $\delta_z$ on the product and 
the usual action on $\cV_{k}$. If we pick an orthonormal basis 
$\{ |j,\nu \rangle \}$ of vectors in $\cV_{j}$ we can represent the 
intertwiner $V$ as an infinite set of operators 
$$V  \vvert{j,\nu}{k}{i}(z) \ := \ V  \vvert{\ j}{k}{i}
[|j,\nu\rangle;\, \cdot\  ](z)\, : \ \cV_{i}\  \rightarrow \ \cV_{k}
\ \ . $$ 
Up to normalization, these operators are uniquely determined by the
intertwining property mentioned above. The latter also restricts their 
operator product expansions to be of the form 
$$ 
V \vvert{j_1,\mu}{k}{r}(z_1)\  V \vvert{j_2,\nu}{r}
{i}(z_2) \ = \ \sum_{s,\rho}\ \Fus{r}{s}{j_2}{j_1}
{i}{k} \ V \vvert{s,\rho}{k}{i}(z_2) \ \langle s, 
\rho| V \vvert{j_2,\nu}{s}{j_1}(z_{12})|j_1,\mu\rangle \ \ , 
$$ 
where $z_{12} = z_1-z_2$. The coefficients $F$ that appear in this 
expansion form the {\em fusing matrix} of the chiral algebra $\cW$.  
Once the operators $V$ have been constructed for all ground states 
$|j,\nu\rangle$, the fusing matrix can be read off from the leading 
terms in the expansion of their products. Explicit formulas can be 
found in the literature. We also note that the defining relation 
for the fusing matrix admits a nice pictorial presentation (see 
Figure 1). It presents the fusing matrix as a close relative of 
the $6J$-symbols which are known from the representation theory 
of finite dimensional Lie algebras.  

\vspace{1cm}
\fig{Graphical description of the fusing matrix. All the lines are
directed as shown in the picture. Reversal of the orientation can 
be compensated by conjugation of the label. Note that in our 
conventions, one of the external legs is oriented outwards. This 
will simplify some of the formulas below.}
{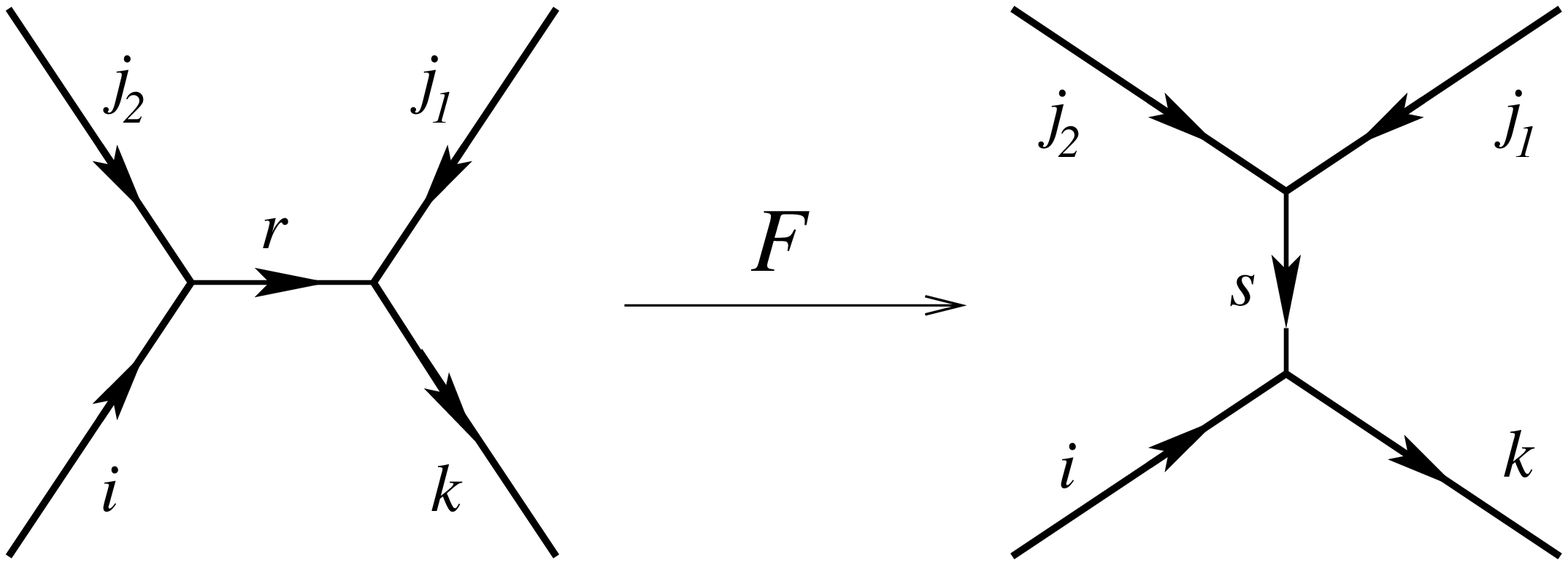}{13truecm}
\figlabel{\basic}
\vspace{1cm}

\paragraph{Example: the U(1)-theory.} The chiral algebra of a single 
free bosonic field is known as U(1)-algebra. It is generated by the 
modes $\a_n$ of the current $J(z)$ with the reality condition $\a_n^* 
= \a_{-n}$. There is only one real zero mode $\a_0 = \a_0^*$ so that 
the zero mode algebra $\cW^0$ is abelian. Hence, all its irreducible 
representations are 1-dimensional and there is one such representation 
for each real number $k$. The vector that spans the corresponding 
1-dimensional space $V_k^0$ is denoted by $|k\rangle$, as before. It 
is easy to see that the space $\cV_k$ which we generate out of 
$|k\rangle$ by the creation operators $\a_{-n}$ admits a positive 
definite bilinear form 
for any choice of $k$. Hence, $\cJ = \mathbb{R}$ coincides with the 
set of irreducible representations of the zero mode algebra in this 
special case. 
\smallskip 

The character $\chi_k$ of the sector $\cV_k$ with conformal weight 
$h_k = \a' k^2/2$ is given by 
$$ \chi_k(q) \ = \ \frac{1}{\eta(q)}\ q^{\a' \frac{k^2}{2}} \ \ .$$
Along with the well known property $\eta(\tilde q) = \sqrt{-i \tau} 
\eta(q)$, the computation of a simple Gaussian integral shows that 
\be  \label{U1S}  
\chi_k(\tq) \ = \ \sqrt{\a'}\, \int dk' \, e^{2\pi i\, \a' kk'}\, 
  \chi_{k'}(q) \ = :\ \sqrt{\a'}\, \int dk' 
  \, S_{kk'} \chi_{k'}(q) \ \ .  
\ee  
This means that the entries of the S-matrix are phases, i.e.\ 
$S_{kk'} = \exp (2 \pi i \a' k k')$. As we have claimed, the S-matrix
is unitary and  symmetric under exchange of $k$ and $k'$. When we 
insert the matrix elements $S_{kk'}$ into the right hand side of 
Verlinde's formula (\ref{Verl}) we find 
\be \label{U1VF} 
 \frac{1}{\sqrt{\a'}}\, \delta(k_1+ k_2-k) \ = \ \sqrt{\a'}\, \int d l 
  \, e^{2 \pi i \a' k_1 l} \, e^{2 \pi i
  \a' k_2 l} \,  e^{-2 \pi i \a' k l} \ \ . \ee 
This is a continuum version of the Verlinde formula. We want to 
demonstrate that the left hand side is indeed related to the fusion 
of representations. In the case of hand, the action of $\delta_z$ on 
the zero mode $\a_0$ is given by 
$$ \delta_z(\a_0) \ = \ \a_0 \o {\bf 1} + {\bf 1} \o \a_0 $$ 
since the current $J$ has conformal weight $h=1$. This shows that 
the fusion product amounts to adding the momenta, i.e.\ $k_1 \circ 
k_2 = k_1 + k_2$. In other words, the product of two sectors $k_1$ 
and $k_2$ contains a single sector $k_1 + k_2$. This indeed agrees 
with the left hand side of the Verlinde formula (\ref{U1VF}). 
\smallskip 

Let us conclude this example with a few comments on the fusing
matrix. In this case it is rather easy to write down an explicit 
formula for the intertwining operators $V$. Once more, they are 
given by the normal ordered exponential, restricted to the spaces 
$\cV_{k}$. When the operator product of two such exponentials 
with momenta $k_1$ and $k_2$ is expanded in the distance $z_1-z_2$,
we find an exponential with momentum $k_1 + k_2$. The coefficient
in front of this term is trivial, implying triviality for the fusing 
matrix.

\subsection{Boundary theory - the closed string sector}

Our goal now is to place a brane into some given background. 
We shall argue that such branes are completely characterized 
by their couplings to closed string modes, i.e.\ by one-point 
functions of bulk fields on the upper half plane. The number 
of couplings one has to specify depends on the exact symmetry 
the brane preserves. From the so-called cluster property we 
shall derive a set of quadratic {\em factorization 
constraints} on the one-point functions which must be 
satisfied by any consistent boundary 
theory.  

\paragraph{Branes - the microscopic setup.}
With some basic notations for the (``parent'') bulk theory set up, we 
can begin our analysis of {\em associated} boundary theories (``open 
descendants''). These are conformal field theories on the upper 
half-plane $\Im z \geq 0$ which, in the interior $\Im z>0$, are 
locally equivalent to the given bulk theory: The state space 
$\cH^{(H)}$ of the boundary CFT is equipped with the action of 
a Hamiltonian $H^{(H)}$ and of bulk fields 
$$ \phi(z,\bar z) \ = \ \Phi^{(H)}(|\varphi\rangle; z,\bz)   
$$ 
-- well-defined for $\Im z >0$ --  which are assigned to the {\em same} 
elements $\varphi \in \cH^{(P)}$ that were used to label fields in the 
bulk theory. Note, however, that the associated fields $\phi$ now act 
on a different space of states $\cH^{(H)}$ and that, for the moment, 
we do not know any fields that are associated with the elements of 
$\cH^{(H)}$. Furthermore, we demand that all the leading terms in 
the OPEs of bulk fields coincide with the OPEs (\ref{POPE}) in the bulk 
theory, i.e.\ for the fields $\phi_{i,\bi}$ one has
\be \label{HOPE} 
 \phi_{i,\bi} (z_1,\bz_1) \phi_{j, \bj}(z_2,\bz_2) \ = \ \sum_{n \bn} 
  \, {C_{i\bi;j\bj}}^{n\bn} \, z_{12}^{h_n-h_i - h_j}  
   \bz_{12}^{\bh_n-\bh_i - \bh_j} 
     \phi_{n,\bn}(z_2,\bz_2) + \dots 
\ee 
These relations express our condition that the brane is placed into 
our given closed string background. At the example of the free bosonic
field we have discussed that the structure of the short distance 
expansion encodes the world-sheet dynamics in the interior of the 
upper half plane. Having the same singularities as in the bulk 
theory means that the boundary conditions do not affect the 
equations of motion in the bulk. 
\smallskip 

In addition, we must require the boundary theory to be conformal. This 
is guaranteed if the Virasoro field obeys the following gluing condition 
\be \label{glueT} 
 T (z) \ = \ \bT (\bz) \ \ \ \ \mbox{ for } \ \ \ \
  z = \bz  \ \ . 
\ee
In the 2D field theory, this condition guarantees that there is 
no momentum flow across the boundary. Note that eq.\ (\ref{glueT}) 
is indeed satisfied for the Virasoro fields in the flat space 
theory (see rel.\ (\ref{glueTflat})).  
\smallskip 

Considering {\em all} possible conformal boundary theories associated to 
a bulk theory whose chiral algebra is a true extension of the Virasoro 
algebra is, at present, too difficult a problem to be addressed 
systematically (see however \cite{GaReWa,GabRec,MaMoSe1,QueSch1} and 
remarks in the final section for some recent progress). For the 
moment, we restrict our considerations to maximally symmetric boundary 
theories, i.e.\  to the class of boundary 
conditions which leave the whole symmetry algebra $\cW$ unbroken. More 
precisely, we assume that all chiral fields $W(z), \bW(\bz)$ can be 
extended analytically to the real line and that there exists a local 
automorphism $\Omega$ -- called the {\em gluing map} -- of the chiral 
algebra $\cW$ such that \cite{RecSch1} 
\be   
   W(z) \ = \ \Omega\bW (\bar z) \ \ \mbox{ for } \ \ z = \bar z\ \ . 
\label{gluecond}
\ee
The condition (\ref{glueT}) is included in equation (\ref{gluecond}) 
if we require $\Omega$ to act trivially on the Virasoro field. Note 
that the boundary conditions we considered in the first lecture are 
maximally symmetric since holomorphic and anti-holomorphic currents 
are glued along the boundary according to eqs.\ (\ref{DBC}) and 
(\ref{glueN}).   
\smallskip

For later use let us remark that the gluing map $\Omega$ on the 
chiral algebra induces a map $\omega$ on the set of sectors. In 
fact, since $\Omega$ acts trivially on the Virasoro modes, and 
in particular on $L_0$, it may be restricted to an automorphism
of the zero modes in the theory. If we pick any representation 
$j$ of the zero mode algebra we can obtain a new representation 
$\omega(j)$ by composition with the automorphism $\Omega$. This
construction lifts from the representations of $\cW^0$ on ground 
states to the full $\cW$-sectors. As a simple example consider 
the U(1) theory with the Dirichlet gluing map $\Omega (\a_n) = 
- \a_n$. We restrict the latter to the zero mode $\a_0$. 
As we have explained above, different sectors are labeled by the 
value $\sqrt{\a'}k$ of $\a_0$ on the ground state $|k\rangle$. If 
we compose the action of $\a_0$ with the gluing map $\Omega$, we 
find $\Omega (\a_0) |k\rangle = - \sqrt{\a'} k |k\rangle$. This 
imitates the action of $\a_0$ on $|\overline{\phantom{,, }}\, k
\rangle$. 
Hence, the map $\omega$ is given by $\omega(k) = - k$.

\paragraph{Ward identities.} 
As an aside, we shall discuss some more technical consequences
that our assumption on the existence of the gluing map $\Omega$ 
brings about. To begin with, it gives rise to an action of one 
chiral algebra $\cW$ on the state space $\cH\equiv \cH^{(H)}$
of the boundary theory. Explicitly, the modes $W_n = W_n^{(H)}$ 
of a chiral field $W$ dimension $h$ are given by 
$$ W_n \ := \ \frac{1}{2\pi i} \int_C \, z^{n+h-1}\, W(z) 
   \, dz + 
   \frac{1}{2\pi i} \int_C \, \bz^{n+h-1} \, \Omega \bW(\bz) \, 
   d \bz $$ 
which generalizes the formulas (\ref{a1}), (\ref{a2}) from the first 
lecture. The operators $W_n$ on the state space $\cH$ are easily 
seen to obey the defining relations of the chiral algebra $\cW$.
Note that there is just one such action of $\cW$ constructed
out of the two chiral bulk fields $W(z)$ and $\Omega \bW(\bar z)$. 
\medskip

In the usual way, the representation of $\cW$ on $\cH$ leads 
to Ward identities for correlation functions of the boundary 
theory. They follow directly from the singular parts of the 
operator product expansions of the fields $W, \Omega \bW$ with 
the bulk fields $\phi(z,\bar z)$. These expansions are fixed 
by our requirement of local equivalence between the bulk theory 
and the bulk of the boundary theory. To make this more precise,  
we introduce the notation $W_>(z) = \sum_{n > -h} W_n z^{-n-h}$. 
The singular part of the OPE is then given by  
\ba 
{\lefteqn{ \left(W(w) \, \phi(z,\bar z)\right)_{\rm sing} \ := \  
[\, W_>(w)\, ,\, \Phi(|\varphi\rangle ;z,\bar z)\, ]}} \label{WOPE}\\[2mm] 
& = & \sum_{n > -h}  \left( \frac{1}{(w-z)^{n+h}} \, \Phi\bigl(W^{(P)}_n 
 |\varphi\rangle ; z, \bar z\bigr) + \frac{1}{(w-\bar z)^{n+h}} \, 
  \Phi\bigl(\Omega\bW{}^{(P)}_n |\varphi\rangle ; z,\bar z\bigr)
  \right) \ \ . \nn 
\ea   
As before, the subscript `sing' reminds us that we only look at 
the singular part of the operator product expansion, and we have 
placed a superscript $(P)$ on the modes $W_n, \bW_n$ to display 
clearly that they act on the elements $|\varphi\rangle \in \cH^{(P)}$ 
labeling the bulk fields in the theory (superscripts  $(H)$, on the 
other hand, are being dropped for most of our discussion). The sum on 
the right hand side of eq.\ (\ref{WOPE}) is always finite because 
$|\varphi\rangle$ is annihilated by all Laurant modes with sufficiently 
large $n$. For $\Im\, w >0$, only the first terms involving $W{}^{(P)}_n$ 
can become singular and the singularities agree with the singular part 
of the OPE between $W(w)$ and $\phi(z,\bar z)$ in the bulk theory. 
Similarly, the singular part of the OPE between $\Omega \bW(w)$ and 
$\phi(z,\bar z)$ in the bulk theory is reproduced by the terms which 
contain $\bW{}^{(P)}_n$, if $\Im\, w < 0$.    
\smallskip

As it stands, the previous formula is rather compressed. So, 
let us spell out at least one more concrete example in which 
the chiral field $W$ has dimension $h=1$  (we shall denote 
any such chiral currents by the letter $J$) and where we 
consider the primaries $\phi_{i,\bi}$ in place of $\phi$. 
Since the corresponding ground states are annihilated 
by all the modes $J_n, \bJ_n$ with $n > 0$, equation (\ref{WOPE}) 
reduces to      
\ba 
\left(J(w) \, \phi_{i,\bi}(z,\bar z)\right)_{\rm sing} \ = \  
 \frac{ X_J^i }{w-z} \, \phi_{i,\bi}(z,\bar z)  \ - \ 
 \varphi_{i,\bi}(z,\bar z) \, \frac{X_{\Omega \bJ}^\bi}{w-\bar z}  \ \ .  
\label{JOPE} 
\ea
The linear maps $X_J^i$ and $X_{\Omega \bJ}^\bi$ were introduced 
in eq.\ (\ref{Xdef}) above; they act on the primary multiplet 
$\phi_{i,\bi}: V_\bi^0 \o \cH \rightarrow V_i^0 \o \cH$ by 
contraction in the first component $V_i^0$ resp.\ 
$V_\bi^0$.         
\smallskip

Ward identities for arbitrary $n$-point functions of fields
$\phi_{i,\bi}$ follow directly from eq.\ (\ref{WOPE}). They have 
the same form as those for chiral conformal blocks in a bulk CFT 
with $2n$ insertions of chiral vertex operators with charges 
$i_1,\dots,i_n,\omega(\bi_1),\dots,\omega(\bi_n)$, see e.g.\ 
\cite{Card84,Card86,RecSch1,FucSch2}. Hence, objects familiar 
from the construction of bulk CFT can be used as building blocks 
of correlators in the boundary theory (``doubling trick''). Note, 
however, that the Ward identities depend on the gluing map 
$\Omega$.

\paragraph{One-point functions.} 
So far we have formalized what it means in world-sheet terms 
to place a brane in a given background (the principle of `local 
equivalence') and how to control its symmetries through gluing 
conditions (\ref{gluecond}) for chiral fields. Now 
it is time to derive some consequences and, in particular, to 
show that a rational boundary theory is fully characterized by 
just a finite set of numbers. 
\smallskip

Using the Ward identities described in the previous paragraph 
together with the OPE (\ref{HOPE}) in the bulk, we can reduce 
the computation of correlators involving $n$ bulk fields to the 
evaluation of one-point functions $\langle \phi_{i,\bi}\rangle_\a$
for the bulk primaries (see Figure 2). Here, the subscript $\a$ 
has been introduced to label different boundary theories that can 
appear for given gluing map $\Omega$.  
\vspace{1cm} 
\fig{With the help of operator product expansions in 
the bulk, the computation of $n$-point functions in a 
boundary theory can be reduced to computing one-point 
functions on the half-plane. Consequently, the latter 
must contain all information about the boundary 
condition.} 
{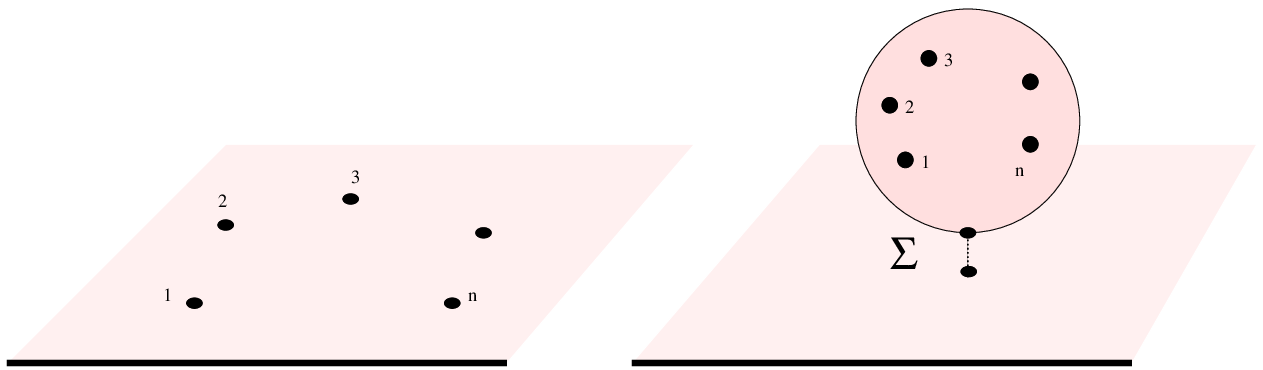}{14truecm} 
\figlabel{\basic}
\vspace{.5cm}

To control the remaining freedom, we notice that the transformation 
properties of $\phi_{i,\bi}$ with respect to $L_n,\ n= 0, \pm 1,$ 
and the zero modes $W_0$,  
$$ [\, W_0\, , \, \phi_{i,\bi}(z,\bar z)\, ] \ = \ 
    X_W^i\ \phi_{i,\bi}(z,\bar z) \ - \ \phi_{i,\bi}(z,\bar z) 
  \ X_{\Omega \bW}^\bi \ \ , $$
\ba 
 [\, L_n\, , \, \phi_{i,\bi}(z,\bar z)\, ] & = & 
   z^n\, ( \, z \partial + h_i(n+1)\, ) \phi_{i,\bi}(z,\bar z)   
  \nn \\[2mm] & & \hspace*{5mm} + \,  
   \bar z^n\, ( \, \bar z {\bar \partial} + \bar h_\bi(n+1)\, ) 
    \phi_{i,\bi}  (z,\bar z) \ \   \nn
\ea
determine the one-point functions up to scalar factors. Indeed, 
an elementary computation using the invariance of the vacuum 
state reveals that the vacuum expectation values $\langle 
\phi_{i,\bi} \rangle_\a$ must be of the form 
\be
 \langle \phi_{i,\bi} (z,\bar z) \rangle_\alpha  =  
 \frac{\cA^\alpha_{i\bi}}{|z - \bar z|^{h_i + h_\bi}}\ \ 
 \label{1ptfct}  
\ee
where 
$$
\cA^\a_{i\bi}: V_\bi^0 \rightarrow V_i^0  \quad\quad \mbox{ obeys } \quad
 \ \ \ X_W^i \ \cA^\a_{i\bi} \ = \ \cA^\a_{i\bi} \ X_{\Omega \bW}^\bi \ \ . 
$$
The intertwining relation in the second line implies $\bi = 
\omega(i^+) \equiv i^\omega$ as a necessary condition for a 
non-vanishing one-point function ($i^+$ denotes the representation 
conjugate to $i$, i.e.\ the unique respresentation which obeys 
$N_{ii^+}^0 = 1$), and since $h_i = h_{i^\omega}$ we can put  $h_i + 
h_\bi= 2 h_i $ in the exponent in eq.\ 
(\ref{1ptfct}). From the irreducibility of the zero mode 
representations on the subspaces $V_i^0$ and Schur's lemma we 
conclude that each non-zero matrix $\cA^\a_{i\bi}$ is determined up 
to one scalar factor $A^\a_i$, 
$$ \cA^\a_{i\bi} \ = \ A^\a_{i} \, \delta_{\bi,i^\omega}\, \cU_{i\bi} $$ 
where $\cU_{i\bi}$ intertwines between two representations of the zero 
mode algebra and is normalized by $\cU_{i\bi}^*\, \cU_{i\bi} = {\bf 1}$. 
In conclusion, we have argued that boundary conditions associated with the 
same bulk theory and the same gluing map $\Omega$, can differ only 
by a set of scalar parameters $A^\a_i$ in the one-point functions. Once we 
know their values, we have specified the boundary theory. This generalizes 
a similar observation we made for branes in flat backgrounds 
(see remark after eq.\ (\ref{D1pt})) and it also agrees with our intuition 
that a brane should be completely characterized by its couplings to closed 
string modes such as the mass and RR charge. 
\smallskip

\paragraph{The cluster property.}  
We are certainly not free to choose the remaining parameters $A^\a_{i}$ 
in the one-point functions arbitrarily. In fact, there exist strong {\em 
sewing constraints} on them that have been worked out by several authors 
\cite{CarLew,Lewe1,PrSaSt1,PrSaSt2,BPPZ1}. These can be derived from the 
following {\em cluster property} of the two-point functions 
\be \lim_{a \ra \infty} \ \langle 
    \phi_{i,i^\omega} (z_1,\bz_1) 
\phi_{j,j^\omega} (z_2+a, \bz_2+a) \rangle 
    \ = \  \langle \phi_{i,i^\omega}(z_1,\bz_1) \rangle \langle 
   \phi_{j,j^\omega} (z_2,\bz_2)\rangle \ \ . \label{cluster}
\ee 
Here, $a$ is a real parameter, and the field $\phi_{j,j^\omega}$ 
on the right hand side can be placed at $(z_2,\bz_2)$ since the 
whole theory is invariant under translations parallel to the 
boundary. 
\smallskip

Let us now see how the cluster property restricts the choice of  
possible one-point functions. We consider the two-point function 
of the two bulk fields as in eq.\ (\ref{cluster}). There are two 
different ways to evaluate this function. On the one hand, we can 
go into a regime where the two bulk fields are very far from each 
other in the direction along the boundary. By the cluster property, 
the result can be expressed as a product of two one-point functions 
and it involves the product of the couplings $A^\a_i$ and $A^\a_j$. 
Alternatively, we can pass into a regime in which the two bulk 
fields are very close to each other and then employ the operator 
product (\ref{HOPE}) to reduce their two-point function to a sum 
over one-point functions. Comparison of the two procedures provides 
the following important relation, 
\be A^\a_i \, A^\a_j \ = \ \sum_k \ \Xi^{k}_{ij} \ 
   A^\alpha_0 \ A^\a_k 
  \ \ .  \label{class} \ee
It follows from our derivation that the coefficient $\Xi^k_{ij}$ 
can be expressed as  a combination 
\be \Xi^k_{ij} \ = \ {C_{i\bi;j\bj}}^{k\bar k} \ \Fus{1}{k}{\ i}{\ j}
{\omega(i^+)}{\omega(j)} \label{XiCF} \ee 
of the coefficients $C$ in the 
bulk OPE and of the fusing matrix. The latter arises when we pass 
from the regime in which the bulk fields are far apart to the regime 
in which they are close together (see Figure 3).  
\vspace{1cm}
\fig{Equations (\ref{class}), (\ref{XiCF}) are derived by comparing 
two limits of the two-point function. The dashed line represents the 
boundary of the world-sheet and we have drawn the left moving sector 
in the lower half-plane (doubling trick). 
%The figure shows that the structure 
%constants $\Xi$ contain a particular element of the fusing 
%matrix along with the OPE coefficients $C$.
}
{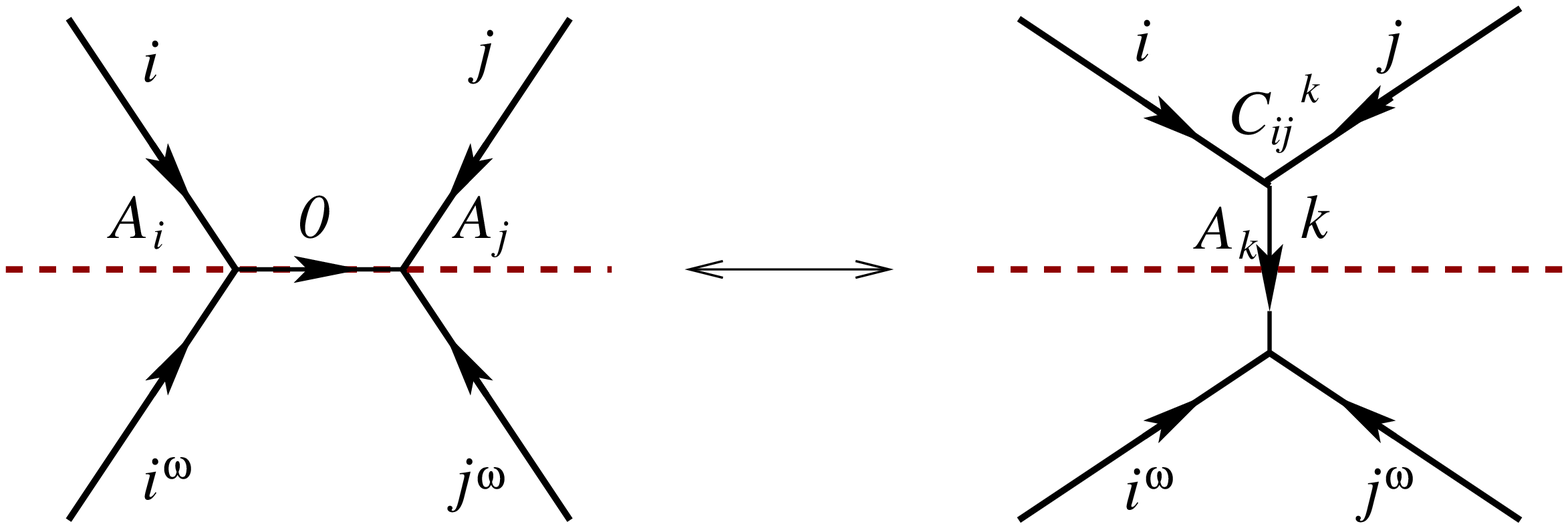}{13truecm}
\figlabel{\basic}
\vspace{1cm}
In some cases, 
$\Xi_{ij}^k$ has been shown to agree with the fusion multiplicities 
or some generalizations thereof (see e.g.\ \cite{PrSaSt1,FucSch1,BPPZ1}). 
The importance of eq.\ (\ref{class}) for a 
classification of boundary conformal field theories has been stressed 
in a number of publications \cite{FucSch1,BPPZ1,RecSch2} and 
is further supported by their close relationship with algebraic 
structures that entered the classification of bulk conformal field 
theories already some time ago (see e.g.\ \cite{Pasq,PetZub1,PetZub2}). 
\smallskip

The algebraic relations (\ref{class}) typically possess several 
solutions which are distiguished by our index $\a$. Hence, maximally 
symmetric boundary conditions are labeled by pairs $(\Omega,\alpha)$. 
The automorphism $\Omega$ is used to glue holomorphic and 
anti-holomorphic fields along the boundary and the consistent choices 
for $\Omega$ are rather easy to classify. Once $\Omega$ has been fixed, 
it determines the set of bulk fields that can have a non-vanishing 
one-point function and it is also referred to as the `type' of the 
boundary theory. For each gluing automorphism $\Omega$, the non-zero 
one-point functions are constrained by algebraic equations (\ref{class}) 
with coefficients $\Xi$ which are determined by the closed string 
background. A complete list of solutions is available in a large 
number of cases. But before presenting them, we want to show 
how one can reconstruct other important information on the brane 
from the couplings $A^\a_i$. In particular, we will be able to 
recover the open string spectrum. Since the derivation makes 
use of {\em boundary states}, we need to introduce this concept 
first.

\paragraph{Boundary states.} 
It is possible to store all information about the couplings $A^\a_i$ 
of closed strings to a brane in a single object, the so-called {\em 
boundary state}. To some extent, such a boundary state can be 
considered as the wave function of a closed string that is sent off 
from the brane $(\Omega,\alpha)$. It is a special linear combinations 
of generalized coherent states (the so-called Ishibashi states). The 
coefficients in this combination are essentially the closed string 
couplings $A^\a_i$. 
\smallskip

One way to introduce boundary states is to equate correlators of bulk 
fields on the half-plane and on the complement of the unit disk in the 
plane. With $z, \bar z$ as before, we introduce coordinates $\zeta, 
\bar\zeta$ on the complement of the unit disk by 
\be
\zeta \ =\  {\frac{1-iz}{1+i z}}\quad\quad {\rm and}  \quad\quad
\bar\zeta \ =\  \frac{1+i\bar z}{1-i \bar z}\ \ . 
\label{coortrsfzeta}\ee
If we use $\,|0\rangle$ to denote the vacuum of the bulk CFT, then 
the boundary state $|\alpha\rangle = |\alpha\rangle_\Omega$ can be 
uniquely characterized by \cite{CarLew,RecSch1}
\be
 \langle\,  \Phi^{(H)}(|\varphi\rangle;z,\bar z) \rangle_{\alpha}
\ =\ \left(\frac{d\zeta}{dz}\right)^h \left(\frac{d\bar \zeta}{d
\bar z}\right)^{\bar h}  \cdot  
\langle 0|\, \Phi^{(P)}(|\varphi\rangle; \zeta,\bar\zeta) | \alpha 
\rangle\label{bstdef}
\ee
for primaries $|\varphi\rangle$ with conformal weights $(h,\bar h)$.   
Note that all quantities on the right hand side are defined in the 
bulk conformal field theory (super-script P), while objects on the 
left hand side live on the half-plane (super-script H).  
\smallskip

In particular, we can apply the coordinate transformation from $(z,\bar z)$ 
to $(\zeta,\bar \zeta)$ on the gluing condition (\ref{gluecond}) to 
obtain
$$ W(\zeta) \ = \ (-1)^h \, \bar \zeta^{2h} \Omega \bW(\bar \zeta) $$ 
along the boundary at $\zeta \bar \zeta = 1$. Expanding this into modes,
we see that the gluing condition (\ref{gluecond}) for chiral fields 
translates into the following linear constraints for the boundary 
state, 
\be
\bigl[\,W_n-(-1)^{h_W}\Omega\bW_{-n}\,\bigr]\,
|\alpha\rangle_{\Omega}\ =\ 0\ .
\label{glueconplane}
\ee
These constraints posses a linear space of solutions. It is spanned 
by generalized coherent (or Ishibashi) states $|i\rangle\! \rangle$.  
Given the gluing automorphism $\Omega$, 
there exists one such solution for each pair $(i,\omega(i^+))$ 
of irreducibles that occur in the bulk Hilbert space \cite{Ishi1}. 
$|i\rangle\!\rangle_{\Omega}$ is unique up to a scalar factor which
can be used to normalize the Ishibashi states such that 
\be
 {}_{\Omega}\langle\!\langle j |\,  \tilde q^{L_0^{(P)}-\frac{c}{24}} 
\,|i\rangle\!\rangle_{\Omega} \ = \ \delta_{i,j}\, \chi_i(\tilde q)
\ \ . \label{ishchar}
\ee
Full boundary states $|\alpha\rangle_{\Omega} \equiv |(\Omega,\alpha)
\rangle$ are given as certain linear combinations of Ishibashi states, 
\be \label{bstf1} 
 |\alpha\rangle_{\Omega} \ = \ \sum_i B^i_{\alpha}\, 
|i\rangle\!\rangle_{\Omega}\ .
\ee 
With the help of (\ref{bstdef}), one can show \cite{CarLew,RecSch1} that 
the coefficients $B^i_\alpha$ are related to the one-point functions of the 
boundary theory by 
\be
A_{i}^{\alpha} \ = \ B^{i^+}_{\alpha}  \ \ . 
\label{AeqBB}\ee
The decomposition of a boundary state into  Ishibashi states contains 
the same information as  the set of one-point functions and therefore 
specifies the ``descendant'' boundary CFT of a given bulk CFT completely. 
\medskip 

Following an idea in \cite{Ishi1}, it is easy to write down an expression 
for the generalized coherent states (see e.g.\ \cite{RecSch1}), but the 
formula is fairly abstract. 
In the case of flat backgrounds, however, their construction can be made very 
explicit. Let us first discuss this for Dirichlet boundary conditions, i.e.\ 
for $\Omega^D \bJ ^a= - \bJ^a$ where $a = d+1, \dots, D,$ labels directions
transverse to the brane as in the first lecturs. Since $k^+_a = -k_a$ (recall 
that fusion of sectors is given by adding momenta) and $\omega (k_a) = -k_a$, 
we have $k_a^\omega = k_a$ and so there exists a coherent state for each 
sector in the bulk theory (\ref{bulkFB}). These states are given by 
$$ 
   |k\rangle\!\rangle_D \ = \ \exp \left(\sum_{n=1}^\infty \frac{g_{ab}}{n} 
     \a^a_{-n} \bar \a^b_{-n} \right)\ |k\rangle \otimes \overline{|k\rangle} 
\ \ .    
$$ 
Using the commutation relation of $\a^a_n$ and $\bar \a^a_n$ it is easy to 
check that $|k\rangle\!\rangle_D$ is annihilated by $\a^a_n - \bar \a^a_{-n}$ 
as we required in eq.\ (\ref{glueconplane}). A special case of our formula 
(\ref{D1pt}) for the one-point function along with the  general rule 
(\ref{AeqBB}) lead to the following boundary state for Dirichlet boundary 
conditions, 
$$ |x_0 \rangle_D \ = \ \int \Pi_{a=d+1}^D (\sqrt{\alpha'} dk_a)\  
    \ e^{-i k_a x^a_0} \ |k\rangle\! \rangle_D \ \ . $$ 
For the directions along the brane, the analysis is different. Here we 
have to use the gluing map $\Omega^B$ from eq. (\ref{glueN}) and a simple 
computation reveals that the condition $\omega^B (k) = - k$ is only solved 
by $k = 0$. This means that we can only construct one coherent state, 
$$ |0\rangle\!\rangle_B \ = \ \exp(- \sum_{n=1}^\infty \frac{G_{ij}}{n} 
     \a^i_{-n} \Omega^B\bar \a^j_{-n} )\ |0\rangle \otimes 
    \overline{|0\rangle} \ \ . 
$$ 
According to eqs.\ (\ref{D1pt}) and (\ref{AeqBB}), this coincides with the 
boundary state $|0\rangle_B = |0\rangle\!\rangle_B$. Hence, the boundary 
state for the branes discussed in the first lecture is given by the 
product $|x_0\rangle_{(B,d)} \equiv |0\rangle_B \otimes |x_0\rangle_D$.      

\subsection{Boundary theory - the open string sector} 

While the one-point functions (or boundary states) uniquely characterize
a boundary conformal field theory, there exist more quantities we are 
interested in. In particular, we shall now see how the coefficients of 
the boundary states determine the spectrum of so-called boundary fields 
which can be inserted along the boundary of the world-sheet. In addition, 
we derive a set of factorization conditions for the operator product 
expansions of these new fields.  

\paragraph{The boundary spectrum.} Our aim is to determine the spectrum 
of open string modes which can stretch between two branes labeled by 
$\a$ and $\beta$, both being of the same type $\Omega$. In world-sheet 
terms, the quantity we want to compute is the partition function on
a strip with boundary conditions $\a$ and $\beta$ imposed along the 
two sides. This is illustrated on the left hand side of Figure 4. The 
figure also illustrates the main idea of the calculation. In fact, 
world-sheet duality allows to exchange space and time and hence to 
turn the one loop open string diagram on the left hand side into a 
closed string tree diagram which is depicted on the right hand side. 
The latter corresponds to a process in which a closed string is created 
on the brane $\a$ and propagates until it gets absorbed by the brane 
$\beta$. Since creation and absorption are controlled by the amplitudes 
$A^\a_i$ and $A^\beta_j$, the right hand side - and hence the partition 
function on the left hand side - is determined by the one-point functions
of bulk fields. 
\vspace{1cm}
\fig{The open string partition function $Z_{\a\b}$ can be computed 
by world-sheet duality. In the figure, the time runs upwards so that 
the left hand side is interpreted as an open string 1-loop diagram 
while the right hand side is a closed string tree diagram.}
{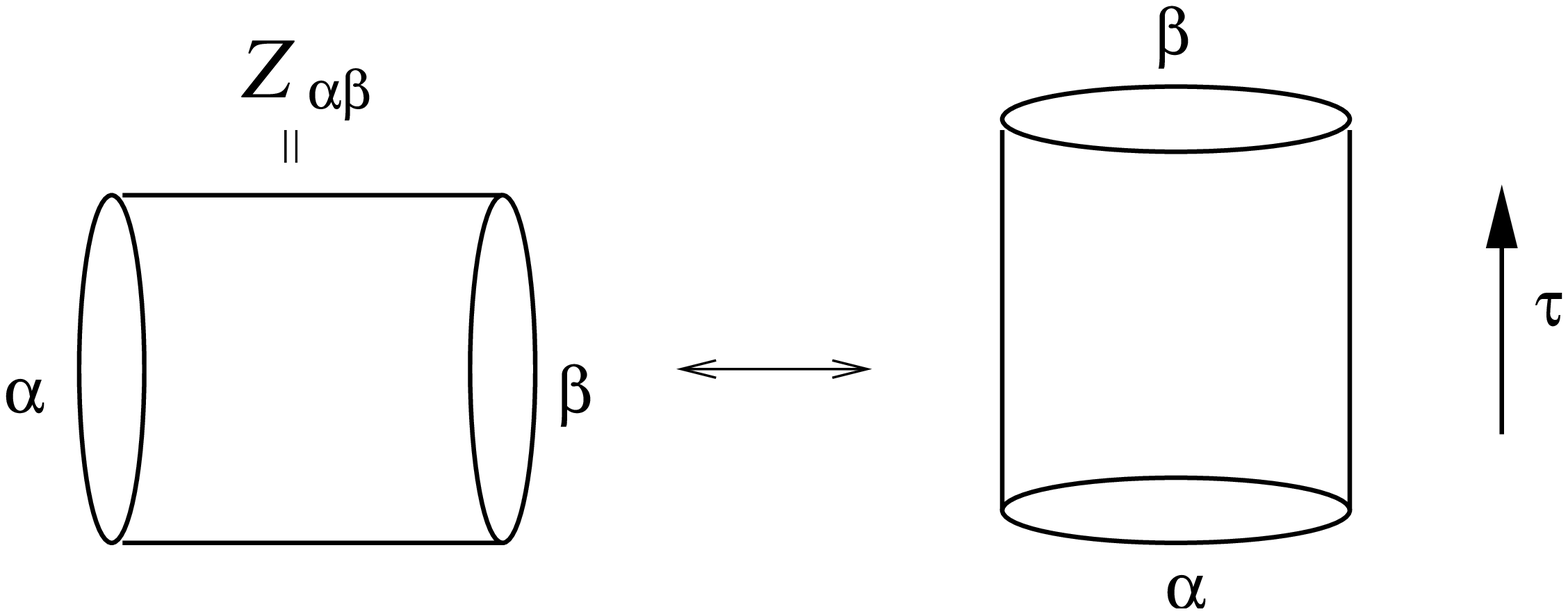}{13truecm}
\figlabel{\basic}
\vspace{1cm}

Let us now become a bit more precise and derive the exact relation 
between the couplings $A$ and the partition function. Reversing the 
above sketch of the calculation, we begin on the left hand side of 
Figure 4 and compute 
$$ \langle \theta \beta| {\tilde q}^{H^{(P)}} |\alpha\rangle \ = \ 
    \sum_j A^\b_{j^+} A^\a_j \langle\!\langle j^+| 
   {\tilde q}^{L_0^{(P)}-\frac{c}{24}}|j^+\rangle \! \rangle 
   \ =  \   \sum_j A^\b_{j^+} A^\a_j \chi_{j^+}(\tilde q)\ \ .  
$$       
Here we have dropped all subscripts $\Omega$ since all the boundary 
and generalized coherent states are assumed to be of the same type. 
The symbol $\theta$ denotes the world-sheet CPT operator in the bulk 
theory. It is a anti-linear map which sends sectors to their conjugate, 
i.e.\ 
$$ \theta A^\b_{j^+}\ |j\rangle\!\rangle \ = \ \left( A^\b_{j^+}
   \right)^* \ |j^+\rangle\! \rangle \ \ . $$ 
Having explained these notations, we can describe the steps we 
performed in the above short computation. To begin with we inserted  
the expansion (\ref{bstf1}),(\ref{AeqBB}) of the boundary 
states in terms of Ishibashi states and the formula $H^{(P)} = 1/2(L_0 
+ \bar L_0) - c/24$ for the Hamiltonian on the plane. With the help of 
the linear relation (\ref{glueconplane}) we then traded $\bar L_0$ for 
$L_0$ before we finally employed the formula (\ref{ishchar}). At this 
point we need to recall the property (\ref{Smatrix}) of characters to 
arrive at 
\be \label{BPF1}  
 \langle \theta \beta| {\tilde q}^{H^{(P)}} |\alpha\rangle
 \ = \ \sum_j A^\b_{j^+} A^\a_j S_{j^+ i} \chi_i(q) \ =:\  
   Z_{\a\b}(q)\ \ .  
\ee
As argued above, the quantity we have computed should be interpreted 
as a boundary partition function and hence as a trace of the operator 
$ \exp(2\pi i \tau H^{(H)})$ over some space $\cH_{\a\b}$ of states 
for the system on a strip with boundary conditions $\a$ and $\beta$ 
imposed along the boundaries. Since our boundary conditions preserve 
the chiral symmetry, the partition function is guaranteed to decompose 
into a sum of the associated characters. Moreover, the coefficients in 
this expansion must be integers and so we conclude\vspace*{2mm}
\be \label{BPF2}  
 Z_{\a\b}(q) \ = \ \sum_i \ {n_{\a\b}}^i\, \chi_i(q) 
  \ \ \ \mbox{ where } \ \ \ {n_{\a\b}}^i \ = \  \sum_j A^{\b}_{j^+} 
 A^\a_j S_{j^+i} \ \in \ \mathbb{N} \ \ .
\ee
This is the desired expression for the partition function in terms
of the couplings $A^\a_i$. Although there exists no general proof, 
it is believed that every solution of the factorization constraints 
(\ref{class}) gives rise to a consistent spectrum with integer 
coefficients ${n_{\a\b}}^i$. A priori, the integrality of the 
numbers ${n_{\a\b}}^i$ provides a strong constraint, known as the 
{\em Cardy condition},  on the set of boundary states and it has 
often been used instead of eqs.\ (\ref{class}) to determine the 
coefficients $A^\a_i$. Note that the Cardy conditions are easier 
to write down since they only involve the modular S-matrix. To 
spell out the factorization constraints (\ref{class}), on the 
other hand, one needs explicit formulas for the fusing matrix and 
the bulk operator product expansion. 
\smallskip

There is one fundamental difference between the Cardy condition 
(\ref{BPF2}) and the factorization constraints (\ref{class}) that 
is worth pointing out. Suppose that we are given a set of solutions
of the Cardy constraint. Then every non-negative integer linear 
combination of the corresponding boundary states defines another 
Cardy-consistent boundary theory. In other words, solutions of the 
Cardy condition form a cone over the integers. The factorization 
constraints (\ref{class}) do not share this property. Geometrically, 
this is easy to understand: we know that it is possible to construct 
new brane configurations from arbitrary superpositions of branes
in the background (though they are often unstable). These brane 
configurations possess a consistent open string spectrum but they 
are not elementary. As long as we are solving the Cardy condition, 
we look for such configurations of branes. The factorization 
constraints (\ref{class}) were derived from the cluster property 
which ensures the system to be in a `pure phase'. Hence, by solving 
eqs.\ (\ref{class}) we search systematically for elementary brane 
configurations that cannot be decomposed any further. Whenever the 
coefficients $\Xi$ are known, solving the factorization constraints 
is clearly the preferable strategy, but sometimes the required 
information is just hard to come by. In such cases, one can still 
learn a lot about possible brane configurations by studying Cardy's 
conditions.

\paragraph{Boundary fields and boundary OPE.} To the partition function 
$Z_{\a\b}$ we have computed in the previous paragraph we can associate 
a state space 
$$ \cH^{(H)}_{\a\b} \ = \ \bigoplus_i {n_{\a\b}}^i \, \cV_i \ \ . $$
The modes of open strings stretching in between the branes $\a$ and 
$\b$ are to be found within this space. Since there should be an open 
string vertex operator for each such mode, we expect that the elements 
in the state space $\cH^{(H)}_{\a\b}$ correspond to boundary operators 
which can be inserted at points $u$ of the boundary where the boundary 
condition jumps from $\a$ to $\beta$. In other words, we have just 
argued for a new state-field correspondence $\Psi^{(H)}$ which 
associates a boundary field with each state in $\cH^{(H)}_{\a\b}$,  
\be 
\psi^{\a\b} (u) \ = \ \Psi^{(H)}(|\psi\rangle;u) \ \ \ \mbox{ for } 
  \ \ \ \ |\psi\rangle \ \in\ \ \cH^{(H)}_{\a\b} \ \ .
\ee
As in the case of bulk fields, we want to introduce a special notation 
for the boundary fields that come with ground states. Once more we fix 
a basis $|e^a_i\rangle$ in the space $V^0_i \subset \cV_i$ and 
introduce the multiplet 
\be\psi^{\a\b}_i (x) \ :=\ \left(\Psi^{(H)}(|e^a_i\rangle;x)\right) \ \ \ 
   \mbox{ with }  \ \ \ |e^a_i\rangle\ \in \ V_i^0 \ \subset\  
  \cH_{\a\b}^{(H)} \ \ .
\label{psii} \ee
Elements of this tuple are numbered by $a = 1, \dots, {\rm dim} V_i^0$. 
We have been a bit sloppy here. In fact, there are certainly cases in 
which the space $\cV_i$ appears with some non-trivial multiplicity 
${n_{\a\b}}^i > 1$. If that is the case, the associated boundary fields
carry an additional index $r = 1, \dots, {n_{\a\b}}^i$. For notational 
reasons we shall omit this extra label, but it is not too difficult to 
add it back into all the following formulas. The multiplet 
$\psi^{\a\b}_i$ carries an irreducible representation of the zero mode 
algebra $\cW_0$. 
\medskip
      
Having introduced boundary fields we are interested in their 
correlation functions. The latter become singular when their world-sheet 
arguments come close and the singularity is again encoded in operator 
product expansions. For the boundary fields $\psi^{\a\b}_i$ these read 
\be \label{HbOPE}
\psi^{\a\b}_i(u_1) \, \psi^{\b\c}_j (u_2)\ = \ 
\sum_k \, (u_1-u_2)^{h_i +h_j - h_k} \, \cC^{\a\b\c}_{ij;k}\,
    \psi^{\a\c}_k(u_2) +\dots \ \ \mbox{ for } \ \ \ 
    u _1 \ > \ u_2 \ \ . 
\ee
The coefficients $\cC^{\a\b\c}_{ij;k} $ are linear maps which intertwine 
the action of the zero mode algebra $\cW_0$ on the field multiplets. We 
can split them into a product of a numerical factor $C^{\a\b\c}_{ij;k}$ 
and an intertwiner $\cU_{ij;k}:V_k^0 \rightarrow V_i^0 \otimes V_j^0$ that 
does no longer depend on the boundary conditions. The intertwiners 
$\cU_{ij;k}$ are normalized by 
$$   
\cU_{ij;l}^* \  \cU_{ij;k} \ = \ \delta_{l,k} \, {\bf 1}_{V_k^0} \ \ . 
$$   
All dynamical information about the scattering of open strings is 
encoded in the numerical factors $C^{\a\b\c}_{ij;k}$ which can be 
non-zero only if ${N_{ij}}^k \neq 0$. 
\smallskip

Four-point functions of these boundary fields must satisfy certain 
factorization constraints (see Figure 5) which allow to derive 
the following constraint on the coefficients of the operator 
product expansion, 
\be \label{HOPEcon} 
C^{\d\a\b}_{ji;r} \, C^{\c\d\b}_{kr;l} \ \sim \ 
 \sum_s \ \Fus{s}{r}{j}{i}{k}{l} \, C^{\c\d\a}_{kj;s} 
\, C^{\c\a\b}_{si;l} \ \ . 
\ee
In writing these equations we have omitted terms associated with 
boundary two-point functions on both sides. The precise condition 
can be found e.g.\ in \cite{Runk1,Runk2,BPPZ2}. Studies of many 
examples and intuition both suggest that these relations possess a 
unique solution, up to some freedom that can be absorbed through the 
normalization of boundary fields. In this sense, the interactions 
of open strings are determined by the multiplicities ${n_{\a\b}}^i$ 
and hence ultimately by the couplings $A^\a_i$ of closed string 
modes to the brane.  
\vspace{1cm}
\fig{The equation (\ref{HOPEcon}) is derived by considering scattering 
amplitudes of four open string modes $(i,j,k,l)$ stretching in between
four different boundary conditions $(\a,\b,\c,\d)$.}
{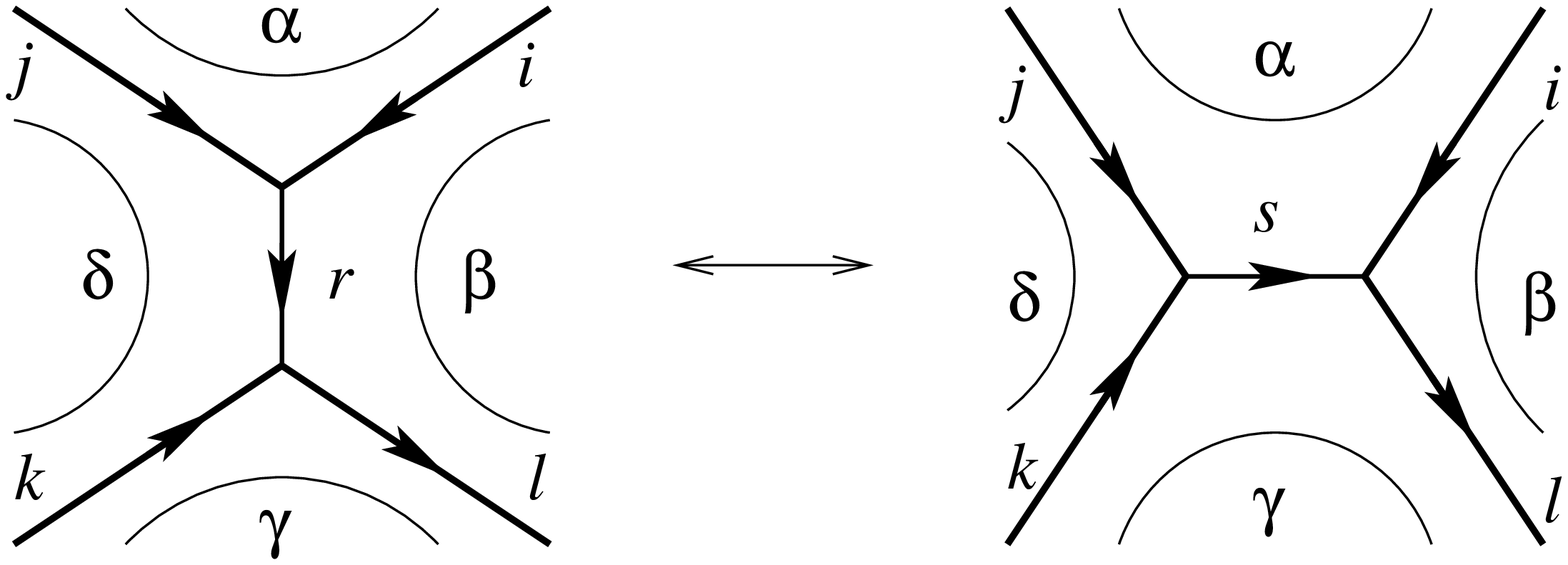}{13truecm}
\figlabel{\basic}
\vspace{1cm}

\paragraph{The Ward identities.} We have already discussed Ward 
identities for correlation functions of bulk fields on the upper 
half-plane $\Sigma$. The extension of such identities to boundary 
fields is straightforward. Using the same notations as in the 
corresponding equation (\ref{WOPE}) for bulk fields, the singular 
part of the operator product expansion between chiral and boundary 
fields reads
\be \label{WbOPE} 
\left(W(w) \, \psi(u)\right)_{\rm sing} \ := \  
[\, W_>(w)\, ,\, \Psi(|\psi\rangle ; u)\, ] 
\ = \ \sum_{n > -h} \ \frac{1}{(w-u)^{n+h}} \, \Psi\bigl(W_n 
 |\psi\rangle ; u\bigr) \ \ . 
\ee
These relations agree with the usual Ward identities for chiral 
vertex operators and so arbitrary correlators in boundary conformal 
field theory can be constructed out of the known formulas for chiral 
blocks. 
 
\subsection{Solution of the theory}   

In the last subsections we derived the three conditions, namely the
relations (\ref{class}), (\ref{BPF2}) and (\ref{HOPEcon}), which are 
fundamental in solving boundary conformal field theory. It is remarkable 
that universal solutions exist for a large class of backgrounds. We will 
now present the basic Cardy solution and then illustrate it in the case 
a single free bosonic field. 

\paragraph{Solution for Cardy case.} 
\def\bj{{\bar \jmath}}
To begin with, let us formulate the main assumption we have to make. 
Suppose we are given some rational bulk conformal field theory with 
a bulk modular invariant partition function of the special form 
\beq 
Z(q,\bar q) \ = \ \sum_j \ 
\chi_j(q) \ \chi_{\bj} (\bar q)\ \ . \label{Cassa}
\eeq 
Here, $j$ runs through the set $\cJ$ of sectors and $\bj$ is
some unique sector that is paired with $j$ in the partition 
function. In the following we describe the solution  for all 
possible possible boundary theories of type $\Omega$, 
provided that the 
\be 
j^\omega \ = \ \omega(j)^+ \ \stackrel{!}{=} \ \bj\ \ . \label{Cass} 
\ee 
Under this condition, Cardy claims that there exist as many 
boundary theories as there are $\cW$-sectors, i.e.\ the number 
of boundary theories is equal to the order of $\cJ$. We shall 
label these boundary theories by $I,J,K \dots \in \cJ$ instead 
of using $\a,\b,\c,\dots$ to remind us that they run through the
same set as the labels $i,j,k,\dots$.  
\smallskip 

We have learned that such boundary theories can be characterized 
by the one-point functions of the primary fields. Cardy proposes 
that the associated couplings $A^J_j$ are simply given by the 
modular S-matrix, i.e.\ 
\be 
\label{1pt} 
 \langle\,  \phi_{j,j^\omega}(z,\bz)\,  \rangle_J \ = \ 
    \frac{S_{Jj}}{\sqrt{S_{0j}}} \ 
    \frac{\cU_{jj^\omega}}{|z-\bz|^{2h_j}} \ \ . 
\ee
Note that this formula makes sense since the boundary label $J$       
runs through the set of $\cW$-sectors. As we explained before, 
$\cU_{jj^\omega}$ is the unitary intertwiner that intertwines 
between the actions of the zero modes $W_0$ on the two spaces 
$V_j^0$ and $V^0_{j^\omega}$ of ground states.  
\smallskip 

The spectrum of open strings stretching between the branes that 
are associated with the labels $I$ and $J$ is encoded in the 
associated partition functions  
\beq Z_{IJ}(q) \ = \ \sum_j {N_{IJ^+}}^{j} \ \chi_j(q) 
\ \ . \label{bpart} \eeq
Here, $J^+$ is defined through the conjugation on sectors of 
the chiral algebra, i.e.\ through ${N_{I J^+}}^0 = \delta_{I,J}$.
Since fusion rules of the chiral algebra are non-negative integers, 
the expression (\ref{bpart}) has the form of a partition function 
for a $\cW$-symmetric system. We will analyze below how equation 
(\ref{bpart}) is related to the formula (\ref{1pt}) for the 
one-point functions.    
\smallskip

The complete solution should include expressions for the boundary 
operator products. According to the expansion (\ref{bpart}), there 
are ${N_{I J^+}}^j$ boundary fields $\psi^{IJ}_j(u)$ (recall that 
they are $\cW_0$ 
multiplets whenever dim$V^0_j > 1$). As in the previous paragraph, 
we suppress the extra index that labels different fields in cases 
when ${N_{I J^+}}^j > 1$. The operator product expansion for two such 
primary fields is claimed to be of the form 
\beq
\psi^{LM}_i(u_1) \ \psi^{MN}_j(u_2) \ = \ 
\sum_{k} \ (u_1- u_2)^{h_i+h_j-h_k}\ \Fus{M}{k}{i}{j}{L}{N}\, 
  \cU_{ij;k} \, \psi^{LN}_k (u_2) \ \ + \dots \ 
\label{bOPE}
\eeq
for $u_1 > u_2$. Here $F$ stands for the fusing matrix of the chiral 
algebra $\cW$. It was introduced at the end of the first subsection. 
The formula (\ref{bOPE}) was originally found for minimal models by 
Runkel \cite{Runk1} and extended to more general cases in 
\cite{FFFS1,FFFS2,zuber}. 
\medskip 

Let us stress that all the important structure constants of the solution, 
namely the list of boundary labels, the one-point functions, the partition 
functions and the boundary operator expansions, have been expressed through 
representation theoretic data of the underlying symmetry. In fact, we have 
used the list of sectors, the modular S-matrix, the fusion rules and 
the fusing matrix to write down the exact solutions. Obviously, the 
rather simple relation between the set of solutions and purely 
representation theoretic quantities only appears for very particular 
choices of bulk modular invariants and gluing maps $\Omega$. This   
manifests itself in our assumption $j^\omega = \bj$. If the assumptions 
(\ref{Cassa}), (\ref{Cass}) are violated, finding solutions is more 
difficult. We will only provide a brief overview on the current status 
of this active field of current research (see last section). As 
restricting as the assumption $j^\omega = \bj$ may appear, it still 
turns out to apply to a large number of interesting situations.

\paragraph{Testing Cardy's solution.} The structure constants $A^\a_j,
{n_{\a\b}}^i$ and $C^{\a\b\c}_{ij;k}$ of any exact solution must 
satisfy our fundamental algebraic constraints (\ref{class}), (\ref{BPF2}),
(\ref{HOPEcon}). Our aim here is to verify at least two of these 
conditions, namely the Cardy condition (\ref{BPF2}) and the factorization 
constraint (\ref{HOPEcon}) for the solution we have provided in eqs.\ 
(\ref{1pt}), (\ref{bpart}) and (\ref{bOPE}). From Cardy's solution we 
read off that $A^J_j = S_{J j}/\sqrt{S_{0j}}$. 
When this is inserted into the formula (\ref{BPF2}) for the integers 
${n_{IJ}}^i$ one finds 
$$ {n_{I J}}^i \ = \ \sum_j \frac{S_{Jj^+} \, S_{I j} \, S_{j^+ i}}
   {S_{0j} } \ = \ N_{I J^+}^i \ \ . $$ 
In the computation we have used the properties $S_{ij^+} = 
S^*_{ij} = S_{i^+ j}$ of the modular S-matrix and the 
Verlinde formula (\ref{Verl}). In fact, the very close relation 
between the Verlinde formula and the expressions in eq.\ 
(\ref{BPF2}) was the main evidence Cardy relied on to support 
his solution. 
\smallskip 

The boundary operator product expansions (\ref{bOPE}) arise from 
the general expression (\ref{HbOPE}) if we equate
$$ C^{IJK}_{ij;k} \ = \ \Fus{J}{k}{i}{j}{I}{K} \ \ . $$ 
When we plug this expression into eq.\ (\ref{HOPEcon}), we end up 
with the following equation for the fusing matrix 
$$ 
\Fus{I}{r}{j}{i}{L}{J} \, \Fus{L}{l}{k}{r}{K}{J} \ = \ 
\sum_s \, \Fus{s}{r}{j}{i}{k}{l} \, \Fus{L}{s}{k}{j}{K}{I}
\, \Fus{I}{l}{s}{i}{K}{J} \ \ . $$ 
This is the famous pentagon equation which holds true for the 
fusing matrix of any chiral algebra (see \cite{MooSei2}). The 
proof of the full factorization constraint is slightly more 
involved since one has to take the contributions from boundary 
two-point functions into account. The latter were neglected in 
our equation (\ref{HOPEcon}).       
 
\paragraph{Example: D0 branes in flat space.} It is nice to see 
how the general formulas allow to recover the solutions of the 
boundary conformal field theory describing a point-like brane in 
a 1-dimensional flat space. The bulk invariant has been spelled 
out in eq.\ (\ref{bulkFB}). It is diagonal in the sense $k = 
\bar k$ and has the special form (\ref{Cassa}), at least if we 
close an eye on the fact that the real line is non-compact and 
hence the theory is not rational. We are interested in boundary 
theories which obey the Dirichlet gluing condition 
$ J(z) = - \bJ(\bz)$. As we have explained earlier, this choice 
of $\Omega$ implies $k^\omega= k$ and hence our main assumption 
(\ref{Cass}) is satisfied. 
\smallskip

Now Cardy assures us that the associated boundary conditions  
are parametrized by a parameter $\a $ which runs through the 
set $\cJ= \mathbb{R}$ of sectors in the theory. Hence, the 
parameter $\a$ must be associated with the transverse position 
of our brane in the 1-dimensional background.      
\smallskip

The precise relation between $\a$ and $x_0$ can be read off from 
the one-point functions. If we insert the formula (\ref{U1S}) for 
the S-matrix of the U(1) theory into the general expression 
(\ref{1pt}) above we obtain 
$$\langle \phi_{k,k} (z,\bar z) \rangle_{\a} \ = \
    \frac{e^{2\pi i\a' k \a}}{|z-\bz|^{\ap k^2}} \ \ . $$
This indeed agrees with a special case of formula (\ref{D1pt}) 
for $D = 1, d= 0$ and $x_0 = 2 \pi \a' \a$.%
\smallskip

Next we want to look at our formula (\ref{bpart}) for the partition 
functions. We determined the fusion rules of the U(1) theory in eq.\ 
(\ref{U1VF}) and plugging them into eq.\ (\ref{bpart}) gives 
$$ Z_{x_0y_0} (q) \ = \ \int d k\ 
\delta(\frac{x_0 - y_0}{2\pi \a'} - k) \chi_k(q) \ = \ \frac{1}{\eta(q)} 
  \ q^{\frac{(x_0-y_0)^2}{8\pi^2 \a'}} \ \ .
$$
In the first lecture we only computed the special case $x_0 
= y_0$ for which we obtain agreement with eq.\ (\ref{flatPF}). 
If $x_0$ and $y_0$ are not the same, an open string must stretch 
over the finite distance $x_0 - y_0$ in between the two branes. 
The energy of such stretched open strings is proportional to 
$(x_0-y_0)^2$ and this explains the exponent of the second 
factor in the partition function. Needless to say that it can 
be found directly by solving the Laplace equation in a strip. 

The field corresponding to the only ground state in $\cH_{x_0y_0}$ 
is given by 
$$ \psi^{x_0 y_0}_p (x) \ = \  \no  e^{ i p X(x)}  \no   
  \ \ \ \ \mbox{ with } \ \ \ p \ =\ \frac{x_0-y_0}{2\pi \a'} 
\ \ \ . $$ 
Since the fusing matrix of the U(1) theory is trivial, the operator 
product expansion of these exponential fields is the one predicted 
by formula (\ref{bOPE}).
\smallskip

Obviously, we did not learn anything new about D0 branes in 
flat space. What we have seen is some kind of high-tech 
derivation of the standard results using lots of complicated 
notions from representation theory of chiral algebras. The point 
is, however, that now we do have a technology ready to be 
applied to more complicated interacting theories with non-linear 
equations of motion for which conventional methods fail. 
We will illustrate the full power of these developments 
in the next lecture.  
\newpage

\section{Application: Branes on group manifolds} 
\setcounter{equation}{0}

We are now in a position to apply the general techniques of 
boundary conformal field theory to the construction of branes
in curved backgrounds. In many ways, group manifolds provide 
an ideal area to illustrate the abstract constructions. Since 
they are homogeneous, they come with a large symmetry that 
facilitates the exact solution. On the other hand, strings 
on group manifolds are described by a non-linear 2D field 
theory and hence are sufficiently non-trivial to demonstrate 
the full power of boundary conformal field theory. Finally, 
the involved models are also fundamental for CFT model 
building. 
\smallskip

We shall first approach branes on group manifolds in a more 
qualitative way, mainly based on our experience from the first 
lecture. Even though the initial reasoning will hardly go beyond
a chain of educated guesses, this can lead us a long way and it 
will help later to present the exact solution in a new light. 
Once the branes on group manifolds have been constructed using
all the formulas from the previous lectures, we compute their 
low-energy effective action and discuss several interesting 
applications to the study of brane dynamics. For simplicity our 
presentation focuses mainly on the group manifold $SU(2) \cong 
S^3$, but many aspects generalize directly to other groups (see 
remarks in the last section).

\subsection{The semi-classical geometry.} 
 \def\rmc{{\rm c}}
\def\Lieg{{\cal G}}
\def\Fun{{\rm Fun}}
\def\rmY{{\rm Y}}
\def\rmhY{\hat{y}}
\def\rmC{{\rm C}} 

In the following introductory paragraph we shall see the basic
contours of a scenario that we are going to derive later through 
our microscopic treatment of branes on group manifolds. From the
preliminary discussion we will extract certain Poisson spaces 
which are argued to describe the semi-classical geometry of the 
relevant branes. Their quantization is the subject of the third
paragraph. 
   
\paragraph{Introductory remarks.} Group manifolds possess a  
non-vanishing constant curvature ${ R} = { R} (g)$ which arises 
from a non-constant metric $g$. It is well known that strings are 
rather picky when it comes to choosing the backgrounds they can 
propagate on. In fact, to lowest order in $\a'$, the background 
metric $g(x)$ and B-field $B(x)$ of a bosonic string have to obey 
the equations 
\be R_{\mu\nu} (g) - \frac{1}{4} \, H_{\mu\rho \sigma} \, 
{H_{\nu}}^{\rho\sigma} + 
 {\cal O}(\a') \ = \ 0 \ \ , \label{SEOM}\ee
where $H_{\mu\nu\sigma}$ are the components of the NSNS 3-form 
$H = dB$ with $B = B_{\mu\nu}(x) dx^\mu \wedge dx^\nu$. In 
writing eq.\ (\ref{SEOM}), the dilaton was assumed to be 
constant. For superstrings the same equations hold as long 
as we set all the RR background fields to zero. This is the 
scenario in which the following discussion is placed. 
\smallskip

From eq.\ (\ref{SEOM}) we conclude immediately that a background 
with non-zero curvature $R_{\mu\nu}$ requires a non-vanishing 
NSNS field $H$ and hence a non-zero B-field. In our analysis 
of branes in flat space, such B-fields caused the coordinates 
along the brane to be quantized and hence they were at the 
origin of the brane's non-commutative geometry. Although the 
details will be different in curved spaces, the basic mechanisms 
are certainly expected to work in the same way. Thus, if we can 
find branes which extend along some directions of a group 
manifold, their world-volume geometry is very likely to be 
quantized. 
\smallskip

To see whether we have a chance to construct extended stable 
branes on group manifolds, let us now restrict to the case of
$SU(2) \cong S^3$. This is also of particular interest in 
string theory because it appears e.g.\ within the background 
$\mathbb{R}^{1,5} \times S^3 \times \mathbb{R}_+$ of NS 
5-branes \cite{CaHaSt1,CaHaSt2} or as part of the geometry 
$AdS_3 \times S^3 \times T^4$. Placing a point-like brane 
somewhere on the $SU(2)$ is obviously not in conflict with 
stability. Higher dimensional objects, however, may seem 
unstable at first sight, since their tension tends to make them 
collapse. Only a brane wrapping the whole $S^3$ could be 
stabilized through the topology, but it is excluded by the 
presence of the non-vanishing NSNS 3-form \cite{WitK,FreWit}. 
Hence, if the tension would be the only factor contributing 
to the stability analysis, our story would be rather short 
and boring. It turns out, however, that the non-vanishing 
B-field also plays an important role and that it can exert 
enough pressure on 2-dimensional spherical branes in $S^3$ 
to balance the tension \cite{BaDoSc,Pawe}. Although the 
initial arguments for this {\em flux stabilization} could 
only be trusted at weak curvature, the statement remains 
correct even deep in the stringy regime \cite{AlReSc3}. Thus, 
we conclude that stable branes on $SU(2)$ can either be 
point-like or they can wrap a 2-dimensional sphere $S^2 
\subset S^3$. From now on we shall consider point-like
branes as degenerate 2-spheres so that we do not have to 
distinguish between the two possibilities. 
\smallskip

Spheres $S^2$ in $S^3$ are parametrized by the location of their 
center and by their radius $r$. Branes wrapping a sphere of radius
$r$ carry a non-vanishing B-field which causes their world-volume
to be quantized. At a fixed scale of non-commutativity, the area 
$A(r)$ of the 2-sphere gets tiled by elementary `Heisenberg cells' 
each of which contributes a single state to the quantized theory. 
Since the number of such states must be integer, we conclude that 
2-spheres are quantizable only for a discrete set of radii $r$. 
Whenever we tune the radius $r$ to one of the allowed values, the
space of wave functions on the corresponding quantized sphere is 
finite dimensional. Coordinates along the brane are observables 
in the quantum theory and hence they are represented by operators
acting on the space of wave functions. If the latter is finite
dimensional, then the coordinates become matrices. 
\smallskip 

This is about as far as our very qualitative discussion can carry. 
There are three main conclusions that we take along. First of all, 
stable branes on $S^3$ are expected to wrap 2-spheres or they can  
be point-like (see Figure 6). Furthermore, 2-spheres can only be 
wrapped for a discrete set of radii and finally, the world-volume 
of such branes has been argued to possess a non-commutative `matrix 
geometry'. We will see all these expectations confirmed by the exact 
treatment.   
\vspace{1cm}
\fig{Stable branes on a 3-sphere are either point-like or they wrap 
a 2-sphere (conjugacy class of $SU(2) \cong S^3$). The 2-dimensional 
branes are stabilized by the background flux \cite{BaDoSc,Pawe}.}
{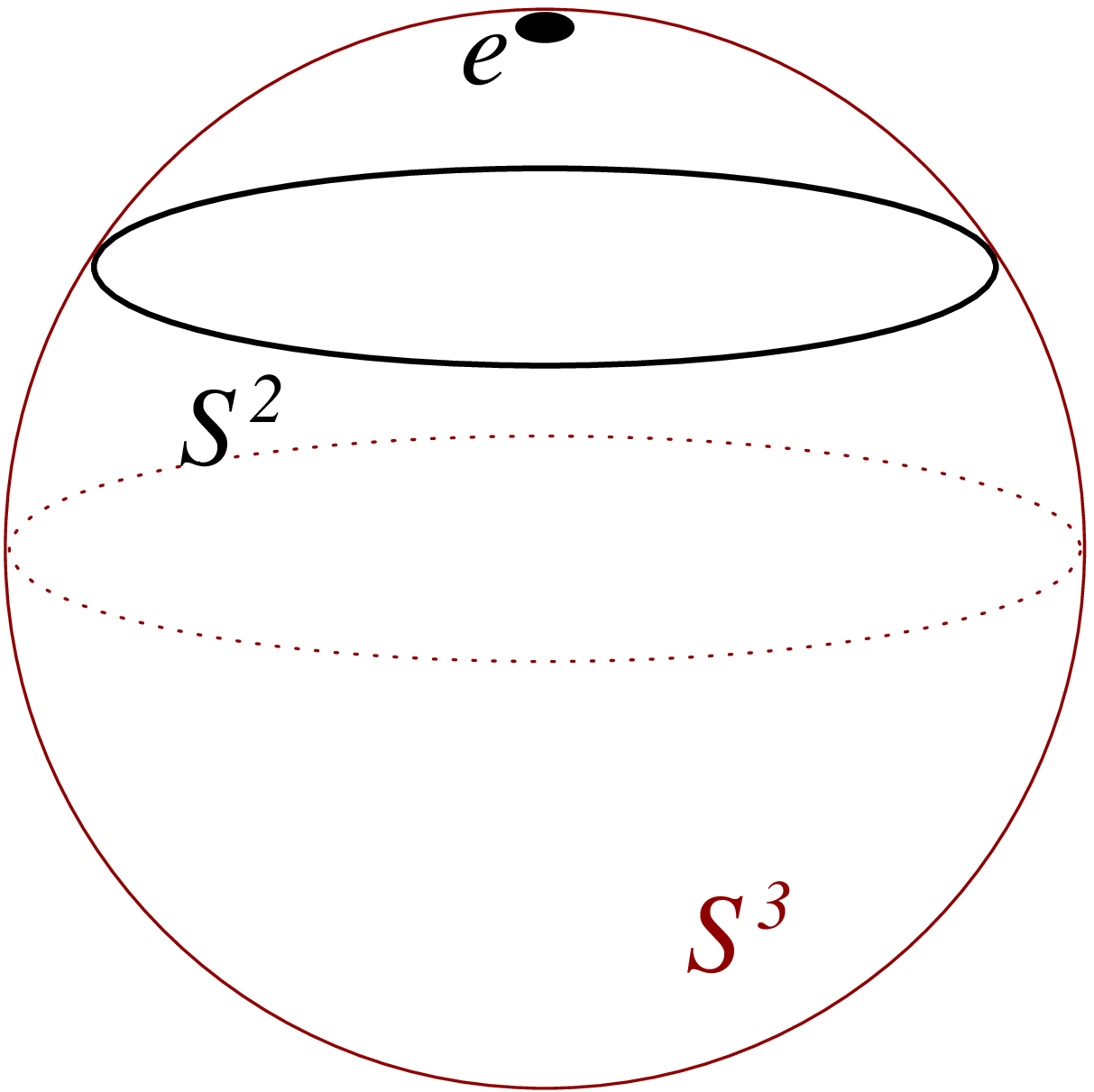}{6truecm}
\figlabel{\basic}
\vspace{1cm}

\paragraph{Gluing condition and brane geometry.} Strings moving on 
a 3-sphere $S^3$ of radius $R \sim \sqrt{\ik}$ are described by the 
SU(2) WZW model at level $\ik$. The world-sheet swept out by an open 
string in $S^3$ is parametrized by a map $g: \Sigma \rightarrow 
{\rm SU(2)}$ which is defined on the upper half-plane $\Sigma$ as 
before. Our aim now is to determine the boundary conditions we need
to impose on $g$ so that we obtain the desired spherical branes. 
Following \cite{AleSch5}, we shall argue that the appropriate choice 
is given by (see also \cite{Gawe1}, \cite{Stan00/1})  
\be  -  \, (\partial g) g^{-1} \ = \   
 \, g^{-1} \bar\partial g\ \  \ \mbox{ for } \ \ \ z \ = \ \bz\ \  . 
\label{GCS}  \ee
To present the findings of \cite{AleSch5}, we split $\partial, 
\bar \partial$ into derivatives $\partial_u, \partial_v$ 
tangential and normal to the boundary and rewrite eq.\ 
(\ref{GCS}) in the form 
\be \left( \Ad(g)-1 \right)  g^{-1} \partial_v g \ =  
     \  i \,\left( \Ad(g)+1 \right)  g^{-1} \partial_u g 
\ \  \ \mbox{ for } \ \ \ z \ = \ \bz \ \ . 
\label{GCS2} \ee 
Here, $\Ad(g)$ denotes the adjoint action (i.e.\ action by 
conjugation) of SU(2) on its Lie algebra {\rm su(2)}. The 
following analysis requires to decompose the tangent space 
$T_h{\rm SU}(2)$ at each point $h \in \,$SU(2) into a part 
$T^{\|}_h{\rm SU}(2)$ tangential to the conjugacy class 
through $h$ and its orthogonal complement $T^\perp_h {\rm SU}(2)$ 
(with respect to the Killing form). Using the simple fact that 
$\Ad(g)|_{T^\perp_g} = 1$ we can now see that with condition 
(\ref{GCS2})  
\begin{enumerate} 
\item  
 the endpoints of open strings on SU(2) are forced to move 
 along conjugacy classes, i.e.\ 
  $$ (g^{-1} \partial_u g)^\perp \ = \ 0 \ \ .$$
 Except for two degenerate cases, namely the points $e$ and 
 $-e$ on the group manifold, the conjugacy classes are 
 2-spheres in SU(2). 
\item the branes wrapping conjugacy classes of SU(2) carry 
 a B-field which is given by 
 \be B \ \sim  \frac{\Ad(g)+1}{\Ad(g)-1} \ \ . 
 \label{Bfield} \ee 
 The associated 2-form is obtained as $\tr (g^{-1} dg \, B \, 
 g^{-1} dg)$ and a short computation shows that it provides a 
 potential for the NSNS 3-form $H \sim \Omega_3$ where $\Omega_3$ 
 denotes  the volume form on $S^3$. 
\end{enumerate} 
The second statement follows from eq.\ (\ref{GCS2}) by comparison 
with the boundary conditions (\ref{BC2}) we used in the flat 
background. Our discussion shows not only that eq.\ (\ref{GCS2}) 
is indeed the desired boundary condition for spherical branes on 
$S^3$ but also it has left us with an exact formula for the 
B-field.  
\medskip

We have learned in the first lecture that the relevant object for 
the brane's non-commutative geometry is not $B$ itself but another 
anti-symmetric tensor $\Theta$ constructed from $B$ through eq.\ 
(\ref{BtoT}). Even though this relation was derived for a flat 
background we may try to apply it naively in the present context. 
For appropriate choice of the metric $g$ and the Regge slope $\a'$, 
we are then led to the expression 
$$
\Theta(g) \ =\ \frac{2}{B-B^{-1}}\ =\ 
\frac{1}{2}\; \bigl(\,\Ad(g^{-1}) - \Ad(g)\,\bigr)\ .
$$
This is a rather complicated object, but it simplifies in the limit 
of large level $\ik$ where the 3-sphere grows and approaches flat 
3-space $\R^3$. One can parametrize points on SU(2) by elements $X$ 
in the Lie algebra su(2), such that near the group unit $g \approx 
1 + X$. Insertion into our formula for $\Theta$ gives
\be \Theta \ = \ - \ad(X) \label{ThX} \ \ . \ee
Here, $\ad$ denotes the adjoint action of su(2) on itself. If we expand 
$X = y^\mu t_\mu$ we can evaluate the matrix elements of $\Theta$ more 
explicitly,  
$$ \Theta_{\mu\nu} \ = \  - \ (\, t_\mu\, ,\,  \ad(X) t_\nu\, ) 
   \ = \ -\, y^\rho\, (\, t_\mu\, , \, {f_{\rho \nu}}^{\sigma}\, 
       t_\sigma\, ) \ = \ f_{\mu\nu\rho}\, y^\rho \ \ , $$       
where $(\cdot,\cdot)$ denotes the Killing form on su(2), the generators
$t_\mu$ are normalized such that $(t_\mu, t_\nu) = \delta_{\mu,\nu}$  
and $f_{\mu\nu\rho}$ are the structure constants of su(2). It is not 
difficult to see that the tensor $\Theta$ gives rise to a Poisson 
structure on $\mathbb{R}^3$, 
\be \{ \, y^\mu \, ,\, y^\nu\} \ = \ \Theta^{\mu\nu}(y) \ = \ 
 {f^{\mu\nu}}_\rho \, y^\rho \ \ .
\label{LPB1}\ee 
In contrast to the Poisson bracket we met in our discussion of flat 
branes, $\Theta$ has a linear dependence on the coordinates. The Poisson 
algebra defined by eq.\ (\ref{LPB1}) possesses a large center. In fact, 
any function of $c(y) = \sum_\mu (y^\mu)^2$ has vanishing Poisson bracket 
with any other function on $\R^3$ so that formula (\ref{LPB1}) induces a 
Poisson structure on the 2-spheres 
\be c(y) \ := \ \sum_\mu \, (y_\mu)^2 \ \stackrel{!}{=} \ \rmc 
\label{LPB2}\ \  .\ee 
If we trust the steps of our reasoning, the two formulas (\ref{LPB1}) and 
(\ref{LPB2}) provide a semi-classical description of the spherical branes
in SU(2) which replaces the simpler formula (\ref{PB}) in the first lecture. 

\paragraph{Quantization and matrices.} Now let us recall that the 
Moyal-Weyl product shows up for brane geometry in flat space with 
constant B-field is obtained from the constant Poisson bracket (\ref{PB}) 
on $\R^d$ through quantization. By analogy, our semi-classical analysis for 
branes on SU(2) $\cong S^3$ suggests that the quantization of 2-spheres 
in $\R^3$ with Poisson bracket (\ref{LPB1}) becomes relevant for branes 
on ${\rm SU(2)}$ in the limit where $\ik \rightarrow \infty$. This is  
sufficient motivation for us to try quantizing the geometry (\ref{LPB1}), 
(\ref{LPB2}).   
\smallskip 

Quantization requires to find some operators $\rmhY^\nu = Q(y^\nu)$ acting 
on a state space $V$ such that \vspace*{-3mm} 
\ba  & & [ \, \rmhY^\mu \, ,\, \rmhY^\nu \, ]\ =  
          \ i\, f^{\mu\nu}_{\ \ \rho} \ \rmhY^\rho 
\label{QLPB1} \\[2mm] 
  & & \rmC \ := \ \sum\   (\rmhY^\mu)^2  \ = \ \rmc\  {\bf 1}  
\label{QLPB2}\ea   
where ${\bf 1} $ denotes the identity operator on the state space $V$. 
These two requirements are the quantum analogues of the classical relations 
(\ref{LPB1}), (\ref{LPB2}) and the quantization problem they pose is easy 
to solve. By the commutation relation (\ref{QLPB1}), the operators 
$\rmhY^\mu$ have to form a representation of ${\rm su(2)}$. Condition 
(\ref{QLPB2}) states that in this representation, the quadratic Casimir 
element $\rmC$ must be proportional to the identity ${\bf 1}$. This is true 
if the representation on $V$ is irreducible. Hence, any irreducible 
representation of ${\rm su(2)}$ can be used to quantize our Poisson 
geometry.        
\smallskip

Irreducible representations of the Lie algebra ${\rm su(2)}$ 
are labeled by one discrete parameter $J = 0, 1/2, 1,$ $\dots $. This 
implies that only a discrete set of 2-spheres in $\R^3$ is quantizable
and their radii increase with the value of the quadratic Casimir in the 
corresponding irreducible representation. For each quantizable 2-sphere 
$S^2_J \subset \R^3$, we obtain a state space $V_J$ of dimension $\dim 
V_J = 2J + 1$ equipped with  an action of the quantized coordinate 
functions $\rmhY^\mu$ on $V_J$. The latter generate the matrix algebra 
$\Mat (2J +1)$. Note that our quantized 2-spheres have all the features 
we anticipated in the introduction above. But this should not be 
considered a derivation since our arguments relied on extending formula 
(\ref{BtoT}) to the present context. While such a step is certainly 
suggestive, we gave no evidence for it to be correct.  

\paragraph{Matrix (fuzzy) geometry.}  
It is useful to go a bit deeper into exploring these 
quantized 2-spheres. Let us start by recalling that 
the space $\Fun(S^2)$ of functions on a 2-sphere is 
spanned by spherical harmonics $Y^j_a \in \Fun (S^2)$ 
where $j$ runs through all integer isospins. A product 
of any two spherical harmonics is again a function on 
the 2-sphere and hence it can be written as a linear 
combination of spherical harmonics, 
\be  Y^i_a \ Y^j_b \ = \ \sum_{k,c} \ 
c_{ijk}\  \CG{i}{j}{k}{a}{b}{c} \ Y^k_c 
\label{SHM} \ee 
with $[:::]$ denoting the Clebsch-Gordan coefficients 
of ${\rm su(2)}$. The structure constants $c_{ijk}$ can 
be found at many places in the literature. We also note 
that elements of the vector multiplet $Y^1_\nu$ may be 
identified with the restriction of the three coordinate 
functions $y^\nu$ to the 2-sphere in $\mathbb{R}^3$.  
\smallskip

The algebra $\Fun(S^2)$ admits an action of ${\rm su(2)}$ 
which is generated by infinitesimal rotations in $\mathbb{R}^3$, 
$$      L_\mu  \ := \ {f_{\mu\rho}}^\nu \, y^\rho \pl_\nu  \ \ .$$
Even though the differential operators $L_\mu$ are initially defined
for arbitrary functions on $\mathbb{R}$, they obviously descend down 
to $\Fun(S^2)$. Under the action of $L_\mu$, spherical harmonics 
$Y^j_a$ transform in the representation $j$ of su(2).    
\medskip

This classical symmetry survives quantization, i.e.\ there exists an 
analogous action of ${\rm su(2)}$ on $\Mat(2J +1)$. It is given by 
\be \tL_\mu \, {\rm A} \ := \ [\, t_\mu^J\, , \, {\rm A}\, ] \ \ \ 
\mbox{ for all } \ \ \ {\rm A} \in \Mat(2J+1)\ . 
\label{Lmu} 
\ee
Here $t^J_\mu$ denote the generators of ${\rm su(2)}$ evaluated 
in the $(2J+1)$-dimensional irreducible representation. One can 
easily decompose this reducible action of su(2) on $\Mat(2J+1)$ 
into its irreducible sub-representations to find the following
equivalence
\be \Mat (2J+1) \  \cong \ \bigoplus_{j=0}^{2J} \, V_j \ \ , 
\label{Matdec}\ee
where the sum on the right hand side runs over integer labels $j$. 
Each irreducible component $V_j$ in this expansion is spanned by 
$(2j+1)$ matrices which we denote by $\rmY^j_a, j \leq 2J$. In 
the case $J=1/2$, only the scalar and the vector multiplet appear 
and explicit expressions for the corresponding $2\times 2$ matrices 
are of course well-known: they are given by the identity and the 
Pauli-matrices, respectively.  
\smallskip

Since $\rmY^j_a$ span $\Mat(2J+1)$, the product of any two such 
matrices may be expressed as a linear combination of matrices 
$\rmY^k_c$, 
\be \rmY^i_a \ \rmY^j_b \ = \ \sum_{k \leq 2J,c}\ 
\SJS{i}{j}{k}{J}{J}{J}\, \CG{i}{j}{k}{a}{b}{c}\ 
 \rmY^k_c \ \ . 
\label{QSHM} \ee 
Here $\{:::\}$ denote the re-coupling coefficients (or 6J-symbols) of 
${\rm su(2)}$. This relation should be considered as a quantization of 
the expansions (\ref{SHM}) and the classical expression is recovered 
from eq.\ (\ref{QSHM}) upon taking the limit $J \rightarrow \infty$
\cite{Hopp}.  
Hence, the matrices $\rmY^j_a$ in the quantized theories  are a proper 
replacement for spherical harmonics. Note, however, that the angular 
momentum $j\leq 2J$ is bounded from above. This may be interpreted as 
`fuzziness' of the quantized 2-spheres on which short distances cannot 
be resolved \cite{Mado1}. We shall eventually refer to  $\rmY^j_a$ as 
`fuzzy spherical harmonics'.

\subsection{The exact CFT solution}     

Having gained some intuition into the main features of branes
on $S^3$, at least in the limit of weak curvature, it is now
time to let the microscopic machinery work for us. Since the 
exact solutions we wrote down in the previous lecture used a
lot of data from the representation theory of the symmetry 
algebra, we will first list some of these data for affine
Lie algebras, which are the chiral algebras relevant 
for strings on group manifolds. Then we present the exact 
formulas for the one-point functions of bulk fields, the 
open string partition function and the boundary OPE, and we 
discuss them in the light of our geometric insights.  
     
\paragraph{Affine Lie algebras.} In the following we 
collect a few basic facts on WZW models and affine Lie 
algebras. Many more details and references to the original 
literature can be found e.g.\ in \cite{GodOli,FrMaSeBook}. The 
fundamental SU(2) valued field $g$ of the WZW model is known 
to satisfy the following classical equation of motion in the 
bulk 
\be \pl \left(g^{-1} (z\,\bz) \bpl g(z,\bz)\right) \ = \ 0 \ \ . 
\label{gle} \ee 
This should be considered as a non-linear version of the Laplace 
equation which governs the string motion in flat backgrounds. It 
follows immediately that the fields 
\be J(z) \ := \ - \ik \, \pl g(z,\bz) \, g^{-1} (z,\bz)
 \ \ \ \ , \ \ \ \     
\bJ(\bz) \ := \ \ik\,  g^{-1}(z,\bz) \bpl g(z,\bz) 
\label{WZWcurr} \ee
are chiral. Since they take values in the Lie algebra su(2), we 
can expand each of these two currents in terms of three component 
fields, i.e.\ $J(z) =: J^\mu(z) t_\mu$ and similarly for the 
anti-holomorphic partner $\bJ$. 
\smallskip 
 
The chiral fields $J^\mu$ of the SU(2) WZW model form an affine 
Lie algebra denoted by $\widehat{\rm SU}(2)_\ik$. It is 
generated by Laurent modes $J^\mu_n$ which obey the following 
commutation relations
\be \label{KMcomm}
[ \, J^\mu_n \, , \, J^\nu_m \, ] \ = \ 
 i {f^{\mu\nu}}_\rho\, J^\rho_{n+m} + {\ik} \, n 
  \, \d^{\mu,\nu} \,   \delta_{n,-m} 
\ee
along with the usual reality property $(J^\mu_n)^* = J^\mu_{-n}$. 
The commutators (\ref{KMcomm}) differ from the corresponding 
relations (\ref{flatcomm1}), (\ref{flatcomm2a}) in the flat space 
theory by the first term on the right hand side. This signals the
presence of a non-vanishing background curvature. 
\smallskip 

Zero-modes $J^\mu_0$ of $\widehat{\rm SU}(2)_\ik$ satisfy the usual 
relations for generators of the finite dimensional Lie algebra su(2). 
Hence, the sectors $\cV_j$ of the theory are created out of the 
$(2j+1)$-dimensional representation spaces $V^0_j= V_j$ of su(2).
But only if $j \leq \ik/2$, these sectors are free of negative 
norm states and hence unitarity leaves us with just a finite  
number of physical representations which we can label through 
$j \in \cJ = \{0,1/2, \dots, \ik/2\}$. Their conformal weights are 
given by $h_j = j(j+1)/(\ik+2)$ and for the modular S-matrix one 
finds
\be \label{Smatrixsu}
S_{ij} \ = \ \sqrt{\frac{2}{\ik+2}} \ \sin \frac{\pi(2i+1)(2j+1)}{\ik+2} 
\ \ . 
\ee
With the help of the Verlinde formula (\ref{Verl}) it is not 
difficult to compute the following fusion rules, 
\be {N_{ij}}^k \ = \ \left\{ \begin{array}{ll} 
   1 \ \ \ \ & \mbox{ for } \ k = |i-j|, \dots, \min (i+j,\ik-i-j)
   \\[2mm] 0 & \mbox{ otherwise } 
   \end{array} \right. \ \ .   
\label{wzwfus} \ee
They are similar to the Clebsch-Gordan multiplicities of su(2),  
apart from the truncation which appears whenever $i+j > \ik/2$.
As in the case of su(2), the trivial representation $k=0$
occurs only in the fusion of $j$ with itself, i.e.\ the 
conjugate $j^+$ of $j$ is given by $j^+ = j$.   
\smallskip 

Formulas for the fusing matrix also exist and they can be found 
in the literature. Since they are rather complicated we will not 
spell them out. Let us only mention one property concerning their 
limiting behavior as we send $\ik \rightarrow \infty$, 
\be \label{fusprop} 
\lim_{\ik \rightarrow \infty} \Fus{J}{k}{i}{j}{I}{K}
\ = \ \SJS{i}{j}{k}{I}{J}{K}       \ \ .     
\ee
This concludes our list of representation theoretic data for 
the affine Lie algebra. We will now use these quantities 
to solve our boundary problem for the SU(2) WZW model.  

\paragraph{The closed string sector.} Our first task is to check 
whether our two basic requirements, namely the gluing condition 
(\ref{gluecond}) and the assumptions (\ref{Cassa}), (\ref{Cass}) 
in Cardy's solution,  
are fulfilled by the boundary condition (\ref{GCS}) we would like 
to impose. In terms of the chiral currents (\ref{WZWcurr}), we can 
rewrite eq.\ (\ref{GCS}) as follows
$$ J^\mu(z) \ = \ \bJ^\mu(\bz) \ \ \ \mbox{ for }  \ \ z \ = \ \bz 
$$ 
and $\mu = 1,2,3$. This indeed has the form of the gluing condition 
(\ref{gluecond}) with $\Omega = \id$ and, moreover, it also implies
the gluing property (\ref{glueT}) for the usual Sugawara-Virasoro 
field   
$$ 
T(z)  \ := \ \frac12\, \frac{\delta_{\mu,\nu}}{\ik+2}\ \no J^\mu(z) J^\nu(z) \no 
\ \ .$$ 
The same formula with $\bJ$ instead of $J$ is used for $\bT$. Hence, 
spherical branes on $S^3$ preserve the full chiral algebra of the WZW 
model, including its conformal symmetry. This puts us into an excellent 
position to succeed with the exact solution. 
\smallskip 

Let us now turn to Cardy's assumptions (\ref{Cassa}), (\ref{Cass}). The 
space of bulk fields for our WZW theory is given by the charge conjugate 
modular invariant (recall that $j^+ = j$),
$$ \cH^{(P)} \ = \ \bigoplus_j \cV_j \otimes \bcV_j \ \ ,  $$ 
which has the necessary form (\ref{Cassa}). 
Since our gluing map $\Omega$ is trivial, it induces the identity map 
$\omega(j) = j$ on the labels $j\in \cJ$. Using our earlier observation 
that sectors of the affine su(2) are self-conjugate, $j^+ = j$, we  
conclude $j^\omega = \omega(j^+) = j = \bj$. This guarantees that 
the WZW model with our choice of boundary conditions (\ref{GCS}) can 
be solved by the formulas we provided at the end of the previous lecture !
\smallskip% 

To begin with, we learn from Cardy's solution that there are 
$\ik +1$ possible boundary theories which we label by $J = 0,1/2, 
\dots,\ik/2$, just as we enumerate the sectors of the corresponding 
affine Lie algebra. In the different boundary theories, the bulk 
fields possess different one-point functions. These are given by 
(see eq.\ (\ref{1pt})) 
\be \langle \phi_{j,j}^{ab}(z,\bz) \rangle_J \ = \ 
    \left(\frac{2}{\ik+2}\right)^{\frac{1}{4}} \,  
    \frac{\sin \frac{\pi(2j+1)(2J+1)}{\ik+2}}
    {\left(\sin\frac{\pi(2j+1)}{\ik+2}\right)^{1/2}}  
    \ \frac{\delta^{a,b}}{|z-\bz|^{2h_j}}
\label{wzw1pt} \ee
where $h_j = j(j+1)/(\ik+2)$. The superscripts $a,b$ placed at 
the symbol $\phi$ label different components within the tensor 
multiplet $\phi_{j,j}$ (see eq.\ (\ref{phiij})). Each of them 
runs through a basis in the representation space $V^0_j$ of su(2).  
Since $\Omega$ is trivial, the intertwiner $\cU_{jj}$ between 
the left and right representation of su(2) is trivial as well, 
i.e.\ $\cU^{ab}_{jj} = \delta^{a,b}$. 
\smallskip
   
When we discussed flat space models, we have described how to read 
off a brane's location from its set of one-point functions. Let us 
now try to repeat the same procedure in the case of branes on SU(2) 
(see \cite{FFFS1,MaMoSe1}). According to the Peter-Weyl theory for 
compact groups, the space of functions on SU(2) is spanned by the 
matrix elements $D^j_{ab}(g)$ of finite dimensional unitary 
representations $D^j$. More precisely, the functions 
\be \phi_j^{ab}(g) \ := \ \sqrt{2j+1}\, D^j_{ab}(g) 
\label{phidef} \ee 
form a complete orthonormal basis of $\Fun(S^3)$ just in the same 
way as the exponentials $\exp(ikx)$ do for $\Fun(\mathbb{R}^D)$. 
Writing $\phi_j^{ab}(g)$ in terms of three Euler angles on $S^3$, 
we find in particular  
$$ \sum_{c = -j}^j \phi_j^{cc}(g)\ = \ \sqrt{2j+1}
    \ \frac{\sin \vartheta (2j+1)}{\sin \vartheta} $$ 
where $\vartheta \in [0,\pi]$ parametrizes the azimuthal angle that is 
transverse to conjugacy classes on SU(2). From the completeness 
of $\sin(n\vartheta)$ on the interval $[0,\pi]$ one then concludes 
\be 
    \frac{1}{\sin\vartheta_0} \delta (\vartheta-\vartheta_0) \ = \ 
    \frac{4}{\pi}\ \sum_j \sum_{c=-j}^j \frac{\sin (\vartheta_0 (2j+1))}
     {\sqrt{2j+1}} \ \phi_j^{cc}(g) \ \ .   
\ee
Except from a numerical factor, the coefficients in this expansion 
agree with the limit of our one-point functions (\ref{wzw1pt}) as 
$\ik$ tends to infinity. We can also phrase this observation in a 
form similar to eq.\  (\ref{flatgeom}),  
\be 
     \langle \phi^{ab}_{j,j} \rangle_{J(\ik)} 
   \ \stackrel{\ik \rightarrow \infty}{\sim}  \ 
   \int_{\rm SU(2)} d\mu(g)\ \rho_0 \, \delta(\vartheta(g)-
     \vartheta_0)\, \phi^{ab}_j(g) 
\ee 
where $\rho_0$ is some constant and $d\mu(g)$ denotes the 
Haar measure on SU(2). In taking the limit, we allowed the 
boundary label $J$ to depend on the level $\ik$ and we defined 
$\vartheta_0 := 2 \pi \lim J(\ik)/\ik$. Since $0\leq J \leq 
\ik/2$, the angle $\vartheta_0$ lies in the interval $[0,\pi]$. 
If we want to obtain a 2-sphere with non-zero angle $\vartheta_0$ 
(and hence with non-zero radius), the boundary label $J$ has to 
be scaled up at the same rate as the level $\ik$. The necessity 
for this rescaling is simple to understand. As we have seen, 
boundary WZW theory at level $\ik$ allows us to fit $\ik+1$ 
different branes on the 3-sphere. Therefore, the difference of 
the azimuthal angles between two neighboring branes is roughly 
$\Delta \vartheta_0 = \pi/\ik$. We numbered these branes with 
$J = 0,1/2,1,\dots$ according to the angle at which they appear
and starting with the smallest one at the group unit. When we 
keep $J$ fixed but increase the level $\ik$, then the $J^{th}$ 
brane moves closer and closer to the unit and in the limit 
it shrinks to a point. This is precisely the behavior that 
forces us to set $J(\ik) \sim \vartheta_0 \ik/2\pi$. Later on 
we will introduce another large $\ik$ limit in which open string 
data are kept fixed rather than the angle $\vartheta_0$.

\paragraph{The open string sector.} As in the first lecture, we 
shall restrict our discussion to open strings on a single brane. 
Strings which have their two ends on different branes present no 
additional difficulty. The state space of the $J^{th}$ boundary 
theory is determined by equation (\ref{bpart}) and it has the 
form    
\def\cH{{\cal H}}
\be \label{partdec}
 \cH_{J} \ = \ \cH_{JJ} \ = \ {\bigoplus}_j \ {N_{JJ}}^j \ \cV_j 
\ee
where $\cV_j$, $j = 0,1/2, \dots,\ik/2$, denote irreducible highest 
weight representations of the affine Lie algebra ${\widehat{{\rm SU}}
(2)}_\ik$, and where ${N_{JJ}}^j$ are the associated fusion rules (see 
eq.\ (\ref{wzwfus})). Note that only integer spins $j$ appear on the right 
hand side of (\ref{partdec}) and that the sum is truncated at $j_{max} 
= \min (2J,\ik-2J)$. In the limit $\ik \rightarrow \infty$, we obtain 
$j_{max} = 2J$ so that the decomposition of $\cH_{J}$ is as close as it 
can be to the decomposition (\ref{Matdec}) of $\Mat(2J+1)$ into su(2) 
multiplets. More precisely, there is a unique correspondence between 
fuzzy spherical harmonics $\rmY^j_a \in \Mat(2J+1)$ and ground states 
in $\cV_j \subset \cH_{J}$. This generalizes the relation between  
the Weyl operators $\exp(ik \hat x)$ and the boundary fields $\no \exp 
(ikX(u)) \no$ that we found for flat branes in the first lecture.      
\smallskip

By the state field correspondence $\Psi$, each state $|\psi\rangle \in 
\cH_{J}$ gives rise to a boundary field $\psi(u) = \Psi(|\psi\rangle;u)$.  
Since we deal with a single boundary condition $J$, we will omit the 
superscripts $J$ that we used in rel.\ (\ref{psii}). Ground states in 
$\cH_{J}$ furnish a muliplet of boundary primary fields
$$ \psi^a_j(u) \ := \ \Psi(| e^a_j\rangle;u)
 \ \ \mbox{ where } \ \ \ |a| \leq j $$ 
and $j = 0,1, \dots, j_{max}$. The operator product expansion 
(\ref{bOPE}) of these open string vertex operators reads 
\cite{AlReSc2}  
\be \label{boundOPE}
    \psi^a_i(u_1)\, \psi_j^b(u_2) \, = \, {\sum}_{k,c}\, 
     u_{12}^{h_k - h_i -
    h_j} \, \CG{i}{j}{k}{a}{b}{c}\, \Fus{J}{k}{j}{i}{J}
    {J} \, \psi_k^c(u_2) + \dots     
\ee
where $h_j = j(j+1)/(\ik+2)$ is the conformal dimension of 
$\psi^a_j$ and the symbol in square brackets stands for the 
Clebsch-Gordan coefficients of su(2). The latter provide the 
intertwiners $\cU_{ij;k}$ which appear in the general formula. 
As we send the level $\ik$ to infinity (while keeping the 
boundary label $J$ fixed), the conformal dimensions $h_j$ tend 
to zero so that the OPE (\ref{boundOPE}) of boundary fields 
becomes regular as in a topological model, 
\be \label{boundOPEtop}
    \left( \psi^a_i(u_1)\, \psi_j^b(u_2)\right)^{\ik \rightarrow 
    \infty} 
    \, = \, {\sum}_{k,c}\, 
    \SJS{j}{i}{k}{J}{J}{J} \, \CG{i}{j}{k}{a}{b}{c}\, 
    \psi_k^c(u_2) + \dots     
\ee
Here we have also used the property (\ref{fusprop}) of the fusing matrix. 
Comparing (\ref{boundOPEtop}) with eq.\ (\ref{QSHM}) we make a 
striking observation \cite{AlReSc2}: In fact, the large $\ik$ limit 
of the operator 
product expansion exactly reproduces the multiplication of matrices~! 
We can use this fact to evaluate arbitrary n-point functions of the 
boundary fields $\psi_j^a$ in the limiting regime. Before we spell 
out the result, let us introduce the notation   
$$ \psi[\rmA](u) \ := \ \sum \ \rma_{jb}\  \psi_j^b(u)
   \ \ \ \mbox{ for all } \ \ \ \rmA \ = \ \sum \rma_{j b} 
    \rmY^j_b \ \in \ \Mat(2J+1) \ \ \ . 
$$    
For an arbitrary set of matrices $\rmA_r \in \Mat(2j+1)$, the operator
product expansion formula (\ref{boundOPEtop}) then implies 
\be \langle \psi[\rmA_1](u_1) \ \psi[\rmA_2](u_2) \ \cdots \ 
  \psi[\rmA_n](u_n) \rangle^{\ik \rightarrow \infty}  \ = \ 
   \tr (\rmA_1 \ \rmA_2 \  \cdots
   \  \rmA_n)\ \ .
\label{wzwcorr} \ee
The trace appears because the vacuum expectation value is SU(2) 
invariant and the trace maps matrices to their SU(2) invariant 
component. 
\smallskip 

Our final expression (\ref{wzwcorr}) for the correlators of open string 
vertex operators shares many features with the formula (\ref{flattopcorr}) 
that we encountered through our investigation of branes in flat space. 
In both cases, the vertex operators are in one-to-one correspondence 
with elements of the non-commutative algebra of `functions' on the 
world-volume of the brane. Furthermore, in a limiting regime, the 
correlators are independent of the insertion points $u_r$ and they 
can be evaluated using the multiplication and integration (trace) on 
the non-commutative world-volume algebra. There remains, however, one  
important difference between the two cases: For branes on SU(2), the 
world-volume algebra is cut off at some angular momentum $2J$ so that 
there are only finitely many linearly independent `functions' (resp.\ 
boundary primary fields).     
\medskip 

Before we conclude our discussion of the solution for the boundary 
problem on SU(2), we would like to make one more remark. The attentive
reader may have wondered already, why all the spherical branes in this
section were centered around the group unit $e \in$ SU(2). Since no 
point on a group manifold is distinguished from any other, there should
exist spherical branes with arbitrary locations. It is not hard to 
spot the place in our analysis at which we broke the SU(2) translation 
symmetry. In fact, it is the gluing condition (\ref{GCS}) that forced 
all the branes to have their center at $e$. This lack of democracy,  
however, is easy to overcome if we admit gluing automorphisms $\Omega$ 
taken from the group of inner automorphisms on SU(2). By definition, an 
inner automorphism is of the form $\Omega = \Ad_g$ with some element 
$g \in$ SU(2). Branes surrounding the point $g$ are obtained with
the boundary condition 
\be       J(z) \ = \ \Ad_g \, \bJ(\bz) \ \ \ \mbox{ for } \ \ \ 
         z \ = \ \bz \ \ . 
\label{rotglue}
\ee
Needless to say that the WZW boundary problem can be solved for any of 
these gluing conditions and the solution is essentially the same as 
above. Only the $\delta^{a,b}$ in the formula (\ref{wzw1pt}) for the bulk 
one-point functions must be replaced through the matrix $D^j_{ab}(g)$.  

\subsection{Fuzzy gauge theory and dynamics} % 
Now that we have an exact CFT-solution for spherical branes on 
$S^3$, we also want to mention some applications. As in the case 
of branes in flat space, a non-commutative gauge theory can be 
associated with stacks of branes on group manifolds. But since 
the underlying world-volume geometry is described by matrix 
algebras, the gauge theories turn out to be matrix models. 
We sketch their derivation in the first subsection and then study 
some classical solutions. The latter possess an interpretation in 
terms of bound state formation. We finish this lecture with some 
brief remarks on brane dynamics in the stringy regime.    

\paragraph{Fuzzy gauge theory.} In the first lecture we have seen 
how information about some limiting behavior of certain boundary 
correlators can be stored in a non-commutative Yang-Mills theory 
on the world-volume of the brane. Our aim here is to repeat this 
analysis for correlators of the fields 
$$   \no J^\mu(u) \psi_j^a(u) \no \ = \ J_<^\mu (u)
   \psi_j^a(u) + \psi_j^a(u) J^\mu_<(u)  \ \ . 
$$ 
Here we have used the same split into raising and lowering modes 
as before. The computation of three- and four-point functions of 
these fields follows exactly the same strategy as in the first 
lecture. In particular, it requires expressions for the operator 
product expansions of the currents among each other and with the 
primary boundary fields (see eqs.\ (\ref{JonJ}), (\ref{Jonpsi}) 
for comparison). From the commutation relations (\ref{KMcomm}) 
in the affine Lie algebra we obtain   
\be \left( J^\mu (u_1) \, J^\nu(u_2) \right)_{sing} \ := \
    [\, J^\mu_>(u_1)\, ,\, J^\nu(u_2) \, ] \ = \ 
    {\ik} \, \frac{\delta^{\mu,\nu}}{(u_1-u_2)^2} 
    + \frac{i {f^{\mu\nu}}_\rho}{(u_1-u_2)} \, J^\rho(u_2) 
    \ \ . 
\label{JOPE1}\ee
The operator product expansion between currents and primary 
fields is a special case of eq.\ (\ref{WbOPE}),   
\be \left( J^\mu (u_1) \, \psi_j^a(u_2) \right)_{sing} 
 \ :=\ [\, J^\mu_>(u_1) \, , \, \psi_j^a(u_2) \, ] 
 \ = \ \frac{(t^j_\mu)_{ab}}{(u_1-u_2)} \, 
      \psi_j^b(u_2) \ \ . 
\label{JbOPE}\ee      
Equipped with all these relations, we are able to calculate the 
low-energy effective action for massless open string modes. With 
respect to the flat space case, there occur three important 
changes during the computation. First of all, it follows from 
eq.\ (\ref{wzwcorr}) that all Moyal-Weyl products get replaced 
by matrix multiplication. Second, there appears a new term 
$f^{\mu\nu}_{\ \ \rho} J^\rho$ in the operator product expansion 
of currents (\ref{JOPE1}). This term leads to an extra 
contribution of the form $f_{\mu\nu\rho} \rmA^\mu \rmA^\nu 
\rmA^\rho$ in the scattering amplitude of three massless open 
string modes. Consequently, the resulting effective action is 
not given by Yang-Mills theory on a fuzzy 2-sphere alone but 
involves also a Chern-Simons like term. Finally, all momenta 
$k_\mu$ that arise from contractions of currents with primaries 
in the flat space computation, must be substituted by the 
representation matrices $t^j_\mu$ (see eq.\ (\ref{JbOPE})). After 
Fourier transformation, the momenta turned into derivatives, and 
similarly our matrices $t^j_\mu$ give rise to the generators 
(\ref{Lmu}) of infinitesimal rotations. 
\smallskip

After these sketchy remarks on the derivation, let us now display 
the final answer (many more details can be found in \cite{AlReSc3}). 
For $M$ branes of type $J$ on top of each other, the results 
of a complete computation can be summarized in the following 
formula (see \cite{AFQS} for normalizations), 
\be  \label{effact}
  \cS_{(M,J)}\ =\ \cS_{{\rm YM}} + \cS_{{\rm CS}}\ = \
   \frac{\pi^2}{\ik^2\, d_J\, M} \ \left( \frac{1}{4}\ \tr 
    \left( \tF_{\mu\nu} \ \tF^{\mu\nu} \right) 
   - \frac{i}{2}\  \tr \left( \tf^{\mu\nu\rho}\; \tCS_{\mu\nu\rho} 
   \right) \right) 
\ee
where $d_J = 2J+1$. We defined the `curvature form' $\tF_{\mu\nu}$ by 
the expression     
\be \label{fieldstr}
  \tF_{\mu\nu}(\tA)\, = \,   
   i\, \tL_\mu \tA_\nu - i\, \tL_\nu \tA_\mu + i \,[ \tA_\mu \astk 
   \tA_\nu] 
   +  \tf_{\mu\nu\rho } \tA^\rho
\ee
and a non-commutative analogue of the Chern-Simons form through 
\be  \label{CSform}
  \tCS_{\mu\nu\rho}(\tA)\, = \, \tL_\mu \tA_\nu \, \tA_\rho 
                   + \frac{1}{3}\; \tA_\mu \, [ \tA_\nu \astk \tA_\rho]
                   - \frac{i}{2}\; \tf_{\mu\nu\sigma}\; \tA^\sigma \, 
                   \tA_\rho \ \ .
\ee
The three fields $\tA^\mu = \sum \rma^\mu_{ja} \rmY^j_a$\ on the fuzzy 
2-sphere $S^2_J\in \mathbb{R}^3$ take values in $\Mat(M)$, i.e.\ all 
their Chan-Paton coefficients $\rma^\mu_{ja}$ should be considered as 
$M\times M$ matrices. Hence, the fields $\tA^\mu$ are elements of \ 
$\Mat(M) \otimes \Mat(2J+1)$. 
Infinitesimal rotations $\tL_\mu$ act exclusively on the fuzzy spherical 
harmonics $\rmY^j_a$ and they commute with the Chan-Paton coefficients 
$\rma^\mu_{ja}$. 
\smallskip 

It follows from a straightforward computation that the action (\ref{effact}) 
is invariant under the gauge transformations 
$$ \tA_\mu  \ \rightarrow \ \tA_\mu + 
   i\, \tL_\mu \lambda \ +\  i\, [\, \tA_\mu\, 
     ,\, \lambda\, ]  \ \ \ \mbox{ for } \ \ \ \lambda \in 
    \Mat(M) \otimes \Mat(2J+1)  \ \ . 
$$
Note that the `mass term' in the Chern-Simons form (\ref{CSform}) 
guarantees the gauge invariance of $\cS_{{\rm CS}}$. On the other hand, 
the effective action (\ref{effact}) is the unique linear combination of 
$\cS_{\rm YM}$ and $\cS_{\rm CS}$ from which mass terms cancel. As we 
shall see below, it is this special feature of our action that allows 
solutions describing translations of the branes on the group manifold. 
The action $\cS_{\rm YM}$ was already considered in the non-commutative 
geometry literature \cite{GrKlPr95/1,GrKlPr95/2,GrKlPr95/3,MadoBook,
WatWat}, where it was derived from a Connes spectral triple and 
viewed as describing Maxwell theory on the fuzzy sphere. Arbitrary 
linear combinations of non-commutative Yang-Mills and Chern-Simons 
terms were considered in \cite{Klimcik:1999uk}.  

\paragraph{Classical solutions and brane dynamics.} 
Stationary points of the action (\ref{effact}) describe condensation 
processes on a brane configuration $Q = (M,J)$ which drive the whole 
system into another configuration $Q'$. To identify the latter, we 
have two different types of information at our disposal. On the one 
hand, we can compare the tension of D-branes in the final configuration 
$Q^\prime$ with the value of the action $\cS_Q(\Lambda)$  at the classical 
solution $\Lambda$. On the other hand, we can look at fluctuations around 
the chosen stationary point and compare their dynamics with the low-energy 
effective theory $\cS_{Q^\prime}$ of the brane configuration 
$Q^\prime$. In formulas, this means that 
\begin{equation}\label{fluctuation}
\cS_{Q} (\Lambda+\delta \tA)\ \overset{!}{=}\  \cS_{Q} (\Lambda) + 
\cS_{Q^\prime} (\delta \tA) \ \ \ \mbox{ with } \ \ \
\cS_Q(\Lambda) \ \overset{!}{=} \ \ln \frac{g_{Q^\prime}}{g_Q} 
\ \ . 
\end{equation}
The second requirement expresses the comparison of tensions in 
terms of the g-factors \cite{AffLud91/1} of the involved conformal 
field theories
\be \label{mass}
g_{(M,J)} \ := \ M \, g_J \ := \ M\, \langle \phi_{0,0}(z,\bz) \rangle_J
   \ = \ M\, \frac{S_{J0}}{\sqrt{S_{00}}}  % \ = \ 
% \left( \frac{2}{\ik+2} \right)^{\frac{1}{4}} 
%  \frac{\sin\frac{(2J+1)\pi}{\ik+2}}{\sin^{\frac12}\frac{\pi}{\ik+2}}
    \ \ \ .   
\ee   
All equalities must hold to the same order in $(1/k)$ that we used when 
we constructed the effective actions. We say that the brane configuration 
$Q$ decays into $Q^\prime$ if $Q^\prime$ has lower mass, i.e.\ whenever 
$g_{Q^\prime} < g_{Q}$. 
\smallskip 

In terms of the world-sheet description, each classical solution of 
the effective action is linked to a conformal boundary perturbation in 
the CFT of the brane configuration $Q$ (cf. results in \cite{AffLud91/1}). 
Adding the corresponding boundary 
terms to the original theory causes the boundary condition to change so 
that we end up with the boundary conformal field theory of another brane 
configuration $Q'$. Recall, however, that all these statements only apply 
to a limiting regime in which the level $\ik$ is sent to infinity. 
\medskip

Let us now become more specific. From eq.\ (\ref{effact}) we obtain 
the following equations of motion for the elements $\tA^\mu \in 
\Mat(M) \otimes \Mat(2J+1)$   
\begin{equation} 
\label{eom} 
\tL_\mu \, \tF^{\mu\nu} \, + \, [\,  \tA_\mu  \;\, , \tF^{\mu\nu}
  \, ]  \ = \ 0 \ \ . 
\end{equation} 
It is easy to find two very different types of solutions. For the first 
one, the gauge fields $\tA_\mu$ are of the special form $\tA^\mu = \rma
^\mu_{00} \rmY^0_0$ with three pairwise commuting Chan-Paton matrices 
$\rma^\mu_{00} \in \Mat (M)$. These solutions come as a $3M$ parameter 
family corresponding to the number of eigenvalues appearing in 
$\{\rma^\mu_{00}\}$. 
The same kind of solutions appears for branes in flat backgrounds. 
They describe rigid translations of the $M$ branes on the 
group manifold. Since each brane's position is specified by $3$ 
coordinates, the number of parameters matches nicely with the 
interpretation. Moving branes around in the background is a rather 
trivial symmetry and the corresponding conformal field theories 
are easy to construct, either directly (see remarks after eq.\ 
(\ref{rotglue})) or through conformal perturbation theory (see 
\cite{RecSch2,RecSch3}). As we have mentioned before, the existence 
of such continuous families of solutions is guaranteed by the absence 
of the mass term in the full effective action.  
\smallskip

There exists a second type of solutions to eqs.\ (\ref{eom}) 
which is a lot more interesting. In fact, any $M(2J + 1)$-dimensional
representation of the Lie algebra su(2) can be used to solve
the equations of motion. Their interpretation was found in 
\cite{AlReSc3}. Here, we describe the answer for a stack of 
$M$ branes of type $J = 0$, i.e.\ of $M$ point-like branes 
at the origin of SU(2). In this case, $\tA_\mu \in \Mat(M) 
\otimes \Mat(1) \cong \Mat(M)$ so that we need an $M$-%
dimensional representation of su(2) to solve the equations
of motion. Let us choose the $M$-dimensional irreducible 
representation of isospin $J_M= (M-1)/2$. Our claim then is 
that this drives the initial stack of $M$ point-like branes 
at the origin into a final configuration containing only a 
single brane wrapping the sphere of type $J_M$, i.e.\ 
$$ (M,\,J = 0 )  \ {\longrightarrow} \ 
   (1,\,J_M = (M-1)/2) \ \ . $$
Support for this statement comes from both the open string 
sector and the coupling to closed strings. 
In the open string sector one can study small fluctuations 
$\delta \tA_\mu$ of the fields $\tA_\mu = \Lambda_\mu + 
\delta \tA_\mu  \in \Mat(M)$ around the stationary point 
$\Lambda_\mu\in \Mat(M)$. If $\Lambda_\mu$ form an irreducible 
representation of su(2), we find 
$$  \cS_{(M,0)} (\Lambda_\mu + \delta \tA_\mu) \ = \ 
    \cS_{(1,J_M)}(\delta \tA_\mu) \ + \  \mbox{const}\ \ \ .  
$$ 
In the closed string channel, the leading term (in the 
1/\ik-expansion) from the exact `mass' formula (\ref{mass}) 
gives \cite{AlReSc3}
$$  \ln \frac{g_{(1,J_M)}}{g_{(M,0)}} \ = \ 
    - \frac{\pi^2}{6} \, \frac{M^2-1}{\ik^2} \ = \ 
    \cS_{(M,0)}(\Lambda_\mu) \ \ . 
$$     
Note that the mass of the final state is lower than the mass of 
the initial configuration. This means that a stack of $M$ 
point-like branes on a 3-sphere is unstable against decay into 
a single spherical brane. Stationary points of the action 
(\ref{effact}) and the formation of spherical branes on $S^3$ 
were also discussed in \cite{HasKra,HiNoSu,JMWY}. Similar 
effects have been described for branes in RR background fields 
\cite{Myers}. The advantage of our scenario with NSNS background 
fields is that it can be treated in perturbative string theory so 
that string effects may be taken into account (see 
\cite{AleSch6,FreSch1,FreSch2} and below). 
\vspace{1cm}
\fig{$M$ point-like branes stacked at the origin of a weakly 
curved $S^3 \sim \mathbb{R}$ are unstable against decay into 
a single spherical brane with label $J_M = (M-1)/2$.}
{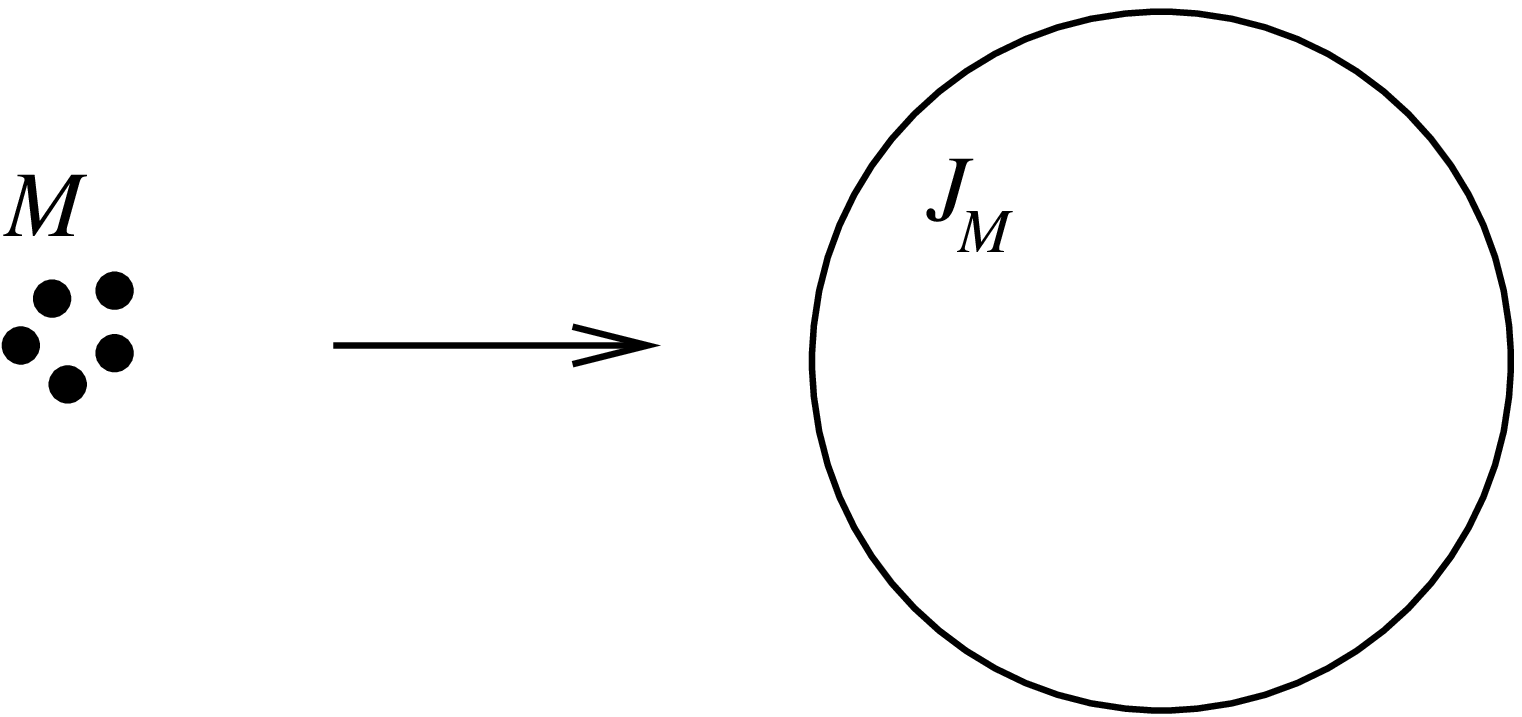}{10truecm}
\figlabel{\basic}
\vspace{1cm}

\paragraph{Dynamics in stringy regime and K-theory.}
\def\Z{\mathbb{Z}}
 
Now we would like to understand the dynamics of branes in the stringy 
regime where $\ik$ is finite. Proceeding  along the lines of the 
previous discussion would force us to include higher order corrections 
to the effective action. Unfortunately, such a complete control of the 
brane dynamics in the stringy regime is out of reach.  
\smallskip

But we could be somewhat less ambitious and ask whether at least some 
of the solutions we found in the large volume limit possess a deformation 
into the small volume theory and if so, which boundary conformal field
theories they correspond to. It turns out that this is possible for all 
the processes that are obtained from constant gauge fields on the brane. 
In this way we may overlook new stationary points of the stringy effective 
action that have no well behaved large $\ik$ limit. On the other hand, the
reduced program has a positive and very beautiful solution that is known 
from the work on the Kondo effect.
\smallskip

The Kondo model is 
designed to understand the effect of magnetic impurities on 
the low-temperature conductance properties of a 3D conductor. 
The latter can have electrons in a number $\ik$ of conduction 
bands. If the impurities are far apart, their effect may be 
understood within an s-wave approximation of scattering events 
between a conduction electron and the impurity. This allows to 
formulate the whole problem on a 2-dimensional world-sheet for 
which the coordinates $(u,v)$ are associated with the time and
the radial distance from the impurity, respectively. One can 
build several currents out of the basic fermionic fields. Among 
them is a {\em spin current} $\vec{J}(u,v)$. It satisfies the 
relations (\ref{JOPE1}) of a $\widehat{\rm SU}(2)_\ik$ current 
algebra. This spin current is the one that couples to the magnetic 
impurity of spin $J_M$ which is sitting at the boundary $v=0$,    
\begin{equation} \label{Kondo} 
 S_{\rm pert} \ \sim \ \lambda \int_{-\infty}^\infty du\, 
\Lambda_\mu J^\mu(u,0)  \ \ . 
\end{equation}
Here, $\vec{\Lambda} = (\Lambda_\mu, \mu=1,2,3)$ is a 
$(2J_M+1)$-dimensional irreducible representation of $su(2)$
and the parameter $\lambda$ controls the strength of the coupling.  
The term (\ref{Kondo}) is identical to the coupling of open 
string ends to a background gauge field $\rmA_\mu = \Lambda_\mu 
\in \Mat(2J_M+1)$. Hence, $\Lambda_\mu$ may be interpreted as a 
constant gauge field on a Chan-Paton bundle of rank $M=2J_M+1$.      
\smallskip 

Fortunately, a lot of techniques have been developed to deal with 
perturbations of the form (\ref{Kondo}) going back even to the work 
of Wilson \cite{Wilson}. In fact, this problem is what Wilson's 
renormalization group techniques were originally designed for. From the 
old analysis we know that there are two different cases to be distinguished. 
When $2 J_M > \ik$ (`under-screening') the low-temperature fixed point 
of the Kondo model appears only at infinite values of $\lambda$. On 
the other hand, the fixed point is reached at a finite value $\lambda 
= \lambda^*$ of the renormalized coupling constant $\lambda$ if $ 2 J_M 
\leq \ik$ (exact- or over-screening resp.). In the latter case, the 
fixed points are described by non-trivial (interacting) conformal 
field theories. One can summarize the results on the spectrum of 
the fixed points through the so-called `absorption of the boundary 
spin'-principle \cite{AffLud91/2,AffLud91/3} 
\begin{equation} \label{charpert}  
  \tr_{V_{J_M} \o \cV_j}\left( q^{H_0 + 
  H_{\rm pert}}\right)_{\lambda = \lambda^*}^{\rm ren} 
  := \sum_l {N_{J_M j}}^{l} \chi_l(q)\ \ .  
\end{equation} 
Here, $H_0 = L_0 + c/24$ is the unperturbed Hamiltonian, the 
superscript $\ ^{\rm ren}$ stands for `renormalized' and $V_{J_M}$ 
denotes the representation space of the representation $J_M$ of 
su(2). 
\smallskip 

In the rule (\ref{charpert}), $\cV_j$ can be any of the sectors in 
the state space $\cH$ of the boundary theory. Formula \eqref{charpert} 
means that our perturbation with some irreducible representation 
with spin $J_M$ interpolates continuously between a building block 
$M \chi_j(q)$ of the partition function of the UV-fixed point (i.e.\ 
$\lambda = 0$) and the sum of characters on the right hand side of 
the  previous formula, 
\begin{equation} \label{abs} M \,  \chi_j (q) \ 
   \longrightarrow \ \sum_l {N_{J_M j}}^{l}\,  \chi_l(q)\ \ , 
 \end{equation}
where $M = 2J_M + 1$. In particular, we can use this formula
to determine the decay product of a stack of $M$ point-like 
branes. Since each of the two string ends can be attached to 
any of the $M$ branes, the partition function of the whole 
stack is $M^2$ times the partition function for a single 
brane. For this system we find    
\ba Z_{(M,0)} (q) & = & M^2 \, Z_{(1,0)}(q) \ = \ M^2 \, \chi_0(q) 
    \nn \\[2mm]  
  & \longrightarrow & M \, \chi_{J_M}(q) \ \longrightarrow \  
  \sum_j {N_{J_M J_M}}^j \, \chi_j(q) \ = \ 
   Z_{(1,J_M)}(q) \ \ . \nn 
\ea 
We applied the rule (\ref{abs}) twice because both endpoints 
of an open string couple to the background field. The result can 
be summarized in the following process  
\be  (M, \, 0 )  \ \longrightarrow \ 
    (1,\, J_M ) \ \ . \label{rule1}\ee
This is formally identical to the decay process we found in the 
large $\ik$ regime, except that this time $2J_M$ is bounded 
from above by the level $\ik$. Our final answer may not seem 
very surprising, but it is still remarkable that there exists 
such a solid derivation even deep in the stringy regime. 

\paragraph{Charges and twisted K-theory.} The analysis of brane 
dynamics on $S^3$  brings us to the last subject of this lecture, 
namely the issue of brane charges. It is a traditional conception 
to measure brane charges through their coupling to closed string 
(RR) modes. When applied to branes on $S^3$, however, this naive 
idea of charge seems to fail. In particular, the couplings of 
branes to closed strings are not quantized \cite{BaDoSc}, at least 
as long as we are not at the limit point of infinite level. For 
this reason, an alternative definition of brane charges was 
proposed in \cite{AleSch6}. There it was suggested to define 
charges as quantities which are invariant under conformal 
perturbations in the world-sheet theory. More precisely, two 
brane configurations $Q$ and $Q'$ are called {\em dynamically 
equivalent} if there exists a conformal boundary perturbation 
that relates the two boundary theories. The space of all 
(anti-)brane configurations modulo this dynamical equivalence 
is the group $C$ of brane charges. The latter is a property of the 
background. 
\smallskip 
 
To determine the brane charges on $S^3$ along with the group they 
generate, let us apply the rule (\ref{rule1}) to a supersymmetric 
theory on a 3-sphere with $K = \ik+2$ units of NSNS flux passing 
through. In this case one can have (anti-)branes wrapping $\ik+1$ 
different conjugacy classes labeled by $J = 0,1/2, \dots,\ik/2$ (see 
e.g. \cite{AleSch6}). If we stack more and more point-like branes at 
the origin, the radius of the sphere that is wrapped by the resulting 
object will first grow, then decrease, and finally a stack of $\ik+1$ 
point-like branes at $e$ will decay to a single point-like object at 
$-e$ (see Figure 8).  By taking orientations into account, one can 
see that the final point-like object is the translate of an 
anti-brane at $e$. Hence, we conclude that the stack of $\ik+1$ 
point-like branes at $e$ has decayed into a single point-like 
anti-brane at $-e$. 
\vspace{1cm} 
\fig{\label{su2branen} Brane dynamics on $S^3$: 
A stack of  point-like branes at $e$ can decay into a single spherical 
object. The distance of the latter increases with the number of branes
in the stack until one obtains a single point-like object at $-e$.}
{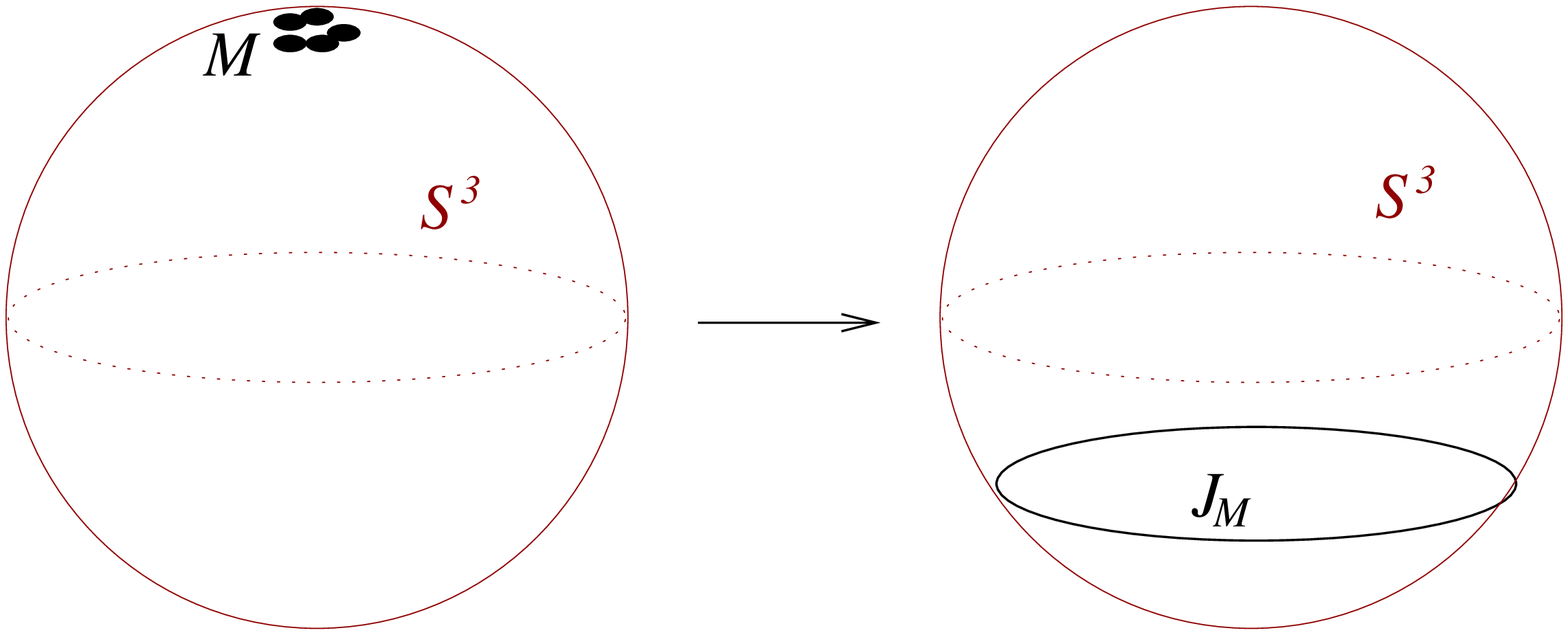}{11truecm}
\figlabel{\basic}
\vspace{1cm}  

In this concrete example, we may assign charge~$1$ to the point-like
branes at $e$ and if we want the charge to be conserved, the decay 
product of $\ik+1$ such point-like branes must have charge $\ik+1$. On 
the other hand we identified the latter with a single anti-brane which 
has charge $-1$. Thus we have to identify $\ik+1$ and $-1$ which means 
that charge is only well-defined modulo $K=\ik+2$ \cite{AleSch6}, i.e.\ 
$$ C(\SU,K) \ = \ \Z_K \ \ . $$
According to a proposal of Bouwknegt and Mathai \cite{BouMat}, the 
brane charges on a background $X$ with non-vanishing NSNS 3-form 
field $H$ take values in some twisted K-groups $K_H^*(X)$ which 
feel the presence of $H \in H^3(X,\Z)$ (see also \cite{WitK,Kapu}
for related proposals when $H$ is torsion). For $S^3$ this twisted 
K-group is known to be $K^*_H(\SU) = \Z_K$ and hence  
$$ C(\SU,K) \ = \ K^*_H(\SU) $$ 
as predicted in \cite{BouMat}. Many more details about this twisted 
K-theory, its computation through spectral sequences and the relation 
with string theory can be found in \cite{MaMoSe2}. Geometric 
construction of brane charges on $S^3$ and other group manifolds have 
also been discussed in \cite{Tay,Stan00/2,AleSch6,FiguStan}. 
\newpage

\section{Some further results and directions} 
\setcounter{equation}{0}
\def\LieG{{\mathfrak{g}}}

During the last years, all ingredients of the technology we have 
used so far were generalized in a variety of different directions. 
Our aim in this final part is to touch upon some of these extensions, 
trying to provide some ideas of the current status in the field along 
with a very incomplete guide to the existing literature. 
\smallskip 

We shall begin with several remarks on the various approaches that more
recent research has followed to generalize the construction of boundary 
conformal field theories beyond the Cardy-case. These include orbifold 
and simple current techniques and the use of conformal embeddings. Once 
more, group manifolds serve as a stage on which we can nicely present
some of the progress that has been made in obtaining exact solutions. 
These lectures finally end with a short summary of some first steps in  
extending the whole program to non-compact backgrounds, focusing on 
some of the main new ingredients and difficulties.

\subsection{Solutions beyond the Cardy case} 

For the exact solutions we have outlined and applied above, we 
had to make several assumptions. Not only were we restricted to 
a very particular class (\ref{Cassa}) of bulk modular invariants, 
but also it was necessary to preserve the maximal chiral symmetry 
$\cW$ of the model through our eqs.\ (\ref{gluecond}). The 
classification of all conformal boundary theories, on the other 
hand, would require to impose the gluing condition (\ref{glueT}) 
for the Virasoro field only. Since chiral symmetries of solvable 
theories with $c\geq 1$ are much larger than the Virasoro algebra, 
there remains a lot of room for new symmetry breaking conformal 
boundary conditions. In fact, constructing boundary theories with 
the minimal Virasoro symmetry tends to lead into non-rational 
models which are notoriously difficult to control. Nevertheless, 
some progress has been made in this direction. Boundary conditions 
with the minimal Virasoro symmetry were systematically investigated 
for 1-dimensional flat targets \cite{GaReWa,GabRec,Janik,CapDAp}. 
In spite of this remarkable progress, such a complete control over 
conformal boundary conditions should be considered exceptional 
and it is probably very difficult to achieve for more complicated 
backgrounds. Less ambitious programs therefore focus on intermediate 
symmetries which are carefully selected so as to render the boundary 
theory rational. We shall sketch the two main approaches in this 
direction and then end this subsection with a few remarks on other 
solution generating techniques.

\paragraph{Orbifold constructions.} In this part we discuss 
some elements of branes in so-called simple current orbifolds. It 
is advantageous at first, to think of the following constructions 
as providing boundary theories for a new class of bulk modular 
invariants, more general than eq.\ (\ref{Cassa}). Later we shall 
then argue that applications of the general formalism also 
include the construction of new boundary theories for backgrounds 
with a bulk invariant of the form (\ref{Cassa}) by analyzing the 
model with respect to a certain orbifold chiral algebra.%
\smallskip 

Investigations of branes in orbifolds have a long history and 
it is not possible to give a complete account here of all the 
existing results. Much of the work was devoted to orbifold 
constructions in flat space (see e.g.\ \cite{DouMoo,AffOsh1,
AffOsh2,Dou,DouFio,DiaGom,Gom,GabSte,Gab,CraGab,BiCrRo}). The 
basis for most of these developments were laid in \cite{DouMoo} 
which uses earlier ideas originating from \cite{PraSag,GimPol}. 
Open string theory in more general conformal field theory 
orbifolds was pioneered by Sagnotti and collaborators starting 
from \cite{PraSag} (see also e.g.\ \cite{PrSaSt1,PrSaSt2,
DMS}). Important contributions were made later by Behrend et al. 
\cite{BPPZ1,BPPZ2} and by Fuchs et al. \cite{FucSchSB1,FucSchSB2,
BiFuSc}. The latter extends the simple current techniques that 
were developed for closed strings in \cite{HalpSC,IntrSC,SchYan1,
SchYan2} to the case of open strings (see also \cite{HuScSo,Sche,
GabGan,IshikTan}). 
\smallskip

Geometrically, branes in an orbifold background are understood 
through branes on its covering space. More specifically, we represent 
an orbifold brane by several pre-images on the covering space which 
are mapped onto each other by the action of the orbifold group. If 
the latter has fixed points, the corresponding branes can be 
resolved so that several different branes are associated with 
the same pre-images on the covering space. These basic ideas 
are common to all orbifolds, and they apply in particular to 
exactly solvable models in which the orbifold action is generated 
by simple currents. Before we state some of the main results that 
have been obtained in this context, we need a few new notions from 
conformal field theory. 
\smallskip

Within the set $\cJ$ of $\cW$ sectors one often finds non-trivial 
elements $\c \in \cJ$ such that the fusion product of $\c$ with any 
other $j \in \cJ$ gives again a single primary $\c \circ j = \c j 
\in \cJ$. Such elements $\c$ are called {\em simple currents} and 
the set of all these simple currents forms an abelian group 
whose product is inherited from the fusion product. From now on, 
let $\Gamma$ denote the group of simple currents or some subgroup 
thereof. Through the fusion of representations, the group $\Gamma$ 
acts on the index set $\cJ$. This action can be diagonalized by 
the S-matrix in the sense that  
\be
S_{\c i\, j} \ = \  e^{2\pi i Q_{\c}(j) }\,  S_{i\, j} \ \ .
\label{Sprop} 
\ee
The quantity $Q_\c(j)$ is known as the {\em monodromy charge}  
and it may be computed from the conformal weights by means of  
the formula $Q_\c(j) = h_j + h_{\c} - h_{\c j}\mod \ 1$. As 
one can see from definition (\ref{Sprop}), the monodromy 
charge gives rise to a character of the group $\Gamma$.  
\smallskip 

Simple current techniques allow to solve the boundary problems 
for bulk partition functions of the following form (see e.g.\ 
\cite{SchYan2})
\be\label{intmod}
      Z^{\orb}(q,\bar q) \ = \ \sum_{[j], Q_\Gamma (j) = 0 } \ 
        |\, \S_{[j]}\, | \
         \bigl(\sum_{j' \in [j]}\ \chi_{j'}(q) \, \bigr)\ 
        \bigl( \sum_{\bj' \in [\bj]}\ \chi_{\bj'}(\bar q) \,
        \bigr) \ \  .  
\ee
Here, we use the symbol $[j]$ to denote the orbit of $j$ under 
the action of $\Gamma$ and we define the stabilizer subgroup 
$\S_{[j]} \subset \Gamma$ by 
\be\label{stab}
 \S_{[j]} \ = \ \{ \ \c \in \Gamma \ \mid\  \c \circ j \ = \ j\ \}
\ \ . \ee
Note that the partition function (\ref{intmod}) does not have the 
simple form (\ref{Cassa}) so that Cardy's theory for the classification and 
construction of branes does not apply directly. 
\smallskip 

But if we choose the gluing map $\Omega$ such that the assumption 
(\ref{Cass}) is satisfied, then the solution to the corresponding 
boundary problem is inherited from Cardy's solution of the theory 
with bulk invariant (\ref{Cassa}) through a simple construction 
that follows very closely the geometric procedure we sketched above
(our presentation follows \cite{BruSch1,BruSch2,MaScSm}). 
In fact, let us recall that the boundary states in Cardy's theory 
are given by 
\be |J\rangle_\Omega \ = \ \sum_i \ \frac{S_{Jj^+}}{\sqrt{S_{0j^+}}}
\, |j \rangle\!\rangle_\Omega \ \ . 
\label{CardyBS} \ee 
Here, $J$ runs through $J \in \cJ$ and we have seen that in some 
sense it encodes the brane's transverse position. The geometric 
ideas suggest to introduce 
\be |[J]\rangle^{\orb}_\Omega \:= \ \frac{1}{\sqrt{|\Gamma|}}\ 
   \sum_{\c \in \Gamma} \ |\c J \rangle_\Omega \ \  .
\label{orbBS}\ee
On the right hand side, we sum over the orbit of `pre-images' of 
the brane $[J]$. To see that the sum (\ref{orbBS}) is a boundary 
state of the orbifold theory (\ref{intmod}), we insert the 
expression (\ref{CardyBS}) into (\ref{orbBS}). Using the relation 
(\ref{Sprop}) we find 
\be |[J]\rangle^{\orb}_\Omega \ = \ \frac{1}{\sqrt{|\Gamma|}}\ 
  \sum_j \,  \bigl(\, \sum_{\c \in \Gamma} \, e^{2\pi i Q_\c(j)} \, \bigr)  
   \, \frac{S_{Jj^+}}{\sqrt{S_{0j^+}}}\, |j \rangle\!\rangle_\Omega
\ \ . 
\label{projBS}\ee
The term in brackets is non-zero, if and only if $Q_\Gamma(j) = 0$. 
Hence, all the generalized coherent states that appear on the right 
hand side of the previous formula are indeed associated with sectors 
in the bulk theory (\ref{intmod}).        
\smallskip 

From the boundary state (\ref{orbBS}) it is easy to work out 
the corresponding boundary partition functions of the orbifold  
model. They are given by
\beq \label{Zproj}  
   Z^{\orb}_{[I][J]}(q) \ = \  
  \sum_{\c,k} \ N_{I \,\, k}^{\ \ \c J} \chi_{k}(q)\ \ .
\end{equation}
This agrees precisely with the prediction from the geometric 
picture of branes on orbifolds. In fact, the $I,J$ can be 
considered as geometric labels specifying the position of the
brane on the covering space. To compute spectrum of two branes 
$[I]$ and $[J]$ of the orbifold theory, we lift $[I]$ to one of 
its pre-images $I$ on the covering space and  include all the 
open strings that stretch between this fixed brane $I$ on the 
cover and an arbitrary pre-image $\c J$ of the second brane $[J]$. 
\smallskip

It is important to notice that in many cases the boundary conditions 
$[I]$ are not elementary and can be further resolved, i.e.\ there 
exists a larger set of boundary theories such that $[I]$ can be 
written as a superposition  of boundary theories with integer 
coefficients. This happens whenever the stabilizer subgroup 
$\S_{[I]}$ is non-trivial. In the absence of discrete torsion, the 
elementary branes resolving the boundary condition $[I]$ are labeled 
by irreducible representations of $\S_{[I]}$.\footnote{More general 
possibilities including discrete torsion have been discussed in 
\cite{Dou,DouFio,Gom,DiaGom,Gab,CraGab}. The extension to conformal 
field theory backgrounds can be found in \cite{FHSSW}} 
Geometrically, this corresponds to the fact that the Chan-Paton 
factors of branes at orbifold fixed points can carry different 
representations of the stabilizer subgroup. General formulas for 
the resolved boundary states and the corresponding boundary partition 
functions can be found e.g.\ in \cite{BPPZ1,FucSchSB1,BPPZ2}.  
\smallskip

Under some additional technical assumptions, operator product 
expansions of boundary fields in simple current orbifolds have 
been studied in \cite{BruSch2,MaScSm}. The results in \cite{MaScSm} 
also cover part of the results on boundary operator product 
expansions for the D-type modular invariant of minimal model 
\cite{Runk2}. There exist several classes of important models 
to which all these findings on branes in simple current orbifolds 
apply. Among them are the Gepner models \cite{Gepn} which are used 
to study aspects of string theory on complete intersection 
Calabi-Yau spaces. The projection method that is encoded in  
formula (\ref{projBS}) was used in \cite{RecSch1} to obtain 
boundary states of Gepner models. Some of these states had to be 
resolved and at least for so-called A-type branes this was done 
explicitly in \cite{BruSch1} (see also \cite{NakNoz,GoJaSa,FuScWa}). 
First geometric interpretations for boundary states in Gepner models 
were found in \cite{BDLR}. There has been a lot of substantial, more 
recent work on branes in Gepner models (see e.g.\ \cite{FKLLSW,FHSSW,
BruDis,Reckperm} and references therein).  
\medskip 

\def\UZ2{{\widehat{\rm U}}{\rm (1)}^{\mathbb{Z}_2}}
As we have briefly mentioned above, there is one rather interesting  
application of the constructions we have presented in this subsection. 
Namely, they can be used to build new branes in backgrounds with an 
invariant of the form (\ref{Cassa}). For branes in compact group 
manifolds, this will be addressed below. Here we shall only take a 
look at the simplest example, namely we show how to recover Neumann 
boundary conditions for a single free bosonic field from Cardy's 
solution. Let us recall that our application of Cardy's solution to the 
flat space theory in section 3.4 only gave theories with Dirichlet 
boundary conditions. Neumann gluing conditions for a 1-dimensional 
flat space, i.e.\ the relations $J(z) = \bJ(\bz)$, preserve the same 
amount of symmetry and they should not be much harder to construct. 
If we compare the two gluing conditions we find that they differ 
only by a reflection $\lambda(J)(z) =  -J(z)$. This defines an 
action of the group $\mathbb{Z}_2$ on the chiral algebra which 
obviously comes from the geometric reflection symmetry of the 
underlying background. Operators which contain an even number of 
oscillators $\a_n$ are invariant under the reflection and we denote 
the corresponding chiral algebra by $\cW = \UZ2$. The idea is now to 
start with the Cardy theory for trivial gluing conditions on $\cW$. 
The above constructions then turn out to furnish all maximally 
symmetric branes for the theory whose target is real line. 

The chiral algebra $\UZ2$ is known to possess the following sectors 
(see e.g.\ \cite{DVVV}). To begin with, there is a continuous family 
of sectors which are labeled by positive momenta $k > 0$. The vacuum 
sector of the U(1) theory splits into two sectors of $\UZ2$ which we 
denote by $0^\pm$. Finally, there are two twisted sectors $\tau^\pm$. 
Even though not all the data from the representation theory of $\UZ2$ 
have been worked out explicitly, we can in principle apply Cardy's 
theory to construct boundary theories for the charge conjugate bulk 
modular invariant. The latter is known to appear when we describe 
closed strings moving on the quotient $\mathbb{R}/ \mathbb{Z}_2$. 
Cardy's theory provides us with a continuum of boundary states $|x_0
\rangle$ which are labeled by some positive real number $x_0 > 0$. 
These can be identified with point-like branes at regular points of 
the quotient space. In addition, there are four discrete boundary 
states $|0^\pm\rangle$ and $|N^\pm\rangle$. Among them are two 
point-like branes sitting at the singularity and a pair of 
1-dimensional branes which extend throughout the whole background. 
\smallskip 

Now we would like to descend from this theory on  $\mathbb{R}/ 
\mathbb{Z}_2$ to some simple current orbifold. Note that $\UZ2$ 
possesses a simple current $\c = 0^-$ which acts on $\cJ$ according 
to $\c \circ\, 0^\pm = 0^\mp, \c \circ\, \tau^\pm = \tau^\mp$ and 
leaves all sectors $k > 0$ from the continuous series invariant. 
If we use this simple current in our formulas above, the bulk modular 
invariant (\ref{intmod}) coincides with the diagonal 
modular invariant (\ref{bulkFB}) of an uncompactified free bosonic field.
According to the general rules we stated above, the simple current 
orbifold possesses two boundary states which are associated with the 
two orbits $[0^+]$ and $[\tau^+]$ of discrete sectors. These boundary 
states describe a point-like brane at $x_0 = 0$ and the 1-dimensional 
brane on $\mathbb{R}$ that we were after. In addition, each sector 
$k > 0$ gives rise to two boundary states labeled by the irreducible 
representations of $\S_{[k]} = \mathbb{Z}_2$. These belong to 
point-like branes at points $\pm x_0$, respectively. In this way, 
all maximally symmetric branes on $\mathbb{R}$ can be constructed 
in terms of representation theoretic data of the chiral 
algebra $\UZ2$.

\paragraph{Twisted branes in WZW models.} The non-abelian generalization 
of the ideas that allowed to obtain Neumann boundary conditions from 
Cardy's solution provides us with new maximally symmetric branes on 
group manifolds. Recall that maximally symmetric boundary conditions 
in the WZW model require to choose some gluing automorphism $\Omega$ 
of the Lie algebra $\LieG$ so that we can glue holomorphic and 
anti-holomorphic currents along the boundary. Even though most of our 
discussion was restricted to $\Omega = \id$, we commented briefly on 
more general situations in which $\Omega = \Ad_g$ is an inner 
automorphism (see remarks around eq.\ (\ref{rotglue})). But this 
does not exhaust all possibilities. In fact, the most general form 
of the gluing condition is     
\begin{equation}
  \label{Lam}
  \Omega \ = \ \lambda \circ \Ad_g \ \ 
  \mbox{ for some } \  g \in G \ \ 
\end{equation}
with some outer automorphism $\lambda$ of $\LieG$, i.e.\ an 
automorphism that does not come from conjugation with some element 
$g$. For simple compact Lie algebras $\LieG$, such outer automorphisms 
come with symmetries $\sigma$ of the Dynkin diagram of $\LieG$ and it 
suffices to let $\lambda$ run through such diagram automorphisms 
$\lambda_\sigma$ (see e.g.\ \cite{KacBook1b}). The Dynkin diagrams 
for $A_{n >1}, D_{n>4}, E_6, E_8$ possess only one non-trivial 
symmetry so that in these cases we shall find one new family of 
so-called {\em twisted} branes. $D_4$ possesses $2$ symmetries, 
while there are no non-trivial actions on all the other Dynkin 
diagrams. The absence of such symmetries for $A_1$ implies that 
we will not find any more maximally symmetric branes on SU(2). 
If the group $G$ 
consists of several simple factors, one can have further maximally 
symmetric branes whenever some of the factors coincide. In such 
cases, holomorphic currents from the different isomorphic factors 
of the same level may be permuted before gluing them to their 
anti-holomorphic partners. The corresponding `permutation branes' 
were studied in \cite{Reckperm} and many of our statements below 
hold for such boundary theories as well. 
\smallskip 

Before we describe the results from boundary conformal field 
theory, let us briefly look at the geometric scenario these
boundary conditions are associated with. We have seen in the 
third lecture that branes constructed with $\lambda= \id$ are 
localized along conjugacy classes or translates thereof (if 
Ad$_g \neq \id$). It turns out that the general case has an
equally simple and elegant interpretation \cite{FFFS1}. Note 
that after exponentiation, any automorphism $\Omega$ of the 
Lie algebra $\LieG$ furnishes an automorphism $\Omega^G$ of 
the group~$G$. Following the same steps as in the third 
lecture, one can then show that the gluing map $\Omega$ 
forces the string ends to stay on one of the following 
$\Omega$-twisted conjugacy classes
$$ \cC^\Omega_u \ := \ \{\,  g \, u\,  \Omega^G(g^{-1})\ |\ g \in G 
\, \} \ \ . $$
The subsets $\cC^\Omega_u \subset G$ are parametrized through 
equivalence classes of group elements $u$ where the equivalence 
relation between two elements $u,v \in G$ is given by: $u \sim_\Omega 
v$ iff $v \in C^\Omega_u$. This parameter space $\U^\Omega$ 
of equivalence classes is not a manifold, i.e.\ it contains singular 
subspaces at which the geometry of the associated twisted conjugacy 
classes changes. In the example of SU(2), the parameter space for 
conjugacy classes is an interval and generic conjugacy classes are 
spherical, but they degenerate to a single point at the two 
end-points of the interval. Similar issues for twisted and untwisted 
branes on SU(3) have been analyzed in great detail in \cite{StanSU3}
(see also \cite{ItoSin1,ItoSin2}). 
For us all this rich structure is of little concern. It suffices 
to know that generic (twisted) conjugacy classes have a transverse 
space of dimension $r_\sigma(G)$ given by the number of orbits that 
the vertices of the Dynkin diagram form under the action of the 
diagram symmetry $\sigma$. This means in particular that $r_\id(G) 
= \rank(G)$. Let us also note that (twisted) conjugacy classes 
come equipped with a B-field. 
\smallskip

As in the SU(2) example, only a discrete set of these (twisted) 
conjugacy classes on $G$ can be wrapped by a brane. This certainly 
comes out of the exact constructions \cite{BiFuSc}, but it can also 
be understood as a quantization condition within a semi-classical 
approach (see \cite{AlScSt,AleSch5} for a related analysis) or 
from the brane's stability \cite{BoRiSc}. All these 
arguments show that there is only a finite set of allowed branes 
whose number depends on the level $\ik$. They are labeled by 
points on a finite $r_\sigma(G)$-dimensional lattice which can 
be considered as a discrete version of the transverse space for 
a generic (twisted) conjugacy class. The precise mathematical 
nature of the labels for non-trivially twisted branes has been 
investigated by several groups (see \cite{BiFuSc,GabGan,PetZub02}). 
Formulas for their boundary states and partition functions were 
originally provided in \cite{BiFuSc} though their expressions are 
not fully explicit. More efficient constructions, at least for a 
large number of cases, have been spelled out in \cite{GabGan,
PetZub02,QuRuSc}. The boundary operator product expansions are known 
for branes wrapping ordinary conjugacy classes. Since these 
theories are of Cardy-type, the coefficients in the operator 
products are obtained from the fusion matrix, as usual (see 
subsection 3.4). For twisted conjugacy classes, similar results
are only known in the limit $\ik \rightarrow \infty$ (see 
\cite{AFQS} and below). 
\smallskip 

\def\iG{{G_\sigma}}
\def\Hom{{\rm Hom}}
Going into more details of the exact solutions would require 
too much of new terminology. But to sketch the general picture, 
we can once more look at the limit of infinite level $\ik$ 
\cite{Quel1,AFQS} where many results may be stated in more  
classical terms. In this regime, the number of branes becomes
infinite and they are labeled by the representations of another 
group $G_\sigma \subset G$. It consists of all elements $g \in 
G$ which are invariant under the diagram automorphism $\lambda_
\sigma^G$ of the group $G$. 
\smallskip 

We can see this group $\iG$ emerging on a semiclassical 
level already. In fact, it is not difficult to show that a generic 
twisted conjugacy class ``close to'' the twisted conjugacy 
class of the group unit can be represented in the form 
$\cC^{\lambda_\sigma} = G \times_{\iG} \cC'$. Here, $\cC'$ is 
an ordinary conjugacy class of $\iG$ and $\iG$ acts on $G$ by 
multiplication from the right. 
In other words, $\cC^{\lambda_\sigma}$ may be considered as 
a bundle over $G/\iG$ with fiber $\cC'$. In the $\ik\to \infty$ 
limit, we keep $\cC'$ small by rescaling its radius. Arguments 
similar to the ones discussed in subsection 4.1 
(see also \cite{AleSch5}) show that $\cC'$ then turns into a 
co-adjoint orbit with its usual linear Poisson structure. At 
the same time, the volume of $G/\iG$ grows with $\ik$ so that 
the corresponding Poisson structure scales down and vanishes 
in the limit. After quantization, we obtain a bundle with 
non-commutative fibers. It possesses a classical base $G/\iG$ 
which is the same for all branes, but the fibers depend on 
$\cC'$ and they are labeled by irreducible representations 
of the group $\iG$. 
\smallskip 

This semi-classical picture for the twisted conjugacy classes 
can also be used to motivate the following  proposal for the 
non-commutative world-volume algebra of twisted branes \cite{AFQS} 
which generalizes the matrix geometries we found for spherical branes
on $S^3$. `Functions' on the quantized co-adjoint orbits $\cC'$ of 
$\iG$ are represented by matrices $\Mat(d_J)$ where $d_J$ is the dimension 
of an irreducible representation of $\iG$. The space of such matrices 
carries an action of $\iG$ or its Lie algebra $\LieG_\sigma$ which 
is defined as in eq.\ (\ref{Lmu}). To built the algebra $\cal F$ of 
`functions' on the entire brane, we combine the matrices with the 
commutative algebra of functions on $G$ and restrict to $\iG$ 
invariants, 
\begin{equation}
  \label{twNCG}
  {\cal F}_J \ \cong \  \Inv_{\LieG_\sigma} \Bigl( \, 
   \Fun(G)\, \otimes \, \Mat(d_J) \, \Bigr) \ \ . 
\end{equation}
Here, the Lie algebra $\LieG_\sigma \subset \LieG$ acts on 
$\Fun(G)$ through right derivatives. If we specialize to $SU(2)$, 
we have $\iG = G$ and hence we recover ${\cal F}_J  \cong \Mat 
(2J+1)$. By construction, ${\cal F}_J$ is an associative non-commutative 
algebra and it comes equipped with an action of $\LieG$ through left 
derivatives on $\Fun (G)$. 
\smallskip 
  
Under the action of $\LieG$, the algebras ${\cal F}$ decompose 
into irreducible multiplets. We met such a decomposition 
in the case of the fuzzy 2-spheres in eq.\ (\ref{Matdec}) and also 
saw by comparison with eq.\ (\ref{partdec}) that they mimic the 
decomposition of boundary partition functions into sectors of the 
chiral algebra, at least up to certain truncations. Similar 
statements can be made in the more general case of $\Omega$-twisted 
branes. The coefficients in the decomposition of the algebra 
(\ref{twNCG}) are easily worked out and they are combinations of 
the so-called branching coefficients for the embedding $G_\sigma 
\rightarrow G$ and the familiar fusion rules of $G$ (see 
\cite{Quel1,AFQS} for concrete formulas). Related expressions 
for finite level were found in \cite{GabGan,PetZub02}. 
\smallskip 
    
With the existing  control over the boundary conformal field
theory of maximally symmetric branes on group manifolds it is 
possible to study their dynamics along the lines of subsection 
4.3. The effective non-commutative field theories were found in 
\cite{AFQS} along with a large number of interesting solutions. 
It turns out that all branes which are localized along ordinary 
conjugacy classes, i.e.\ for which $\lambda$ is trivial, are 
obtained from condensates on a stack of point-like branes and 
hence they carry only a single charge (see \cite{FreSch1}). One new 
charge appears with each diagram automorphism. Once more, all 
branes associated with the same symmetry $\sigma$ of the Dynkin 
diagram appear as bound states of stacks built from a single 
`generating' twisted brane. Among the many processes that have 
been studied in \cite{AFQS} the ones that are generated by 
constant gauge fields admit a deformation into the stringy 
regime \cite{FreSch1}. As in the case of SU(2), one can employ 
the absorption of the boundary spin principle (cf.\ the rule 
(\ref{abs})) to determine the final brane configurations after 
condensation. The set of processes one obtains in this way 
suffice to determine the order of the charge carried by the 
generating branes (see \cite{FreSch1}) and the results agree 
nicely with the findings from twisted $K$-theory (see 
\cite{MaMoSe2} and references therein). 

\paragraph{Conformal embeddings and cosets.} The construction 
of all maximally symmetric branes on group manifolds is a 
remarkable achievement of orbifold and simple current methods. 
We shall now see that symmetry breaking branes on group manifolds 
(and in other backgrounds) can be obtained with another classical 
technique, namely through the systematic use of so-called conformal 
embeddings. A chiral algebra $\cW'$ is said to be conformally 
embedded into $\cW$ if the respective Virasoro elements are 
mapped onto each other. Such an embedding is called rational 
if all $\cW$-sectors decompose into a {\em finite} sum of sectors 
for the conformally embedded algebra $\cW'$. Hence, if a rational 
model with maximal chiral algebra $\cW$ is analyzed with respect 
to $\cW'$, it stays rational. 
\smallskip 

The first application of rational conformal embeddings to the 
construction of boundary theories can be found in \cite{AfOsSa2} where 
they  were used to break the symmetry of a $c=2$ torus compactification 
at some particular radius. A related idea then appeared later in 
the work \cite{MaMoSe1} to build symmetry breaking branes on the 
group manifold SU(2). Actually, Maldacena et al.\ employ the 
parafermionic coset SU(2)/U(1), a free bosonic U(1) model and some 
orbifold ideas in their construction of (unstable) 1- and 3- 
dimensional branes on SU(2). A generalization to SU(N) which 
uses the abelian cosets SU(N)/U(1)$^{N-1}$ was proposed in 
\cite{MaMoSe2}. Our presentation here will follow the approach of 
\cite{QueSch1}. The latter allows to incorporate non-abelian 
cosets and leads to a very large set of new symmetry breaking 
branes on group manifolds.     
\smallskip   
   
In \cite{MaMoSe1,MaMoSe2,QueSch1}, coset chiral algebras are a 
central ingredient of the whole procedure. Let us therefore 
briefly outline a few basics from the coset or GKO construction 
\cite{GoKeOl}. Suppose we are given some affine Lie algebra 
$\asg$ or its associated chiral algebra $\cW(G)$ along with  
some affine subalgebra $\cW(U) \subset \cW(G)$. Then we can look for 
the maximal chiral subalgebra $\cW(G/U)$ within $\cW(G)$ which 
commutes with $\cW(U)$. As one can easily check, the algebra 
$\cW(G/U)$ contains a Virasoro field $T^{G/H} = T^{G}-T^{H}$
of central charge $c^{G/U} = c^G-c^U$. By construction, the 
chiral algebra  $\cW' = \cW(G/U) \otimes \cW(U)$ is conformally 
embedded into $\cW(G)$. In fact, the Virasoro field $T^{G/U} +
T^H$ of $\cW'$ coincides with the Virasoro field $T^G$. Moreover, 
each representation of $\cW^G$ can be shown to decompose into a 
finite number of representations for $\cW' \subset \cW(G)$ (see
e.g.\ \cite{FrMaSeBook}). The construction of $\cW'$ that we 
have just outlined can be applied repeatedly if we set $U = U_1$ 
and select a chiral algebra $\cW(U_2) \subset \cW(U_1)$ etc. In 
this way, we obtain a large number of conformally embedded 
chiral algebras 
$$ \cW' \ = \ \cW(G/U_1) \otimes \cW(U_1/U_2) \otimes \cdots 
   \otimes \cW(U_{N-1}/U_N) \otimes \cW(U_N) \ \subset \cW(G) 
\ \ .$$ 
Except from some subtleties that may arise from the centers of the 
groups $U_s$ (see e.g.\ \cite{Gepn89,SchYancos}), the formulas in 
\cite{QueSch1} provide a set of boundary theories which preserve 
such chiral algebras $\cW'$. 
\smallskip

Here we shall content ourselves with a description of their geometry 
\cite{Quel2}. As in our construction of $\cW'$, we must choose a 
chain of groups $U_s, s = 1, \dots, N,$ along with homomorphism 
$\epsilon_s: U_s \rightarrow U_{s-1}$ (we set $U_0 = G$). The latter 
are assumed to induce embeddings of the corresponding Lie algebras. 
Furthermore, we select an automorpism $\Omega_s$ on each group $U_s$. 
Given these data, it is possible to construct a set of branes  which 
preserve an $U_N$ group symmetry. These are localized along the 
following sets   
\ba 
 \cC^{\underline \Omega}_{\underline \epsilon; \underline u}
  &  = &  \cC^0_{u_0} \, \cdot\,  \cC^1_{u_1}\,  \cdot \, \dots\,  
     \cdot\,  \cC^N_{u_N} \ \subset \ G \ \ \ \mbox{ where } 
    \label{prodC}   \\[2mm]  
     \cC^s_{u_s} & = & \Omega_0 \circ \epsilon_1 \circ \dots 
     \circ \Omega_{s-1} \circ  
     \epsilon_s ( \cC^{\Omega_s}_{u_s} ) \ \subset \ 
     G \ \ \ \mbox{ for } \ \ 
     u_s \ \in \ U_s   \nn 
\ea
and $\cC^0_{u_0} = \cC^{\Omega_0}_{u_0}$ for $u_0 \in G$. The 
$\cdot$ indicates that we consider the set of all points in $G$ 
which can be written as products (with group multiplication) of 
elements from the various subsets. One should stress that branes
may be folded onto the subsets (\ref{prodC}) such that a given
point is covered several times. This phenomenon has been observed
for a special case in \cite{MaMoSe1}. In some examples, depending 
on the choice of $\underline{u}$, several different branes can 
wrap the same set (\ref{prodC}). Such a situation occurs e.g.\ 
for the volume filling brane on $S^3$ (which requires an even 
level $\ik$) but is it not understood in general.              
\medskip 

The new symmetry breaking boundary theories have various applications, 
in particular when dealing with groups $G = G_1 \times \dots \times 
G_n$ which factorize into several simple factors. Let us note 
that many interesting solvable string backgrounds are factorizable 
or orbifolds of factorizable backgrounds. Some boundary states for 
such theories can be factored accordingly so that they are simply 
products of boundary states for each of the individual factors. 
But this does certainly not exhaust all possibilities. Generically, 
branes preserving the maximal chiral symmetry are factorizing. Only 
the permutation branes that exist for backgrounds $G$ with several 
identical simple factors are non-factorizable and maximally symmetric 
at the same  time. Many more interesting examples of non-factorizable 
branes are obtained from constructions of symmetry breaking branes 
(see \cite{QueSch1,Quel2}).  
\smallskip 
 
There exists another - superficially very different - setup which 
leads exactly to the same type of problems. It arises by considering 
a one-dimensional quantum system with a defect (see e.g.\ \cite{Wong:pa,
AffOsh2,LeClair:1997gz,Nayak,Salelec,Saleur:2000gp} and 
\cite{McAvity:1995zd,ErGuKi} for higher dimensional analogues), 
or, more generally, two different systems on the half-lines $v<0$ and 
$v>0$ which are in contact at the origin. The defect or contact at $v=0$ 
could be totally reflecting, or more interestingly it could be partially 
(or fully) transmitting. To fit such a system into our general discussion, 
we apply the usual folding trick (see Figure 9). After such a folding, 
the defect or contact is located at the boundary of a new system on 
the half-line. In the bulk, the new theory is simply a product of the 
two models that were initially placed to both sides of the contact at 
$v=0$.   
\vspace{10mm} 
\fig{The folding trick relates a system on the real line with a 
     defect to a tensor product theory on the half line.}
{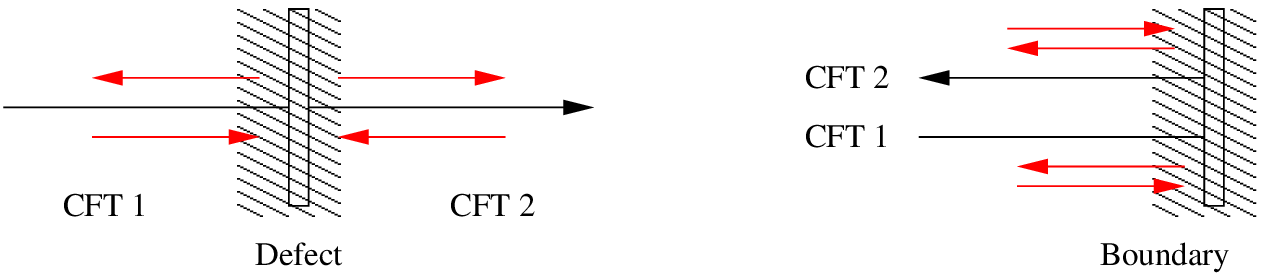}{12truecm}
\figlabel{\basic} \hspace*{-3mm} 
\vspace{10mm}

Factorizing boundary states for the new product theory on the 
half-line correspond to totally reflecting defects or contacts.
With our new boundary states we can go further and couple the 
two systems in a non-trivial way. Since we always start with 
conformal field theories with chiral algebras $\cW_1= \cW_<$ 
and $\cW_2= \cW_>$ on either side of $v=0$, it is natural to look 
for contacts that preserve conformal invariance. This requires to 
preserve the sum of the two Virasoro algebras of the individual 
theories. After folding the system, the preserved Virasoro algebra 
is diagonally embedded into the product theory $\cW = \cW_1 \otimes 
\cW_2$. Of course, one can often embed a larger chiral algebra $\cW'$ 
and then look for defects that preserve the extended symmetry. This 
is exactly the setup to which our general ideas apply. Note that they
are capable of constructing defect lines which join two conformal 
field theories with different central charge. Such defects are known 
to appear on the boundary of an AdS-space when there is a brane in 
the bulk which extends all the way to the boundary \cite{KarRan1,
KarRan2,DWFrOo,BdBDO}. Simpler examples without 
jumps in the central charge have also been analyzed in \cite{PetZubtw,
FuRuSccat}. 
\smallskip

Finally, let us briefly mention that the branes we have discussed 
here and in the pervious sections descend to (asymmetric) coset 
models and therefore many of our insights directly apply to this 
very large class of backgrounds. Coset theories possess a maximal 
chiral algebra of the form $\cW(G/H)$ and they describe strings 
moving on the space $G/H$ with non-constant background fields (see 
e.g.\ \cite{Nahm:1988sn,BaRaSa,GawKup1,GawKup2,TseygWZW} for some 
early work on the geometry of the bulk theories). The geometry of 
Cardy-type branes in such cosets was exhibited in \cite{Gawe2} (see 
also \cite{FreSch1,MaMoSe1,EliSar,FalGaw01,FreSch3}). Non-commutative 
gauge theories for branes in coset models were found and studied in 
\cite{FreSch3} (see also \cite{ReRoSc,Graham:2001pp} for related work 
in a more traditional conformal field theory framework). These 
effective actions have interesting implications on brane bound state 
formation in some limiting regime. The resulting structure of possible 
condensates is much richer than for group manifolds since there are 
more conserved charges \cite{FreSch3}. Many of the processes in 
\cite{FreSch3} admit a deformation into the stringy regime. In fact, 
an extension of the `absorption or the boundary spin'-principle 
(\ref{charpert}), (\ref{abs}) to coset models was formulated in 
\cite{FreSch4} and it was tested 
against known results on the boundary flows in minimal models
\cite{Chim:1996kf,Lesage:1998qf,Ahn:1998xm,Dorey:1999cj,Graham:2001tg}
(see also \cite{FredPhD} for additional examples). More recent work on 
the construction of branes in coset models and related issues includes 
\cite{Ishik,Kubota:2001ai,Walton:2002db,Sarkissian:2002ie,IshikTan,
QueSch2}.

\paragraph{Other approaches to exact solutions.}  
In these notes we presented a conventional approach to the 
construction of exact solutions in which 
we start from the bulk and then work our way through to the 
boundary by first solving the factorization constraints 
(\ref{class}) for the one-point functions, then computing the 
boundary partition function through eq.\ (\ref{BPF2}) and 
finally solving the relations (\ref{HOPEcon}) for the structure 
constants of the boundary operator product expansions. It is 
interesting, however, to turn the whole procedure upside down 
and to start from the boundary. This may seem almost hopeless 
at first, partly because solving the complicated non-linear 
equations (\ref{HOPEcon}) requires us to guess some appropriate 
multiplicities ${n_{\a\b}}^j$ for the sectors of the boundary 
theory. A closer look reveals, however, that much of this problem  
can be linearized. Once eqs.\ (\ref{HOPEcon}) have been solved for 
a certain choice of the multiplicities ${n_{\a\b}}^j$, one may 
go back and determine the coefficients $A^\a_j$ 
from eqs.\ (\ref{BPF2}). Finally, even the structure constants of 
the bulk operator products can be calculated through formula 
(\ref{XiCF}). Such an approach has been suggested by Petkova 
and Zuber \cite{PetZubbtoB}. It was then rephrased and extended 
systematically in \cite{FuRuSc1,FuRuSccat}. This whole program 
is quite elegant and it is somewhat similar in spirit to recent 
attempts in string field theory to reconstruct the closed from 
open strings (see e.g.\ \cite{Shat}).  
\medskip 

There is another very interesting solution generating technique 
that has been applied very successfully in the past, namely the 
use of boundary deformation theory (see \cite{RecSch2,RecSch3} 
for some general results). Unlike the ideas we presented above, 
boundary deformation 
theory is capable of constructing non-rational boundary theories 
from rational ones. The idea is to start from some rational boundary 
model and to look for exactly marginal operators among its boundary 
fields. If they exist, they generate a continuous family of new 
boundary theories. The latter correspond to translates of the 
original branes if the marginal field is taken from the chiral 
algebra itself, but they typically break much of the symmetry 
otherwise. At least in some examples \cite{CKLM,PolThodef,RecSch2} 
even such symmetry breaking deformations can be constructed 
perturbatively, to all orders in perturbation theory. The 
corresponding boundary theories play an important role 
for our understanding of open string tachyon condensation and time 
dependent open string backgrounds (see e.g.\ \cite{Senbtach1,
Senbtach2}).

\subsection{Towards non-compact backgrounds}   
\def\H3p{H_3^+}
\def\raa{{\varrho_0}}
\def\raas{{\varrho_\ast}} 
Our presentation above was mainly tailored towards compact curved 
backgrounds, or, in world-sheet terms, rational (boundary) conformal 
field theories. Technically, our assumptions implied that there were 
only finitely many primary fields and hence the various constraining
equations  (see (\ref{class}), (\ref{BPF2}), (\ref{HOPEcon})) on the 
structure constants $A^\a_i$ and $C$ had to be solved for a finite 
number of unknowns.  All this changes drastically when we deal with 
non-compact backgrounds and even though some of the general ideas do 
carry over, at least after appropriate modifications, there is no 
generic exact construction to replace Cardy's solution for rational 
models. This is mainly due to the fact that analogues of the Cardy 
condition (\ref{BPF2}) are less restrictive and therefore one cannot 
get away without deriving and solving some factorization constraints 
(see below). 
\smallskip 

The same problems are certainly present in the bulk theories already
so that there is only a small number of exactly solved models to begin 
with. Most attention in the past has been devoted to Liouville theory. 
The exact solution of this model (in a certain regime) was proposed in 
\cite{DorOtt,ZamZam} (the proposal was partly based on \cite{GouLi}) 
and then more thoroughly analyzed in \cite{TeschL1,PonTes1,PonTes2,
Teschrev}. The boundary problem for Liouville theory was treated by 
several authors \cite{CreGer,FaZaZa,TeschbL,Hoso,PonTes3}. 

Except from the supersymmetric versions of the Liouville model (see 
e.g.\ \cite{RasSta,Pogh} and \cite{FukHos,AhRiSt} for the boundary 
problem), there exists only one other non-rational model that has 
been solved in the bulk. It describes strings moving on a Euclidean 
analogue of $AdS_3$. Since this background is a very close relative 
of the 3-sphere that we studied extensively in the third lecture, 
we shall use it here to explain some of the similarities and 
differences between solving rational and non-rational boundary 
conformal field theories. We begin with a brief introduction to 
the bulk theory and then turn to the possible branes in this 
background. After a short review of their classical geometry, 
it is explained how to master the various subtleties that arise 
when we try to obtain factorization constraints for the one-point 
functions. Then we analyze the open string sector. In particular, 
we shall motivate and explain the concept of a {\em relative 
partition function} and its relation to the {\em reflection 
amplitude}.

\paragraph{The bulk of the $\H3p$ WZW model.} The model we are 
interested in describes strings moving on the space $\H3p$ of 
hermitian unimodular $2\times 2$ matrices with positive trace, 
$$ \H3p = \{ h \in \SLC\, |\,  h^*\, = \, h\, ,\,  \tr h > 0 \} 
\ \ . $$
It is easy to see that $\H3p$ is a non-compact coset $\SLC/\SU$. 
Following the standard rules, one can write down the classical 
WZW model for this geometry. The associated quantum theory has been 
solved by Teschner in a series of papers \cite{TeschH31,TeschH32,
TeschH33,TeschH34} after Gawedzki computed its bulk partition 
function in \cite{GawNC}. 
\smallskip

There are several good reasons to study the $\H3p$ model. As 
we mentioned before, $\H3p$ is a Euclidean version of $AdS_3$ (with 
NSNS 3-form) and much of the recent progress towards the construction 
of perturbative closed string theory for $AdS_3$, see \cite{MaOo1,MaOo2,
MaOo3} and references therein, has been based on the Euclidean 
background. Furthermore, one can descend from $\H3p$ to a coset 
describing the 2D Euclidean black hole \cite{Wit1} which is part 
of many interesting string backgrounds (see e.g. \cite{HoSt,MaSt,
OoVa}). We shall not return to this coset below, but we want to 
mention that its partition function was recently computed in 
\cite{HaPrTr} and the results confirm expectations which go back 
to the work of Dijkgraaf et al.\ \cite{DVV}. The theory was 
conjectured by Fateev, Zamolodchikov and Zamolodchikov to be  
T-dual to sine-Liouville theory (see \cite{KaKoKu} for a more 
precise description of the conjecture). The bulk operator product 
expansions of sine-Liouville were studied \cite{FukHossL}. A 
supersymmetric version of the T-duality \cite{GivKut} which 
involves the $N=2\ $ $\SLR/U(1)$ Kazama-Suzuki quotient on one side 
and $N=2$ Liouville theory on the other was proven in \cite{HorKap}.    
\smallskip 

To proceed with our outline of the $\H3p$ model, it is convenient 
to parametrize this space through coordinates $(\phi, \c,\bc)$ such 
that 
\begin{equation}\label{pggpar} 
 h \ = \ \left( \begin{array}{cc} e^\phi \ & \ e^\phi \bc \\
                   e^\phi \c \ & \  e^\phi \c \bc + e^{-\phi} 
                  \end{array} \right) \ \ . 
\end{equation}
Here, $\phi$ runs through the real numbers and $\c$ is 
a complex coordinate with conjugate $\bc$. We can visualize the 
geometric content of these coordinates most easily by expressing 
them in terms of the more familiar coordinates $(\rho,
\tau,\theta)$ (see Figure 10), 
$$ \c \ = \ e^{\tau + i \theta} \, \tanh \rho \ \ \ \ \ 
   \mbox{ and } \ \ \ \ \ e^\phi \ = \ e^{-\tau} \cosh \rho 
   \ \ \ . 
$$  
At fixed $\ga,\bc$, the boundary of $\H3p$ is reached in the limit 
of infinite $\phi$. The boundary is now represented as the complex 
plane with coordinates $\c,\bc$. \\[10mm]
\vspace{10mm} 
\parbox[h]{8.1cm}{\fig{The coordinates $(\rho,\tau,\theta)$ parametrize $\H3p$
as shown. The boundary of $\H3p$ appears at $\rho = \infty$.}
{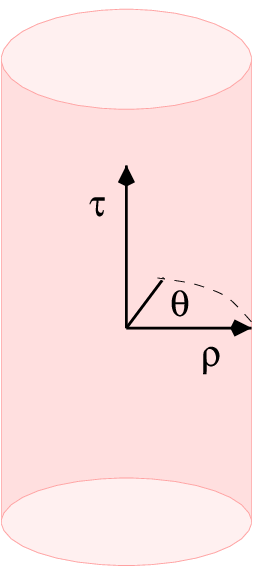}{2truecm}
\figlabel{\basic}} \hspace*{-3mm} 
\parbox[h]{8cm}{\fig{$AdS_2$-branes in $AdS_3$ are parametrized 
by a parameter $\varrho_0$ measuring the distance from $\rho=0$.}
{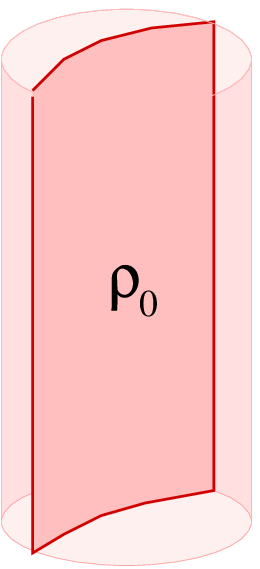}{2truecm}   
\figlabel{\basic}}
\vspace{0mm}

Any wave function on $\H3p$ can be expanded in terms of eigen-functions 
of the Laplace operator on $\H3p$. Since there exists an action of $\SLC$ 
on $\H3p$ which commutes with the Laplace operator, each eigen-space must 
carry some representation of $\SLC$. It is not difficult to show that the 
possible eigenvalues of the Laplacian are given by $j(j+1), j = - \frac12 
+ iP,$ where $P$ is a non-negative real number and that the associated 
eigen-spaces carry the irreducible representation $D^j$ from the principal 
continuous series. Explicitly, the eigen-functions are given by the 
following formula 
\begin{eqnarray} 
\varphi_j(w,\bw|\phi,\c,\bc)  & = & - \frac{2j+1}{\pi} \  
  (v_w \, h \, v_w^*)^{2j} \label{phief} \\[2mm] 
& = & - \frac{2j+1}{\pi} \left( |w - \c|^2 e^\phi + e^{-\phi} 
      \right)^{2j}. \nonumber 
\end{eqnarray}  
Here, $v_w = (-w,1)$ depends on a complex coordinate $w$. The latter is a
continuous analogue of the discrete labels $a,b,\dots$ we used for 
functions $\Phi^{ab}_j$ on $S^3$ (see eq.\ (\ref{phidef})). In fact, $\SLC$ 
acts on $w$ by the usual rational transformations and thereby it generalizes 
the role played by the group $\SU \times \SU$ of left and right translations 
on $S^3$. One may consider the functions $\varphi_j(w,\bw|h)$ as wave-function
of some particle that was created with `radial momentum' $j$ at the boundary 
point with coordinates $w,\bw$ \cite{BORT}. They form a basis in the space 
of square integrable functions on $\H3p$ and are in one-to-one correspondence 
with the ground states of the bulk conformal field theory on $\H3p$ (see 
\cite{GawNC}). The state-field correspondence then provides us with the 
following set of bulk fields 
$$ \phi_j(w,\bw;z,\bz) \ = \ \Phi^{(P)}(\, |\varphi_j(w,\bw)\rangle\, ;\, 
z,\bz\,  ) \ \ . $$ 
We will not spell out the coefficients of their operator product expansions, 
but they can be found in \cite{TeschH31,TeschH33}.

\paragraph{Introduction to branes in $\H3p$.} The microscopic study 
of brane geometries in the Lorentzian model began with the work of 
Stanciu \cite{StanAdS} who used the relation between $AdS_3$ and the 
group $\SLR$ to apply the known results about branes in group manifolds. 
It was later shown by Bachas and Petropoulos \cite{BacPet} that the most 
interesting branes on $AdS_3$ are associated with twisted conjugacy 
classes in the sense of \cite{FFFS1} (cf.\ our discussion in the 
previous subsection). These are localized along \ $AdS_2\, \subset\,  
AdS_3$ (see Figure 11) and they are parametrized by a single real 
parameter $\varrho_0$. In addition one can have branes localized 
along $H_2$, $dS_2$, the light cone, as well as point-like branes. 
Not all of these geometric possibilities correspond to physical 
brane configurations, though: The branes localized along $dS_2$, 
for example, were found to have a supercritical electric field 
on their world-volume \cite{BacPet}. 
\smallskip 

Most of the branes we have just listed possess a Euclidean counterpart. 
Here we will be concerned mainly with the Euclidean $AdS_2$ branes in 
$\H3p$ which are localized along the surfaces 
\be \tr (\, \varpi \, h\, ) \ = \ 2\sinh (\varrho_0) \ \ \ 
  \mbox{ where }  \ \ \ \varpi \ = \ \left( 
  \begin{array}{ll} 0 & 1 \\ 1 & 0 \end{array} \right) 
  \ \    \label{ads2} 
\ee 
and parametrized by $\varrho_0 \in \mathbb{R}$. After rotation with 
a particular $\SLC$ symmetry transformation of $\H3p$ these branes 
are localized along one of the connected component $H^\pm_2$ of the 
Euclidean $H_2$ brane, so that we do not have to treat these two 
types of branes seperately. While the non-compact hyper-surfaces 
(\ref{ads2}) preserve an $\SLR \subset \SLC$ symmetry, a family 
of compact SU(2) symmetric 2-spheres is obtained through the 
equations 
\be \tr (\, h\, ) \ = \ 2\cosh (\varsigma_0) \ \label{s2} 
\ee  
with $\varsigma_0 \geq 0$. These spheres degenerate to a single 
point for $\varsigma_0 =0$. The boundary conformal field theory 
analysis shows that there exist related (unstable) branes with a 
spherical symmetry, but they seem to have an imaginary 
radius $\varsigma_0$. Otherwise, these boundaries behave very much in 
the way one would expect from spherical branes. In particular, they 
possess a finite dimensional space of boundary primary fields. We 
will make a few more remarks about these spherical boundary 
conditions as we proceed, but will focus mainly on the non-compact 
branes. 
\smallskip 

The space of ground states in the boundary theory for a single 
$AdS_2$ brane consists of eigen-functions for the Laplacian on 
$AdS_2$. As in the case of $AdS_3$ (see eq.\ \ref{phief}), it 
is easy to find an exact expression for the eigen-functions with 
eigen-value $j = j(j+1), j \in  -\frac12 + i \mathbb{R}^+_0$ by 
restricting 
\be 
\psi_j(w|h)  \  := \   \  
  (v'_w \, h \, {v'}_w^*)^{j} \label{psief}  
\ee
to the 2-dimensional surface (\ref{ads2}). These functions are 
parametrized through some real coordinate $w$ which appears in 
$v'_w = (iw,1)$. The eigen-functions transform according to the 
(infinite dimensional) irreducible representations from the 
principal continuous series of the symmetry group $\SLR$. Once 
more, we obtain a continuous spectrum of momenta $j$ which run 
through the same values as for the bulk fields, but this time 
$j$ labels representations of $\SLR$ rather than $\SLC$. The 
state field correspondence $\Psi$ associates a boundary field 
to each function (\ref{psief}), 
$$ \psi_j(w;u) \ = \ \Psi(\, |\psi_j(w)\rangle\, ; \, u\, ) \ \ .    
$$

\paragraph{The closed string sector.} 
In constructing the exact boundary CFTs that describe branes in 
$\H3p$ we follow the same strategy as before, i.e.\ we try to find 
the one-point functions of the bulk fields $\phi_j(w,\bw;z,\bz)$  
by solving appropriate factorization constraints. Once again the 
simplest factorization constraint arises from the two-point 
functions of the theory and even the main idea behind its
derivation is similar to the compact situation (see Figure 3). If 
we imagine the two bulk 
fields close to each other it is most natural to use the bulk operator 
product expansion to get a factorization in the closed string channel, 
leading to a representation of the two-point function as an integral 
over one-point functions. This is very much the same as in the rational 
models, apart from the fact that the bulk operator product expansion 
contains a continuum of primary fields. 

The second regime, however, in which the fields are far apart from 
each other differs more drastically from what we have seen in the 
second lecture. In fact, for compact backgrounds we projected this
regime onto the channel in which the identity field with label $0$ 
propagates along the boundary. But now such a projection would 
vanish for a very simple reason: our bulk fields $\phi_j$ correspond
to normalizable states of the model. The boundary identity field, on 
the other hand, can certainly not be associated with a normalizable 
state simply because the constant function on a non-compact brane is
not normalizable. Since it is impossible to create a non-normalizable
excitation on the boundary with a normalizable excitation in the 
bulk, the strategy of our derivation of the constraint (\ref{class}) 
breaks down for non-compact branes. 

Fortunately, there exists a way out \cite{TeschH31,FaZaZa}. In fact, 
the fields $\phi_j$ we have considered so far are not the only ones 
in the theory. They are the fields that are in one-to-one 
correspondence with the normalizable states of the model. By 
analytic continuation in $j$, however, we obtain additional fields 
which are still perfectly well defined even though they do not 
correspond to any normalizable state (see \cite{TeschH33,TeschH34} 
for a rigorous justification). For certain discrete values of $j$, 
the new fields are associated with degenerate representations 
of the current algebra. This implies that the operator product of 
these degenerate fields with any other field of the theory contains 
only finitely many primary fields and, more importantly, that the 
factorization in the open string channel includes a contribution 
from the identity boundary field.
\smallskip 

The factorization constraints that arise from the degenerate field
$\phi_{1/2}$ have been worked out and solved in \cite{PoScTe,LeOoPa}.   
For the one-point functions one finds 
\ba\label{res1}  
 \langle \phi_j(w,\bw;z,\bz)\rangle_\raa&=&\pi\, 2^{\frac{3}{4}}\,
 \sqrt{b}  \ \nu_b^{j+\frac{1}{2}}\,\Ga(1+b^2(2j+1))\,  
\ \frac{|w+\bw|^{2j}\, e^{-\raa(2j+1)\sgn(w+\bw)}}{|z-\bz|^{2h_j}}\ \ 
\\[2mm] 
\mbox{where }  & & \nu_b \ = \ \frac{\Ga(1-b^2)}{\Ga(1+b^2)} 
 \ \ \ , \ \ \ h_j \ = \ - b^2 j (j+1) 
\ea 
and $b$ is related to the level $\ik$ by $b^2 = (\ik-2)^{-1}$. A short 
analysis shows that the $w$-dependent terms $\sgn(w+\bw)^\epsilon 
|w+\bw|^{2j}$ for $\epsilon = 0,1$ are analogues of the intertwiner 
$\cU$ in the general formula (\ref{1ptfct}) for one-point functions. 
We should stress here that the factorization constraints obtained with 
the field $\phi_{1/2}$ do not fix the solution uniquely. Therefore, 
it would be quite interesting to investigate further conditions that 
arise e.g.\ from the degenerate field with $j = 1/2b^2$. These 
relations have not been worked out yet, but the one-point function
was shown to pass further consistency conditions that come with the 
open string sector (see \cite{PoScTe} and below).  

\paragraph{The open string sector.} If we could follow the same 
strategy as in the compact case, the next step would be to compute 
the boundary partition function from the coefficients of the 
one-point function. Once more, things are not that simple for 
non-compact branes. In fact, if we would naivly copy the old 
computation we would end up with a divergent result. The reasons
for this are very general and we will explain them in a simple 
quantum mechanical setup first before returning to our $AdS_2$ 
branes. 
\smallskip 
 
In systems with a continuous energy spectrum, the spectral set itself 
does not contain much information about the dynamics. Consider, for 
example, a 1-dimensional quantum system with a positive potential $V(x)$ 
which vanishes at $x \rightarrow \infty$ and diverges as we approach 
$x =- \infty$. Such a system has a continuous spectrum which is 
bounded from below by $E=0$ and under some mild assumptions, the 
spectrum does not depend at all on details of the potential $V$
(see e.g.\ \cite{Mess}). There is much more dynamical information 
stored in the so-called reflection amplitude of the system. Recall 
that for each value $E>0$ of the energy, our system admits a unique 
(up to normalization) wave function. It has the form 
\begin{equation}
\psi_E(x)\ = \ \; e^{-ipx}\;+ \; R(p) \; e^{+ipx}\   
  \ \ \ \mbox{ where }
  p \ =\ \sqrt{E} \ \ . 
\end{equation}
The phase in front of the second term is called the {\em reflection 
amplitude}. It is a functional of the potential which is  
very sensitive to small changes of $V$. In fact, it even encodes 
enough data to reconstruct the whole potential.  
\smallskip 

From the reflection amplitude $R(p)$ we can extract some spectral 
density function $\rho$. To this end, let us regularize the 
system by placing a reflecting wall at $x=L$, with large positive 
$L$. Later we will remove the cutoff $L$, i.e.\ send it to infinity. 
But as long as $L$ is finite, our system has a discrete spectrum 
so that we can count the number of energy or momentum levels in 
each interval of some fixed size and thereby we define a density 
of the spectrum. Its expansion around $L = \infty$ starts with the 
following two terms 
\begin{equation}\label{dos_asym}
\rho^{L}(p) \ =\ \frac{L}{\pi}+\frac{1}{2\pi i}\frac{\pa}{\pa p}
\ln R(p)\ + \ \dots \ \ 
\end{equation} 
where the first one diverges for $L\ra \infty$. This divergence is 
associated with the infinite region of large $x$ in which the whole 
system approximates a free theory and consequently it is universal, 
i.e.\  under only mild assumptions it is independent of the 
potential $V(x)$. The sub-leading term, however, is much more 
interesting and we can extract it from the regularized theory if 
we compute relative spectral densities before taking the limit 
$L \rightarrow \infty$. This can be done by fixing one reference 
potential $V_\ast$ whose regularized spectral density we denote by 
$\rho^L_\ast$. The relative spectral density for a potential $V$ is 
then given by 
$$ \rho_{\rm rel} (p) \ := \ \lim_{L\rightarrow \infty} 
  \left(\, \rho^L(p) - \rho^L_\ast(p) \right) \ = \ 
   \frac{1}{2\pi i} \, \frac{\partial}{\partial p} 
   \, \ln \frac{R(p)}{R_\ast(p)} \ \ . $$ 
It is not difficult to transfer these observations from quantum 
mechanics to the investigation of non-compact branes. As in the toy 
model, the naive partition functions of our boundary theories for 
$AdS_2$-branes diverge. But it is possible to introduce a cutoff $L$ 
and to construct partition functions relative to a fixed reference 
brane with parameter $\raas$ \cite{PoScTe},     
\ba
Z^{}_{\rm rel}(q|\raa;\raas) & = & \lim_{L\rightarrow \infty}\,    
\left( Tr_{\CH^{(H)}_{\raa}}\bigl(\, q^{\SH^{(H)}_\raa}\, \bigr)^L \,
   - \, Tr_{\CH^{(H)}_{\raa}}\bigl(\, q^{\SH^{(H)}_{\raas}}\, 
   \bigr)^L \right) \nn \\[2mm] 
\label{Zrel}  
 & \sim & \int\limits_{0}^{\infty} dP \;\frac{1}{2\pi i}\frac{\pa}{\pa P}
  \log \frac{R\bigl(-\frac{1}{2}+iP|\raa\bigr)}{R\bigl(-\frac{1}{2}+iP|
  \raas\bigr)}\; \chi^j (q) 
\ea
Here we have used the regularized characters $\chi^j(q)\ =\ q^{b^2P^2}
\eta^{-3}(q)$. Formulas for the reflection amplitude $R(j|\raa) = 
R(-1/2 + iP)$ can be obtained in two different ways. One 
possibility is to employ world-sheet duality to derive the partition 
function from the one-point functions (\ref{res1}), just as it was 
done in rational theories. This leads to the expressions \cite{PoScTe} 
\begin{equation} \label{res2} 
R(j|\raa)\;=\; \nu_b^{-iP} 
\frac{\Ga^2_\ik(b^{-2}-iP+\frac12)}{\Ga^2_\ik(b^{-2}+iP+\frac12)}
\frac{\Ga_\ik^{}(b^{-2}+2iP)}{\Ga_\ik(b^{-2}-2iP)}
\frac{S_\ik(2R+P)}{S_\ik(2R-P)}\ \ ,
\end{equation}
where $R\equiv \raa/ 2\pi b^2$ and the two special functions 
$S_\ik(x)$ and $\Gamma_\ik(x)$ are defined through
\begin{eqnarray} \label{Sk}
\log S_\ik(x) & = & i\int\limits_{0}^{\infty}\frac{dt}{t}
\Biggl(\frac{\sin 2tb^2x}{2\sinh b^2t\sinh t}-\frac{x}{t}
\Biggr) \ \ , \\[2mm] 
\label{Gk} 
\Ga_\ik(x) & = & b^{b^2x(x-b^{-2})}(2\pi)^{\frac{x}{2}}
\Ga_2^{-1}(x|1,b^{-2}) \ \ .
\end{eqnarray}
Here, $\Ga_2(x|\omega_1,\omega_2)$ denotes Barnes Double Gamma 
function.  If we follow this route to compute the partition 
function from the boundary states, then world-sheet duality 
does not give rise to a constraint on one-point functions 
simply because for systems with a continuous spectrum there 
are no a priori integrality conditions. 
\smallskip 

But there is another way of obtaining the relative partition function 
through a direct construction of the stringy reflection amplitude 
$R(j|\raa)$. Just as in rational models, open string states may be 
created by boundary operators $\psi_j(w|u)$ where $u$ is the usual 
coordinate for the boundary of the world-sheet. One can then study 
the scattering amplitude for an open string that is sent in with 
momentum $j_1$ from the boundary of $AdS_3$ into an outgoing open 
string with momentum $j_2$. The reflection amplitude is obtained 
from the two-point function of boundary operators through
\begin{equation} \label{intro3} 
\langle \psi_{j_1} (w_1|u_1) \, \psi_{j_2}(w_2|u_2) \rangle_\raa   
\ \sim \ \delta(j_1 - j_2)  \ R(j_1|\raa)  \, 
\frac{1}{|u_1 - u_2|^{h_j}} \ \ . 
\end{equation}
Here, $j_i \in  -1/2 + i \QR^+$ and we omitted some $(j_i,w_i)$ 
dependent factor that is determined by the symmetry. This leaves 
us with the problem to find an expression for the boundary
two-point functions. The latter are subject to factorization 
constraints for the boundary three point functions which have 
been worked out and solved in \cite{PoScTe}. The resulting 
expression for the reflection amplitude is the one we have 
spelled out in eq.\ (\ref{res2}). Let us stress once more that 
world-sheet duality only provides us with a consistency 
condition for the one-point functions of bulk fields after 
some boundary factorization problem has been solved. 
\smallskip 

Now we are only missing one more piece of data, namely the 
boundary operator product expansions. Except from some very 
special subset which is encoded in the boundary 2-point 
function, these operator products of boundary fields are 
not yet known. But since the few coefficients that have been 
found in \cite{PoScTe} are still related to elements of 
the fusing matrix (though with an interesting shift between 
boundary and representation labels), it is very likely that 
the solution comes once again from the representation 
theory of chiral algebras. Let us briefly note that in the case 
of spherical branes the situation is better. Their couplings to 
closed string modes were obtained in \cite{PoScTe} (correcting 
earlier formulas in \cite{GiKuSc,ParSah}) and they were  
shown to possess a discrete set of boundary primary fields 
\cite{GiKuSc}. Operator products for the latter have been 
spelled out in \cite{Ponsot}.
\smallskip 

This brings us to the end of our discussion of non-compact
branes. We have tried to sketch how some of the fundamental 
ideas we developed during the investigation of compact 
backgrounds need to be modified. But as we have seen, there 
appear many new problems that so far have only been solved 
in a few models. Finding model independent exact solutions 
as in the case of compact backgrounds remains one of the 
many challenging problems for future research. 

\baselineskip=15pt
\begingroup\raggedright\endgroup


\begin{thebibliography}{100}

\bibitem{Doug3}
M.~R. Douglas, {\it {D-branes, categories and $N = 1$ supersymmetry}},
  \href{http://xxx.lanl.gov/abs/hep-th/0011017}{{\tt hep-th/0011017}}.

\bibitem{Doug4}
M.~R. Douglas, {\it {D-branes and $N = 1$ supersymmetry}},
  \href{http://xxx.lanl.gov/abs/hep-th/0105014}{{\tt hep-th/0105014}}.

\bibitem{AGMOO}
O.~Aharony, S.~S. Gubser, J.~M. Maldacena, H.~Ooguri, and Y.~Oz, {\it Large {N}
  field theories, string theory and gravity}, {\em Phys. Rept.} {\bf 323}
  (2000) 183--386, \href{http://xxx.lanl.gov/abs/hep-th/9905111}{{\tt
  hep-th/9905111}}.

\bibitem{Pollec}
J.~Polchinski, {\it {TASI} lectures on {D}-branes},
  \href{http://xxx.lanl.gov/abs/hep-th/9611050}{{\tt hep-th/9611050}}.

\bibitem{Ginslec}
P.~Ginsparg, {\it Applied conformal field theory}, Lectures given at 
  Les Houches
  Summer School in Theoretical Physics, Les Houches, France, Jun 28 - Aug 5,
  1988.

\bibitem{Cardlec}
J.~L. Cardy, {\it Conformal invariance and statistical mechanics},
Lectures given at Les Houches Summer School in Theoretical Physics, 
Les Houches, France, Jun 28 - Aug 5. 

\bibitem{FrMaSeBook}
P.~Di~Francesco, P.~Mathieu, and D.~Senechal, {\em {Conformal Field Theory}}.
\newblock Graduate Texts in Contemporary Physics. Springer, New York, 1999.

\bibitem{BPZ}
A.~A. Belavin, A.~M. Polyakov, and A.~B. Zamolodchikov, {\it Infinite conformal
  symmetry in two-dimensional quantum field theory}, {\em Nucl. Phys.} {\bf
  B241} (1984) 333--380.

\bibitem{Card84}
J.~L. Cardy, {\it Conformal invariance and surface critical behaviour}, {\em
  Nucl. Phys.} {\bf B240} (1984) 514--532.

\bibitem{Card86}
J.~L. Cardy, {\it Effect of boundary conditions on the operator content of
  two-dimensional conformally invariant theories}, {\em Nucl. Phys.} {\bf
  B275} (1986) 200--218.

\bibitem{Card89}
J.~L. Cardy, {\it Boundary conditions, fusion rules and the {Verlinde} formula},
  {\em Nucl. Phys.} {\bf B324} (1989) 581.

\bibitem{Lewe1}
D.~C. Lewellen, {\it Sewing constraints for conformal field theories on surfaces
  with boundaries}, {\em Nucl. Phys.} {\bf B372} (1992) 654--682.

\bibitem{BiaSag1}
M.~Bianchi and A.~Sagnotti, {\it On the systematics of open string theories},
  {\em Phys. Lett.} {\bf B247} (1990) 517--524.

\bibitem{BiPrSa1}
M.~Bianchi, G.~Pradisi, and A.~Sagnotti, {\it Planar duality in the discrete
  series}, {\em Phys. Lett.} {\bf B273} (1991) 389--398.

\bibitem{BiaSag2}
M.~Bianchi and A.~Sagnotti, {\it Twist symmetry and open string {Wilson} lines},
  {\em Nucl. Phys.} {\bf B361} (1991) 519--538.

\bibitem{PrSaSt1}
G.~Pradisi, A.~Sagnotti, and Y.~S. Stanev, {\it Planar duality in {SU(2)} {WZW}
  models}, {\em Phys. Lett.} {\bf B354} (1995) 279--286,
  \href{http://xxx.lanl.gov/abs/hep-th/9503207}{{\tt hep-th/9503207}}.

\bibitem{PrSaSt2}
G.~Pradisi, A.~Sagnotti, and Y.~S. Stanev, {\it The open descendants of
  nondiagonal {SU(2)} {WZW} models}, {\em Phys. Lett.} {\bf B356} (1995)
  230--238, \href{http://xxx.lanl.gov/abs/hep-th/9506014}{{\tt
  hep-th/9506014}}.

\bibitem{RecSch1}
A.~Recknagel and V.~Schomerus, {\it D-branes in {Gepner} models}, {\em Nucl.
  Phys.} {\bf B531} (1998) 185--225,
  \href{http://xxx.lanl.gov/abs/hep-th/9712186}{{\tt hep-th/9712186}}.

\bibitem{FucSch2}
J.~Fuchs and C.~Schweigert, {\it Branes: {From} free fields to general
  backgrounds}, {\em Nucl. Phys.} {\bf B530} (1998) 99--136,
  \href{http://xxx.lanl.gov/abs/hep-th/9712257}{{\tt hep-th/9712257}}.

\bibitem{GrScWi1Book}
M.~B. Green, J.~H. Schwarz, and E.~Witten, {\em {Superstring Theory I}}.
\newblock Cambridge Monographs on Mathematical Physics. Cambridge University
  Press, Cambridge, 1987.

\bibitem{GrScWi2Book}
M.~B. Green, J.~H. Schwarz, and E.~Witten, {\em {Superstring Theory II}}.
\newblock Cambridge Monographs on Mathematical Physics. Cambridge University
  Press, Cambridge, 1987.

\bibitem{Pol1Book}
J.~G. Polchinski, {\em {String Theory : An Introduction to the Bosonic
  String}}.
\newblock Cambridge Monographs on Mathematical Physics. Cambridge University
  Press, Cambridge, 1998.

\bibitem{Pol2Book}
J.~G. Polchinski, {\em {String Theory : Superstring Theory and Beyond}}.
\newblock Cambridge Monographs on Mathematical Physics. Cambridge University
  Press, Cambridge, 1998.

\bibitem{PetZublec}
V.~B. Petkova and J.-B. Zuber, {\it Conformal boundary conditions and what they
  teach us}, \href{http://xxx.lanl.gov/abs/hep-th/0103007}{{\tt
  hep-th/0103007}}.

\bibitem{Salelec}
H.~Saleur, {\it Lectures on non perturbative field theory and quantum impurity
  problems}, \href{http://xxx.lanl.gov/abs/cond-mat/9812110}{{\tt
  cond-mat/9812110}}.

\bibitem{Gablec}
M.~R. Gaberdiel, {\it Lectures on {non-BPS Dirichlet} branes}, {\em Class. Quant.
  Grav.} {\bf 17} (2000) 3483--3520,
  \href{http://xxx.lanl.gov/abs/hep-th/0005029}{{\tt hep-th/0005029}}.

\bibitem{AngSaglec}
C.~Angelantonj and A.~Sagnotti, {\it Open strings},
  \href{http://xxx.lanl.gov/abs/hep-th/0204089}{{\tt hep-th/0204089}}.

\bibitem{Weyl}
H.~Weyl, {\it Quantum mechanics and group theory}, {\em Z. Phys.} {\bf 46} (1927)
  1.

\bibitem{KonSch2}
A.~Konechny and A.~Schwarz, {\it Introduction to {M(atrix)} theory and
  noncommutative geometry, {Part II}}, {\em Phys. Rept.} {\bf 360} (2002)
  353--465, \href{http://xxx.lanl.gov/abs/hep-th/0107251}{{\tt
  hep-th/0107251}}.

\bibitem{Moyal}
J.~E. Moyal, {\it Quantum mechanics as a statistical theory}, {\em Proc.
  Cambridge Phil. Soc.} {\bf 45} (1949) 99--124.

\bibitem{SeiWit99}
N.~Seiberg and E.~Witten, {\it String theory and noncommutative geometry}, {\em
  JHEP} {\bf 09} (1999) 032, \href{http://xxx.lanl.gov/abs/hep-th/9908142}{{\tt
  hep-th/9908142}}.

\bibitem{DouHul}
M.~R. Douglas and C.~Hull, {\it D-branes and the noncommutative torus}, {\em
  JHEP} {\bf 02} (1998) 008, \href{http://xxx.lanl.gov/abs/hep-th/9711165}{{\tt
  hep-th/9711165}}.

\bibitem{CheKro}
Y.-K.~E. Cheung and M.~Krogh, {\it Noncommutative geometry from 0-branes in a
  background {B-field}}, {\em Nucl. Phys.} {\bf B528} (1998) 185--196,
  \href{http://xxx.lanl.gov/abs/hep-th/9803031}{{\tt hep-th/9803031}}.

\bibitem{ArArSh}
F.~Ardalan, H.~Arfaei, and M.~M. Sheikh-Jabbari, {\it Noncommutative geometry from
  strings and branes}, {\em JHEP} {\bf 02} (1999) 016,
  \href{http://xxx.lanl.gov/abs/hep-th/9810072}{{\tt hep-th/9810072}}.

\bibitem{ChuHo1}
C.-S. Chu and P.-M. Ho, {\it Noncommutative open string and {D-brane}}, {\em
  Nucl. Phys.} {\bf B550} (1999) 151,
  \href{http://xxx.lanl.gov/abs/hep-th/9812219}{{\tt hep-th/9812219}}.

\bibitem{Scho}
V.~Schomerus, {\it D-branes and deformation quantization}, {\em JHEP} {\bf 06}
  (1999) 030, \href{http://xxx.lanl.gov/abs/hep-th/9903205}{{\tt
  hep-th/9903205}}.

\bibitem{CallThor}
C.~G. Callan and L.~Thorlacius, {\it Open string theory as dissipative quantum
  mechanics}, {\em Nucl. Phys.} {\bf B329} (1990) 117.

\bibitem{NekSch}
N.~Nekrasov and A.~Schwarz, {\it Instantons on noncommutative {R**4} and (2,0)
  superconformal six dimensional theory}, {\em Commun. Math. Phys.} {\bf 198}
  (1998) 689--703, \href{http://xxx.lanl.gov/abs/hep-th/9802068}{{\tt
  hep-th/9802068}}.

\bibitem{GoMiSt}
R.~Gopakumar, S.~Minwalla, and A.~Strominger, {\it Noncommutative solitons}, {\em
  JHEP} {\bf 05} (2000) 020, \href{http://xxx.lanl.gov/abs/hep-th/0003160}{{\tt
  hep-th/0003160}}.

\bibitem{GroNek1}
D.~J. Gross and N.~A. Nekrasov, {\it Monopoles and strings in noncommutative gauge
  theory}, {\em JHEP} {\bf 07} (2000) 034,
  \href{http://xxx.lanl.gov/abs/hep-th/0005204}{{\tt hep-th/0005204}}.

\bibitem{Poly}
A.~P. Polychronakos, {\it Flux tube solutions in noncommutative gauge theories},
  {\em Phys. Lett.} {\bf B495} (2000) 407--412,
  \href{http://xxx.lanl.gov/abs/hep-th/0007043}{{\tt hep-th/0007043}}.

\bibitem{AgGoMiSt}
M.~Aganagic, R.~Gopakumar, S.~Minwalla, and A.~Strominger, {\it Unstable solitons
  in noncommutative gauge theory}, {\em JHEP} {\bf 04} (2001) 001,
  \href{http://xxx.lanl.gov/abs/hep-th/0009142}{{\tt hep-th/0009142}}.

\bibitem{HaKrLa}
J.~A. Harvey, P.~Kraus, and F.~Larsen, {\it Exact noncommutative solitons}, {\em
  JHEP} {\bf 12} (2000) 024, \href{http://xxx.lanl.gov/abs/hep-th/0010060}{{\tt
  hep-th/0010060}}.

\bibitem{KonSch1}
A.~Konechny and A.~Schwarz, {\it Introduction to {M(atrix)} theory and
  noncommutative geometry}, {\em Phys. Rept.} {\bf 360} (2002) 353--465,
  \href{http://xxx.lanl.gov/abs/hep-th/0012145}{{\tt hep-th/0012145}}.

\bibitem{Harvlec}
J.~A. Harvey, {\it Komaba lectures on noncommutative solitons and {D-branes}},
  \href{http://xxx.lanl.gov/abs/hep-th/0102076}{{\tt hep-th/0102076}}.

\bibitem{DouNek}
M.~R. Douglas and N.~A. Nekrasov, {\it Noncommutative field theory}, {\em Rev.
  Mod. Phys.} {\bf 73} (2002) 977--1029,
  \href{http://xxx.lanl.gov/abs/hep-th/0106048}{{\tt hep-th/0106048}}.

\bibitem{SchBou}
P.~Bouwknegt and K.~Schoutens, {\em {W - symmetry: reprints}}.
\newblock World Scientific, 1995.

\bibitem{Bonn}
R.~Blumenhagen {\em et.~al.}, {\it W algebras with two and three generators},
  {\em Nucl. Phys.} {\bf B361} (1991) 255--289.

\bibitem{Verl}
E.~Verlinde, {\it Fusion rules and modular transformations in {2-D} conformal
  field theory}, {\em Nucl. Phys.} {\bf B300} (1988) 360.

\bibitem{MooSei2}
G.~W. Moore and N.~Seiberg, {\it Classical and quantum conformal field theory},
  {\em Commun. Math. Phys.} {\bf 123} (1989) 177.

\bibitem{GaReWa}
M.~R. Gaberdiel, A.~Recknagel, and G.~M.~T. Watts, {\it The conformal boundary
  states for {SU(2)} at level 1}, {\em Nucl. Phys.} {\bf B626} (2002)
  344--362, \href{http://xxx.lanl.gov/abs/hep-th/0108102}{{\tt
  hep-th/0108102}}.

\bibitem{GabRec}
M.~R. Gaberdiel and A.~Recknagel, {\it Conformal boundary states for free bosons
  and fermions}, {\em JHEP} {\bf 11} (2001) 016,
  \href{http://xxx.lanl.gov/abs/hep-th/0108238}{{\tt hep-th/0108238}}.

\bibitem{MaMoSe1}
J.~Maldacena, G.~W. Moore, and N.~Seiberg, {\it Geometrical interpretation of
  {D-branes} in gauged {WZW} models}, {\em JHEP} {\bf 07} (2001) 046,
  \href{http://xxx.lanl.gov/abs/hep-th/0105038}{{\tt hep-th/0105038}}.

\bibitem{QueSch1}
T.~Quella and V.~Schomerus, {\it Symmetry breaking boundary states and defect
  lines}, {\em JHEP} {\bf 06} (2002) 028,
  \href{http://xxx.lanl.gov/abs/hep-th/0203161}{{\tt hep-th/0203161}}.

\bibitem{CarLew}
J.~L. Cardy and D.~C. Lewellen, {\it Bulk and boundary operators in conformal
  field theory}, {\em Phys. Lett.} {\bf B259} (1991) 274--278.

\bibitem{BPPZ1}
R.~E. Behrend, P.~A. Pearce, V.~B. Petkova, and J.-B. Zuber, {\it On the
  classification of bulk and boundary conformal field theories}, {\em Phys.
  Lett.} {\bf B444} (1998) 163,
  \href{http://xxx.lanl.gov/abs/hep-th/9809097}{{\tt hep-th/9809097}}.

\bibitem{FucSch1}
J.~Fuchs and C.~Schweigert, {\it A classifying algebra for boundary conditions},
  {\em Phys. Lett.} {\bf B414} (1997) 251--259,
  \href{http://xxx.lanl.gov/abs/hep-th/9708141}{{\tt hep-th/9708141}}.

\bibitem{RecSch2}
A.~Recknagel and V.~Schomerus, {\it Boundary deformation theory and moduli spaces
  of {D}-branes}, {\em Nucl. Phys.} {\bf B545} (1999) 233,
  \href{http://xxx.lanl.gov/abs/hep-th/9811237}{{\tt hep-th/9811237}}.

\bibitem{Pasq}
V.~Pasquier, {\it Operator content of the {ADE} lattice models}, {\em J. Phys.}
  {\bf A20} (1987) 5707.

\bibitem{PetZub1}
V.~B. Petkova and J.~B. Zuber, {\it On structure constants of sl(2) theories},
  {\em Nucl. Phys.} {\bf B438} (1995) 347--372,
  \href{http://xxx.lanl.gov/abs/hep-th/9410209}{{\tt hep-th/9410209}}.

\bibitem{PetZub2}
V.~B. Petkova and J.~B. Zuber, {\it From {CFT's} to graphs}, {\em Nucl. Phys.}
  {\bf B463} (1996) 161--193,
  \href{http://xxx.lanl.gov/abs/hep-th/9510175}{{\tt hep-th/9510175}}.

\bibitem{Ishi1}
N.~Ishibashi, {\it The boundary and crosscap states in conformal field theories},
  {\em Mod. Phys. Lett.} {\bf A4} (1989) 251.

\bibitem{Runk1}
I.~Runkel, {\it Boundary structure constants for the {A-series} {Virasoro} minimal
  models}, {\em Nucl. Phys.} {\bf B549} (1999) 563,
  \href{http://xxx.lanl.gov/abs/hep-th/9811178}{{\tt hep-th/9811178}}.

\bibitem{Runk2}
I.~Runkel, {\it Structure constants for the {D-series} {Virasoro} minimal
  models}, {\em Nucl. Phys.} {\bf B579} (2000) 561,
  \href{http://xxx.lanl.gov/abs/hep-th/9908046}{{\tt hep-th/9908046}}.

\bibitem{BPPZ2}
R.~E. Behrend, P.~A. Pearce, V.~B. Petkova, and J.-B. Zuber, {\it Boundary
  conditions in rational conformal field theories}, {\em Nucl. Phys.} {\bf
  B570} (2000) 525--589, \href{http://xxx.lanl.gov/abs/hep-th/9908036}{{\tt
  hep-th/9908036}}.

\bibitem{FFFS1}
G.~Felder, J.~{Fr\"ohlich}, J.~Fuchs, and C.~Schweigert, {\it The geometry of
  {WZW} branes}, {\em J. Geom. Phys.} {\bf 34} (2000) 162--190,
  \href{http://xxx.lanl.gov/abs/hep-th/9909030}{{\tt hep-th/9909030}}.

\bibitem{FFFS2}
G.~Felder, J.~{Fr\"ohlich}, J.~Fuchs, and C.~Schweigert, {\it Conformal boundary
  conditions and three-dimensional topological field theory}, {\em Phys. Rev.
  Lett.} {\bf 84} (2000) 1659,
  \href{http://xxx.lanl.gov/abs/hep-th/9909140}{{\tt hep-th/9909140}}.

\bibitem{zuber}
J.-B. Zuber, {\it {CFT, BCFT, ADE} and all that},
  \href{http://xxx.lanl.gov/abs/hep-th/0006151}{{\tt hep-th/0006151}}.

\bibitem{CaHaSt1}
C.~G. Callan, J.~A. Harvey, and A.~Strominger, {\it World sheet approach to
  heterotic instantons and solitons}, {\em Nucl. Phys.} {\bf B359} (1991)
  611--634.

\bibitem{CaHaSt2}
C.~G. Callan, J.~A. Harvey, and A.~Strominger, {\it Supersymmetric string
  solitons}, \href{http://xxx.lanl.gov/abs/hep-th/9112030}{{\tt
  hep-th/9112030}}.

\bibitem{WitK}
E.~Witten, {\it {D-branes} and {K-theory}}, {\em JHEP} {\bf 12} (1998) 019,
  \href{http://xxx.lanl.gov/abs/hep-th/9810188}{{\tt hep-th/9810188}}.

\bibitem{FreWit}
D.~S. Freed and E.~Witten, {\it Anomalies in string theory with {D-branes}},
  \href{http://xxx.lanl.gov/abs/hep-th/9907189}{{\tt hep-th/9907189}}.

\bibitem{BaDoSc}
C.~Bachas, M.~R. Douglas, and C.~Schweigert, {\it Flux stabilization of
  {D-branes}}, {\em JHEP} {\bf 05} (2000) 048,
  \href{http://xxx.lanl.gov/abs/hep-th/0003037}{{\tt hep-th/0003037}}.

\bibitem{Pawe}
J.~Pawelczyk, {\it {SU(2)} {WZW} {D-branes} and their noncommutative geometry from
  {DBI} action}, {\em JHEP} {\bf 08} (2000) 006,
  \href{http://xxx.lanl.gov/abs/hep-th/0003057}{{\tt hep-th/0003057}}.

\bibitem{AlReSc3}
A.~Yu.~Alekseev, A.~Recknagel, and V.~Schomerus, {\it Brane dynamics in background
  fluxes and non-commutative geometry}, {\em JHEP} {\bf 05} (2000) 010,
  \href{http://xxx.lanl.gov/abs/hep-th/0003187}{{\tt hep-th/0003187}}.

\bibitem{AleSch5}
A.~Yu.~Alekseev and V.~Schomerus, {\it {D-branes} in the {WZW} model}, {\em Phys.
  Rev.} {\bf D60} (1999) 061901,
  \href{http://xxx.lanl.gov/abs/hep-th/9812193}{{\tt hep-th/9812193}}.

\bibitem{Gawe1}
K.~Gawedzki, {\it Conformal field theory: {A} case study},
  \href{http://xxx.lanl.gov/abs/hep-th/9904145}{{\tt hep-th/9904145}}.

\bibitem{Stan00/1}
S.~Stanciu, {\it D-branes in group manifolds}, {\em JHEP} {\bf 01} (2000) 025,
  \href{http://xxx.lanl.gov/abs/hep-th/9909163}{{\tt hep-th/9909163}}.

\bibitem{Hopp}
J.~Hoppe, {\it Diffeomorphism groups, quantization and {$SU(\infty)$}}, {\em Int.
  J. Mod. Phys.} {\bf A4} (1989) 5235.

\bibitem{Mado1}
J.~Madore, {\it The fuzzy sphere}, {\em Class. Quant. Grav.} {\bf 9} (1992)
  69--88.

\bibitem{GodOli}
P.~Goddard and D.~Olive, {\it {Kac-Moody} and {Virasoro} algebras in relation to
  quantum physics}, {\em Int. J. Mod. Phys.} {\bf A1} (1986) 303.

\bibitem{AlReSc2}
A.~Yu.~Alekseev, A.~Recknagel, and V.~Schomerus, {\it Non-commutative world-volume
  geometries: {Branes} on {SU(2)} and fuzzy spheres}, {\em JHEP} {\bf 09}
  (1999) 023, \href{http://xxx.lanl.gov/abs/hep-th/9908040}{{\tt
  hep-th/9908040}}.

\bibitem{AFQS}
A.~Y. Alekseev, S.~Fredenhagen, T.~Quella, and V.~Schomerus, {\it Non-commutative
  gauge theory of twisted {D-branes}},
  \href{http://xxx.lanl.gov/abs/hep-th/0205123}{{\tt hep-th/0205123}}.

\bibitem{GrKlPr95/1}
H.~Grosse, C.~Klimcik, and P.~Presnajder, {\it Towards finite quantum field theory
  in noncommutative geometry}, {\em Int. J. Theor. Phys.} {\bf 35} (1996)
  231--244, \href{http://xxx.lanl.gov/abs/hep-th/9505175}{{\tt
  hep-th/9505175}}.

\bibitem{GrKlPr95/2}
H.~Grosse, C.~Klimcik, and P.~Presnajder, {\it Field theory on a supersymmetric
  lattice}, {\em Commun. Math. Phys.} {\bf 185} (1997) 155,
  \href{http://xxx.lanl.gov/abs/hep-th/9507074}{{\tt hep-th/9507074}}.

\bibitem{GrKlPr95/3}
H.~Grosse, C.~Klimcik, and P.~Presnajder, {\it Simple field theoretical models on
  noncommutative manifolds},
  \href{http://xxx.lanl.gov/abs/hep-th/9510177}{{\tt hep-th/9510177}}.

\bibitem{MadoBook}
J.~Madore, {\em An introduction to noncommutative differential geometry and its
  physical applications}.
\newblock Cambridge University Press, 1999.

\bibitem{WatWat}
U.~Carow-Watamura and S.~Watamura, {\it Noncommutative geometry and gauge theory
  on fuzzy sphere}, {\em Commun. Math. Phys.} {\bf 212} (2000) 395,
  \href{http://xxx.lanl.gov/abs/hep-th/9801195}{{\tt hep-th/9801195}}.

\bibitem{Klimcik:1999uk}
C.~Klimcik, {\it A nonperturbative regularization of the supersymmetric
  {Schwinger} model}, {\em Commun. Math. Phys.} {\bf 206} (1999) 567,
  \href{http://xxx.lanl.gov/abs/hep-th/9903112}{{\tt hep-th/9903112}}.

\bibitem{AffLud91/1}
I.~Affleck and A.~W.~W. Ludwig, {\it Universal noninteger `ground state
  degeneracy' in critical quantum systems}, {\em Phys. Rev. Lett.} {\bf 67}
  (1991) 161--164.

\bibitem{RecSch3}
A.~Recknagel and V.~Schomerus, {\it Moduli spaces of {D-branes} in
  {CFT-backgrounds}}, {\em Fortsch. Phys.} {\bf 48} (2000) 195,
  \href{http://xxx.lanl.gov/abs/hep-th/9903139}{{\tt hep-th/9903139}}.

\bibitem{HasKra}
K.~Hashimoto and K.~Krasnov, {\it D-brane solutions in non-commutative gauge
  theory on fuzzy sphere}, {\em Phys. Rev.} {\bf D64} (2001) 046007,
  \href{http://xxx.lanl.gov/abs/hep-th/0101145}{{\tt hep-th/0101145}}.

\bibitem{HiNoSu}
Y.~Hikida, M.~Nozaki, and Y.~Sugawara, {\it Formation of spherical {D2-brane} from
  multiple {D0-branes}}, {\em Nucl. Phys.} {\bf B617} (2001) 117--150,
  \href{http://xxx.lanl.gov/abs/hep-th/0101211}{{\tt hep-th/0101211}}.

\bibitem{JMWY}
D.~P. Jatkar, G.~Mandal, S.~R. Wadia, and K.~P. Yogendran, {\it Matrix dynamics of
  fuzzy spheres}, {\em JHEP} {\bf 01} (2002) 039,
  \href{http://xxx.lanl.gov/abs/hep-th/0110172}{{\tt hep-th/0110172}}.

\bibitem{Myers}
R.~C. Myers, {\it Dielectric-branes}, {\em JHEP} {\bf 12} (1999) 022,
  \href{http://xxx.lanl.gov/abs/hep-th/9910053}{{\tt hep-th/9910053}}.

\bibitem{AleSch6}
A.~Yu.~Alekseev and V.~Schomerus, {\it {RR} charges of {D2-branes} in the {WZW}
  model}, \href{http://xxx.lanl.gov/abs/hep-th/0007096}{{\tt hep-th/0007096}}.

\bibitem{FreSch1}
S.~Fredenhagen and V.~Schomerus, {\it Branes on group manifolds, gluon
  condensates, and twisted {K-theory}}, {\em JHEP} {\bf 04} (2001) 007,
  \href{http://xxx.lanl.gov/abs/hep-th/0012164}{{\tt hep-th/0012164}}.

\bibitem{FreSch2}
S.~Fredenhagen and V.~Schomerus, {\it Brane dynamics in {CFT} backgrounds},
  \href{http://xxx.lanl.gov/abs/hep-th/0104043}{{\tt hep-th/0104043}}.

\bibitem{Wilson}
K.~G. Wilson, {\it The renormalization group: {C}ritical phenomena and the {K}ondo
  problem}, {\em Rev. Mod. Phys.} {\bf 47} (1975) 773.

\bibitem{AffLud91/2}
I.~Affleck and A.~W.~W. Ludwig, {\it The {Kondo} effect, conformal field theory
  and fusion rules}, {\em Nucl. Phys.} {\bf B352} (1991) 849--862.

\bibitem{AffLud91/3}
I.~Affleck and A.~W.~W. Ludwig, {\it Critical theory of overscreened {Kondo} fixed
  points}, {\em Nucl. Phys.} {\bf B360} (1991) 641--696.

\bibitem{BouMat}
P.~Bouwknegt and V.~Mathai, {\it {D-branes}, {B-fields} and twisted {K-theory}},
  {\em JHEP} {\bf 03} (2000) 007,
  \href{http://xxx.lanl.gov/abs/hep-th/0002023}{{\tt hep-th/0002023}}.

\bibitem{Kapu}
A.~Kapustin, {\it D-branes in a topologically nontrivial {B-field}}, {\em Adv.
  Theor. Math. Phys.} {\bf 4} (2000) 127--154,
  \href{http://xxx.lanl.gov/abs/hep-th/9909089}{{\tt hep-th/9909089}}.

\bibitem{MaMoSe2}
J.~Maldacena, G.~W. Moore, and N.~Seiberg, {\it D-brane instantons and {K-theory}
  charges}, {\em JHEP} {\bf 11} (2001) 062,
  \href{http://xxx.lanl.gov/abs/hep-th/0108100}{{\tt hep-th/0108100}}.

\bibitem{Tay}
W.~Taylor, {\it D2-branes in {B-fields}}, {\em JHEP} {\bf 07} (2000) 039,
  \href{http://xxx.lanl.gov/abs/hep-th/0004141}{{\tt hep-th/0004141}}.

\bibitem{Stan00/2}
S.~Stanciu, {\it A note on {D-branes} in group manifolds: Flux quantization and
  {D0-charge}}, {\em JHEP} {\bf 10} (2000) 015,
  \href{http://xxx.lanl.gov/abs/hep-th/0006145}{{\tt hep-th/0006145}}.

\bibitem{FiguStan}
J.~M. Figueroa-O'Farrill and S.~Stanciu, {\it D-brane charge, flux quantization
  and relative (co)homology}, {\em JHEP} {\bf 01} (2001) 006,
  \href{http://xxx.lanl.gov/abs/hep-th/0008038}{{\tt hep-th/0008038}}.

\bibitem{Janik}
R.~A. Janik, {\it Exceptional boundary states at c = 1}, {\em Nucl. Phys.} {\bf
  B618} (2001) 675--688, \href{http://xxx.lanl.gov/abs/hep-th/0109021}{{\tt
  hep-th/0109021}}.

\bibitem{CapDAp}
A.~Cappelli and G.~D'Appollonio, {\it Boundary states of {$c = 1$} and {$3/2$}
  rational conformal field theories}, {\em JHEP} {\bf 02} (2002) 039,
  \href{http://xxx.lanl.gov/abs/hep-th/0201173}{{\tt hep-th/0201173}}.

\bibitem{DouMoo}
M.~R. Douglas and G.~W. Moore, {\it D-branes, {Quivers, and ALE Instantons}},
  \href{http://xxx.lanl.gov/abs/hep-th/9603167}{{\tt hep-th/9603167}}.

\bibitem{AffOsh1}
M.~Oshikawa and I.~Affleck, {\it Defect lines in the {Ising} model and boundary
  states on orbifolds}, {\em Phys. Rev. Lett.} {\bf 77} (1996) 2604--2607,
  \href{http://xxx.lanl.gov/abs/hep-th/9606177}{{\tt hep-th/9606177}}.

\bibitem{AffOsh2}
M.~Oshikawa and I.~Affleck, {\it Boundary conformal field theory approach to the
  critical two-dimensional {Ising} model with a defect line}, {\em Nucl.
  Phys.} {\bf B495} (1997) 533--582,
  \href{http://xxx.lanl.gov/abs/cond-mat/9612187}{{\tt cond-mat/9612187}}.

\bibitem{Dou}
M.~R. Douglas, {\it D-branes and discrete torsion},
  \href{http://xxx.lanl.gov/abs/hep-th/9807235}{{\tt hep-th/9807235}}.

\bibitem{DouFio}
M.~R. Douglas and B.~Fiol, {\it D-branes and discrete torsion. {II}},
  \href{http://xxx.lanl.gov/abs/hep-th/9903031}{{\tt hep-th/9903031}}.

\bibitem{DiaGom}
D.-E. Diaconescu and J.~Gomis, {\it Fractional branes and boundary states in
  orbifold theories}, {\em JHEP} {\bf 10} (2000) 001,
  \href{http://xxx.lanl.gov/abs/hep-th/9906242}{{\tt hep-th/9906242}}.

\bibitem{Gom}
J.~Gomis, {\it D-branes on orbifolds with discrete torsion and topological
  obstruction}, {\em JHEP} {\bf 05} (2000) 006,
  \href{http://xxx.lanl.gov/abs/hep-th/0001200}{{\tt hep-th/0001200}}.

\bibitem{GabSte}
M.~R. Gaberdiel and J.~Stefanski, Bogdan, {\it Dirichlet branes on orbifolds},
  {\em Nucl. Phys.} {\bf B578} (2000) 58--84,
  \href{http://xxx.lanl.gov/abs/hep-th/9910109}{{\tt hep-th/9910109}}.

\bibitem{Gab}
M.~R. Gaberdiel, {\it Discrete torsion orbifolds and {D-branes}}, {\em JHEP} {\bf
  11} (2000) 026, \href{http://xxx.lanl.gov/abs/hep-th/0008230}{{\tt
  hep-th/0008230}}.

\bibitem{CraGab}
B.~Craps and M.~R. Gaberdiel, {\it Discrete torsion orbifolds and {D-branes.
  II}}, {\em JHEP} {\bf 04} (2001) 013,
  \href{http://xxx.lanl.gov/abs/hep-th/0101143}{{\tt hep-th/0101143}}.

\bibitem{BiCrRo}
M.~Billo, B.~Craps, and F.~Roose, {\it Orbifold boundary states from {Cardy's}
  condition}, {\em JHEP} {\bf 01} (2001) 038,
  \href{http://xxx.lanl.gov/abs/hep-th/0011060}{{\tt hep-th/0011060}}.

\bibitem{PraSag}
G.~Pradisi and A.~Sagnotti, {\it Open string orbifolds}, {\em Phys. Lett.} {\bf
  B216} (1989) 59.

\bibitem{GimPol}
E.~G. Gimon and J.~Polchinski, {\it Consistency conditions for orientifolds and
  {D}-manifolds}, {\em Phys. Rev.} {\bf D54} (1996) 1667--1676,
  \href{http://xxx.lanl.gov/abs/hep-th/9601038}{{\tt hep-th/9601038}}.

\bibitem{DMS}
E.~Dudas, J.~Mourad, and A.~Sagnotti, {\it Charged and uncharged {D-branes} in
  various string theories}, {\em Nucl. Phys.} {\bf B620} (2002) 109--151,
  \href{http://xxx.lanl.gov/abs/hep-th/0107081}{{\tt hep-th/0107081}}.

\bibitem{FucSchSB1}
J.~Fuchs and C.~Schweigert, {\it Symmetry breaking boundaries. {I: General}
  theory}, {\em Nucl. Phys.} {\bf B558} (1999) 419,
  \href{http://xxx.lanl.gov/abs/hep-th/9902132}{{\tt hep-th/9902132}}.

\bibitem{FucSchSB2}
J.~Fuchs and C.~Schweigert, {\it Symmetry breaking boundaries. {II: More}
  structures, examples}, {\em Nucl. Phys.} {\bf B568} (2000) 543,
  \href{http://xxx.lanl.gov/abs/hep-th/9908025}{{\tt hep-th/9908025}}.

\bibitem{BiFuSc}
L.~Birke, J.~Fuchs, and C.~Schweigert, {\it Symmetry breaking boundary conditions
  and {WZW} orbifolds}, {\em Adv. Theor. Math. Phys.} {\bf 3} (1999) 671--726,
  \href{http://xxx.lanl.gov/abs/hep-th/9905038}{{\tt hep-th/9905038}}.

\bibitem{HalpSC}
M.~B. Halpern, {\it Direct approach to operator conformal constructions: {From}
  fermions to primary fields}, {\em Ann. Phys.} {\bf 194} (1989) 247.
  Erratum-ibid. {\bf 200} (1989) 422.

\bibitem{IntrSC}
K.~A. Intriligator, {\it Bonus symmetry in conformal field theory}, {\em Nucl.
  Phys.} {\bf B332} (1990) 541.

\bibitem{SchYan1}
A.~N. Schellekens and S.~Yankielowicz, {\it Extended chiral algebras and modular
  invariant partition functions}, {\em Nucl. Phys.} {\bf B327} (1989) 673.

\bibitem{SchYan2}
A.~N. Schellekens and S.~Yankielowicz, {\it Simple currents, modular invariants
  and fixed points}, {\em Int. J. Mod. Phys.} {\bf A5} (1990) 2903--2952.

\bibitem{HuScSo}
L.~R. Huiszoon, A.~N. Schellekens, and N.~Sousa, {\it Open descendants of
  non-diagonal invariants}, {\em Nucl. Phys.} {\bf B575} (2000) 401--415,
  \href{http://xxx.lanl.gov/abs/hep-th/9911229}{{\tt hep-th/9911229}}.

\bibitem{Sche}
A.~N. Schellekens, {\it Open strings, simple currents and fixed points},
  \href{http://xxx.lanl.gov/abs/hep-th/0001198}{{\tt hep-th/0001198}}.

\bibitem{BruSch1}
I.~Brunner and V.~Schomerus, {\it D-branes at singular curves of {Calabi-Yau}
  compactifications}, {\em JHEP} {\bf 04} (2000) 020,
  \href{http://xxx.lanl.gov/abs/hep-th/0001132}{{\tt hep-th/0001132}}.

\bibitem{BruSch2}
I.~Brunner and V.~Schomerus, {\it On superpotentials for {D-branes in Gepner}
  models}, {\em JHEP} {\bf 10} (2000) 016,
  \href{http://xxx.lanl.gov/abs/hep-th/0008194}{{\tt hep-th/0008194}}.

\bibitem{MaScSm}
K.~Matsubara, V.~Schomerus, and M.~Smedback, {\it Open strings in simple current
  orbifolds}, {\em Nucl. Phys.} {\bf B626} (2002) 53--72,
  \href{http://xxx.lanl.gov/abs/hep-th/0108126}{{\tt hep-th/0108126}}.

\bibitem{FHSSW}
J.~Fuchs, L.~R. Huiszoon, A.~N. Schellekens, C.~Schweigert, and J.~Walcher,
  {\it Boundaries, crosscaps and simple currents}, {\em Phys. Lett.} {\bf B495}
  (2000) 427--434, \href{http://xxx.lanl.gov/abs/hep-th/0007174}{{\tt
  hep-th/0007174}}.

\bibitem{Gepn}
D.~Gepner, {\it Space-time supersymmetry in compactified string theory and
  superconformal models}, {\em Nucl. Phys.} {\bf B296} (1988) 757.

\bibitem{NakNoz}
M.~Naka and M.~Nozaki, {\it Boundary states in {Gepner} models}, {\em JHEP} {\bf
  05} (2000) 027, \href{http://xxx.lanl.gov/abs/hep-th/0001037}{{\tt
  hep-th/0001037}}.

\bibitem{GoJaSa}
S.~Govindarajan, T.~Jayaraman, and T.~Sarkar, {\it Worldsheet approaches to
  {D-branes} on supersymmetric cycles}, {\em Nucl. Phys.} {\bf B580} (2000)
  519--547, \href{http://xxx.lanl.gov/abs/hep-th/9907131}{{\tt
  hep-th/9907131}}.

\bibitem{FuScWa}
J.~Fuchs, C.~Schweigert, and J.~Walcher, {\it Projections in string theory and
  boundary states for {Gepner} models}, {\em Nucl. Phys.} {\bf B588} (2000)
  110--148, \href{http://xxx.lanl.gov/abs/hep-th/0003298}{{\tt
  hep-th/0003298}}.

\bibitem{BDLR}
I.~Brunner, M.~R. Douglas, A.~E. Lawrence, and C.~Romelsberger, {\it D-branes on
  the quintic}, {\em JHEP} {\bf 08} (2000) 015,
  \href{http://xxx.lanl.gov/abs/hep-th/9906200}{{\tt hep-th/9906200}}.

\bibitem{FKLLSW}
J.~Fuchs {\em et.~al.}, {\it Boundary fixed points, enhanced gauge symmetry and
  singular bundles on {K3}}, {\em Nucl. Phys.} {\bf B598} (2001) 57--72,
  \href{http://xxx.lanl.gov/abs/hep-th/0007145}{{\tt hep-th/0007145}}.

\bibitem{BruDis}
I.~Brunner and J.~Distler, {\it Torsion {D-branes} in nongeometrical phases},
  \href{http://xxx.lanl.gov/abs/hep-th/0102018}{{\tt hep-th/0102018}}.

\bibitem{Reckperm}
A.~Recknagel, {\it Permutation branes},
  \href{http://xxx.lanl.gov/abs/hep-th/0208119}{{\tt hep-th/0208119}}.

\bibitem{DVVV}
R.~Dijkgraaf, C.~Vafa, E.~Verlinde, and H.~Verlinde, {\it The operator algebra of
  orbifold models}, {\em Commun. Math. Phys.} {\bf 123} (1989) 485.

\bibitem{KacBook1b}
V.~G. Kac, {\em Infinite Dimensional {Lie} Algebras}.
\newblock Cambridge University Press, Cambridge, 1990.

\bibitem{StanSU3}
S.~Stanciu, {\it An illustrated guide to {D-branes in SU(3)}},
  \href{http://xxx.lanl.gov/abs/hep-th/0111221}{{\tt hep-th/0111221}}.

\bibitem{ItoSin1}
T.~Itoh and S.-J. Sin, {\it A note on singular {D-branes} in group manifolds},
  \href{http://xxx.lanl.gov/abs/hep-th/0206238}{{\tt hep-th/0206238}}.

\bibitem{ItoSin2}
T.~Itoh and S.-J. Sin, {\it Classification and quantum moduli space of {D-branes}
  in group manifolds}, \href{http://xxx.lanl.gov/abs/hep-th/0207077}{{\tt
  hep-th/0207077}}.

\bibitem{AlScSt}
A.~Y. Alekseev, P.~Schaller, and T.~Strobl, {\it The topological {G/G WZW model}
  in the generalized momentum representation}, {\em Phys. Rev.} {\bf D52}
  (1995) 7146--7160, \href{http://xxx.lanl.gov/abs/hep-th/9505012}{{\tt
  hep-th/9505012}}.

\bibitem{QuRuSc}
T.~Quella, I.~Runkel and C.~Schweigert,
{\it An algorithm for twisted fusion rules},
{\tt math.qa/0203133}.
%%CITATION = MATH-QA 0203133;%%

\bibitem{BoRiSc}
P.~Bordalo, S.~Ribault, and C.~Schweigert, {\it Flux stabilization in compact
  groups}, {\em JHEP} {\bf 10} (2001) 036,
  \href{http://xxx.lanl.gov/abs/hep-th/0108201}{{\tt hep-th/0108201}}.

\bibitem{GabGan}
M.~R. Gaberdiel and T.~Gannon, {\it Boundary states for {WZW} models},
  \href{http://xxx.lanl.gov/abs/hep-th/0202067}{{\tt hep-th/0202067}}.

\bibitem{PetZub02}
V.~B. Petkova and J.~B. Zuber, {\it Boundary conditions in charge conjugate
  {sl(N)} {WZW} theories}, \href{http://xxx.lanl.gov/abs/hep-th/0201239}{{\tt
  hep-th/0201239}}.

\bibitem{Quel1}
T.~Quella, {\it Branching rules of semi-simple {Lie} algebras using affine
  extensions}, \href{http://xxx.lanl.gov/abs/math-ph/0111020}{{\tt
  math-ph/0111020}}.

\bibitem{AfOsSa2}
I.~Affleck, M.~Oshikawa, and H.~Saleur, {\it Quantum {Brownian} motion on a
  triangular lattice and c = 2 boundary conformal field theory}, {\em Nucl.
  Phys.} {\bf B594} (2001) 535--606,
  \href{http://xxx.lanl.gov/abs/cond-mat/0009084}{{\tt cond-mat/0009084}}.

\bibitem{MaMoSe3}
J.~Maldacena, G.~W. Moore, and N.~Seiberg, {\it D-brane charges in five-brane
  backgrounds}, {\em JHEP} {\bf 10} (2001) 005,
  \href{http://xxx.lanl.gov/abs/hep-th/0108152}{{\tt hep-th/0108152}}.

\bibitem{GoKeOl}
P.~Goddard, A.~Kent, and D.~I. Olive, {\it Virasoro algebras and coset space
  models}, {\em Phys. Lett.} {\bf B152} (1985) 88.

\bibitem{Gepn89}
D.~Gepner, {\it Field identification in coset conformal field theories}, {\em
  Phys. Lett.} {\bf B222} (1989) 207.

\bibitem{SchYancos}
A.~N. Schellekens and S.~Yankielowicz, {\it Field identification fixed points in
  the coset construction}, {\em Nucl. Phys.} {\bf B334} (1990) 67.

\bibitem{Quel2}
T.~Quella, {\it Hierarchy of symmetry breaking branes on group manifolds},
  \href{http://xxx.lanl.gov/abs/hep-th/}{{\tt hep-th/0209157}}.

\bibitem{Wong:pa}
E.~Wong and I.~Affleck,
{\it Tunneling In Quantum Wires: A Boundary Conformal Field Theory Approach},
{\em Nucl.\ Phys.\ } {\bf B417} (1994) 403.
%%CITATION = NUPHA,B417,403;%%

\bibitem{LeClair:1997gz}
A.~LeClair and A.~W.~W. Ludwig, {\it Minimal models with integrable local
  defects}, {\em Nucl. Phys.} {\bf B549} (1999) 546--562,
  \href{http://xxx.lanl.gov/abs/hep-th/9708135}{{\tt hep-th/9708135}}.

\bibitem{Nayak}
C.~Nayak, M.~Fisher, A.~W. Ludwig, and H.~H. Lin, {\it Resonant multilead
  point-contact tunneling}, {\em Phys. Rev.} {\bf B59} (1999) 15694.

\bibitem{Saleur:2000gp}
H.~Saleur, {\it Lectures on non perturbative field theory and quantum impurity
  problems. {II}}, \href{http://xxx.lanl.gov/abs/cond-mat/0007309}{{\tt
  cond-mat/0007309}}.

\bibitem{McAvity:1995zd}
D.~M. McAvity and H.~Osborn, {\it Conformal field theories near a boundary in
  general dimensions}, {\em Nucl. Phys.} {\bf B455} (1995) 522--576,
  \href{http://xxx.lanl.gov/abs/cond-mat/9505127}{{\tt cond-mat/9505127}}.

\bibitem{ErGuKi}
J.~Erdmenger, Z.~Guralnik, and I.~Kirsch, {\it Four-dimensional superconformal
  theories with interacting boundaries or defects},
  \href{http://xxx.lanl.gov/abs/hep-th/0203020}{{\tt hep-th/0203020}}.

\bibitem{KarRan1}
A.~Karch and L.~Randall, {\it Locally localized gravity}, {\em JHEP} {\bf 05}
  (2001) 008, \href{http://xxx.lanl.gov/abs/hep-th/0011156}{{\tt
  hep-th/0011156}}.

\bibitem{KarRan2}
A.~Karch and L.~Randall, {\it Open and closed string interpretation of {SUSY
  CFT's} on branes with boundaries}, {\em JHEP} {\bf 06} (2001) 063,
  \href{http://xxx.lanl.gov/abs/hep-th/0105132}{{\tt hep-th/0105132}}.

\bibitem{DWFrOo}
O.~DeWolfe, D.~Z. Freedman, and H.~Ooguri, {\it Holography and defect conformal
  field theories}, \href{http://xxx.lanl.gov/abs/hep-th/0111135}{{\tt
  hep-th/0111135}}.

\bibitem{BdBDO}
C.~Bachas, J.~de~Boer, R.~Dijkgraaf, and H.~Ooguri, {\it Permeable conformal walls
  and holography}, \href{http://xxx.lanl.gov/abs/hep-th/0111210}{{\tt
  hep-th/0111210}}.

\bibitem{PetZubtw}
V.~B. Petkova and J.~B. Zuber, {\it Generalised twisted partition functions},
  {\em Phys. Lett.} {\bf B504} (2001) 157--164,
  \href{http://xxx.lanl.gov/abs/hep-th/0011021}{{\tt hep-th/0011021}}.

\bibitem{FuRuSccat}
J.~Fuchs, I.~Runkel, and C.~Schweigert, {\it {TFT construction of RCFT
  correlators. I: Parition} functions},
  \href{http://xxx.lanl.gov/abs/hep-th/0204148}{{\tt hep-th/0204148}}.

\bibitem{Nahm:1988sn}
W.~Nahm, {\it Gauging symmetries of two-dimensional conformally invariant
  models},. UCD-88-02.

\bibitem{BaRaSa}
K.~Bardakci, E.~Rabinovici, and B.~Saering, {\it String models with {$c<1$}
  components}, {\em Nucl. Phys.} {\bf B299} (1988) 151.

\bibitem{GawKup1}
K.~Gawedzki and A.~Kupiainen, {\it {G/H} conformal field theory from gauged {WZW}
  model}, {\em Phys. Lett.} {\bf B215} (1988) 119--123.

\bibitem{GawKup2}
K.~Gawedzki and A.~Kupiainen, {\it Coset construction from functional integrals},
  {\em Nucl. Phys.} {\bf B320} (1989) 625.

\bibitem{TseygWZW}
A.~A. Tseytlin, {\it Conformal sigma models corresponding to gauged {Wess-Zumino-
  Witten} theories}, {\em Nucl. Phys.} {\bf B411} (1994) 509--558,
  \href{http://xxx.lanl.gov/abs/hep-th/9302083}{{\tt hep-th/9302083}}.

\bibitem{Gawe2}
K.~Gawedzki, {\it Boundary {WZW}, {G/H}, {G/G} and {CS} theories},
  \href{http://xxx.lanl.gov/abs/hep-th/0108044}{{\tt hep-th/0108044}}.

\bibitem{EliSar}
S.~Elitzur and G.~Sarkissian, {\it D-branes on a gauged {WZW} model}, {\em Nucl.
  Phys.} {\bf B625} (2002) 166--178,
  \href{http://xxx.lanl.gov/abs/hep-th/0108142}{{\tt hep-th/0108142}}.

\bibitem{FalGaw01}
F.~Falceto and K.~Gawedzki, {\it Boundary {G/G} theory and topological
  {Poisson-Lie} sigma model},
  \href{http://xxx.lanl.gov/abs/hep-th/0108206}{{\tt hep-th/0108206}}.

\bibitem{FreSch3}
S.~Fredenhagen and V.~Schomerus, {\it D-branes in coset models}, {\em JHEP} {\bf
  02} (2002) 005, \href{http://xxx.lanl.gov/abs/hep-th/0111189}{{\tt
  hep-th/0111189}}.

\bibitem{ReRoSc}
A.~Recknagel, D.~Roggenkamp, and V.~Schomerus, {\it On relevant boundary
  perturbations of unitary minimal models}, {\em Nucl. Phys.} {\bf B588}
  (2000) 552--564, \href{http://xxx.lanl.gov/abs/hep-th/0003110}{{\tt
  hep-th/0003110}}.

\bibitem{Graham:2001pp}
K.~Graham, {\it On perturbations of unitary minimal models by boundary condition
  changing operators}, {\em JHEP} {\bf 03} (2002) 028,
  \href{http://xxx.lanl.gov/abs/hep-th/0111205}{{\tt hep-th/0111205}}.

\bibitem{FreSch4}
S.~Fredenhagen and V.~Schomerus, {\it On boundary {RG-flows} in coset conformal
  field theories}, \href{http://xxx.lanl.gov/abs/hep-th/0205011}{{\tt
  hep-th/0205011}}.

\bibitem{Chim:1996kf}
L.~Chim, {\it Boundary {S-matrix} for the tricritical {Ising} model}, {\em Int.
  J. Mod. Phys.} {\bf A11} (1996) 4491--4512,
  \href{http://xxx.lanl.gov/abs/hep-th/9510008}{{\tt hep-th/9510008}}.

\bibitem{Lesage:1998qf}
F.~Lesage, H.~Saleur, and P.~Simonetti, {\it Boundary flows in minimal models},
  {\em Phys. Lett.} {\bf B427} (1998) 85--92,
  \href{http://xxx.lanl.gov/abs/hep-th/9802061}{{\tt hep-th/9802061}}.

\bibitem{Ahn:1998xm}
C.~Ahn and C.~Rim, {\it Boundary flows in the general coset theories}, {\em J.
  Phys.} {\bf A32} (1999) 2509--2525,
  \href{http://xxx.lanl.gov/abs/hep-th/9805101}{{\tt hep-th/9805101}}.

\bibitem{Dorey:1999cj}
P.~Dorey, I.~Runkel, R.~Tateo, and G.~Watts, {\it {g}-function flow in perturbed
  boundary conformal field theories}, {\em Nucl. Phys.} {\bf B578} (2000)
  85--122, \href{http://xxx.lanl.gov/abs/hep-th/9909216}{{\tt hep-th/9909216}}.

\bibitem{Graham:2001tg}
K.~Graham, I.~Runkel, and G.~M.~T. Watts, {\it Minimal model boundary flows and c
  = 1 {CFT}}, {\em Nucl. Phys.} {\bf B608} (2001) 527--556,
  \href{http://xxx.lanl.gov/abs/hep-th/0101187}{{\tt hep-th/0101187}}.

\bibitem{FredPhD}
S.~Fredenhagen, {\it Dynamics of {D-branes} in curved backgrounds},. PhD thesis,
  unpublished.

\bibitem{Ishik}
H.~Ishikawa, {\it Boundary states in coset conformal field theories},
  \href{http://xxx.lanl.gov/abs/hep-th/0111230}{{\tt hep-th/0111230}}.

\bibitem{Kubota:2001ai}
T.~Kubota, J.~Rasmussen, M.~A. Walton, and J.-G. Zhou, {\it Maximally symmetric
  {D-branes in gauged WZW } models},
  \href{http://xxx.lanl.gov/abs/hep-th/0112078}{{\tt hep-th/0112078}}.

\bibitem{Walton:2002db}
M.~A. Walton and J.-G. Zhou, {\it D-branes in asymmetrically gauged {WZW} models
  and axial- vector duality},
  \href{http://xxx.lanl.gov/abs/hep-th/0205161}{{\tt hep-th/0205161}}.

\bibitem{Sarkissian:2002ie}
G.~Sarkissian, {\it Non-maximally symmetric {D-branes on group manifold in the
  Lagrangian approach}}, {\em JHEP} {\bf 07} (2002) 033,
  \href{http://xxx.lanl.gov/abs/hep-th/0205097}{{\tt hep-th/0205097}}.

\bibitem{IshikTan}
H.~Ishikawa and T.~Tani, {\it Novel construction of boundary states in coset
  conformal field theories},
  \href{http://xxx.lanl.gov/abs/hep-th/0207177}{{\tt hep-th/0207177}}.

\bibitem{QueSch2}
T.~Quella and V.~Schomerus, {\it Asymmetric cosets},
  \href{http://xxx.lanl.gov/abs/hep-th/}{{\tt hep-th/}}.

\bibitem{PetZubbtoB}
V.~B. Petkova and J.~B. Zuber, {\it {BCFT: From} the boundary to the bulk},
  \href{http://xxx.lanl.gov/abs/hep-th/0009219}{{\tt hep-th/0009219}}.

\bibitem{FuRuSc1}
J.~Fuchs, I.~Runkel, and C.~Schweigert, {\it Conformal correlation functions,
  {Frobenius} algebras and triangulations}, {\em Nucl. Phys.} {\bf B624}
  (2002) 452--468, \href{http://xxx.lanl.gov/abs/hep-th/0110133}{{\tt
  hep-th/0110133}}.

\bibitem{Shat}
S.~L. Shatashvili, {\it On field theory of open strings, tachyon condensation and
  closed strings}, \href{http://xxx.lanl.gov/abs/hep-th/0105076}{{\tt
  hep-th/0105076}}.

\bibitem{CKLM}
C.~G. Callan, I.~R. Klebanov, A.~W.~W. Ludwig, and J.~M. Maldacena, {\it Exact
  solution of a boundary conformal field theory}, {\em Nucl. Phys.} {\bf B422}
  (1994) 417--448, \href{http://xxx.lanl.gov/abs/hep-th/9402113}{{\tt
  hep-th/9402113}}.

\bibitem{PolThodef}
J.~Polchinski and L.~Thorlacius, {\it Free fermion representation of a boundary
  conformal field theory}, {\em Phys. Rev.} {\bf D50} (1994) 622--626,
  \href{http://xxx.lanl.gov/abs/hep-th/9404008}{{\tt hep-th/9404008}}.

\bibitem{Senbtach1}
A.~Sen, {\it Descent relations among bosonic {D-branes}}, {\em Int. J. Mod.
  Phys.} {\bf A14} (1999) 4061--4078,
  \href{http://xxx.lanl.gov/abs/hep-th/9902105}{{\tt hep-th/9902105}}.

\bibitem{Senbtach2}
A.~Sen, {\it Time evolution in open string theory},
  \href{http://xxx.lanl.gov/abs/hep-th/0207105}{{\tt hep-th/0207105}}.

\bibitem{DorOtt}
H.~Dorn and H.~J. Otto, {\it Two and three point functions in {Liouville}
  theory}, {\em Nucl. Phys.} {\bf B429} (1994) 375--388,
  \href{http://xxx.lanl.gov/abs/hep-th/9403141}{{\tt hep-th/9403141}}.

\bibitem{ZamZam}
A.~B. Zamolodchikov and A.~B. Zamolodchikov, {\it Structure constants and
  conformal bootstrap in {Liouville} field theory}, {\em Nucl. Phys.} {\bf
  B477} (1996) 577--605, \href{http://xxx.lanl.gov/abs/hep-th/9506136}{{\tt
  hep-th/9506136}}.

\bibitem{GouLi}
M.~Goulian and M.~Li, {\it Correlation functions in {Liouville} theory}, {\em
  Phys. Rev. Lett.} {\bf 66} (1991) 2051--2055.

\bibitem{TeschL1}
J.~Teschner, {\it On the {Liouville} three point function}, {\em Phys. Lett.}
  {\bf B363} (1995) 65--70, \href{http://xxx.lanl.gov/abs/hep-th/9507109}{{\tt
  hep-th/9507109}}.

\bibitem{PonTes1}
B.~Ponsot and J.~Teschner, {\it Liouville bootstrap via harmonic analysis on a
  noncompact quantum group},
  \href{http://xxx.lanl.gov/abs/hep-th/9911110}{{\tt hep-th/9911110}}.

\bibitem{PonTes2}
B.~Ponsot and J.~Teschner, {\it {Clebsch-Gordan and Racah-Wigner} coefficients for
  a continuous series of representations of {$U_q({\rm sl(2,R)})$}}, {\em
  Commun. Math. Phys.} {\bf 224} (2001) 613--655,
  \href{http://xxx.lanl.gov/abs/math.qa/0007097}{{\tt math.qa/0007097}}.

\bibitem{Teschrev}
J.~Teschner, {\it Liouville theory revisited}, {\em Class. Quant. Grav.} {\bf 18}
  (2001) R153--R222, \href{http://xxx.lanl.gov/abs/hep-th/0104158}{{\tt
  hep-th/0104158}}.

\bibitem{CreGer}
E.~Cremmer and J.-L. Gervais, {\it The quantum strip: {Liouville} theory for open
  strings}, {\em Commun. Math. Phys.} {\bf 144} (1992) 279--302.

\bibitem{FaZaZa}
V.~Fateev, A.~B. Zamolodchikov, and A.~B. Zamolodchikov, {\it Boundary {Liouville
  field theory. I: Boundary } state and boundary two-point function},
  \href{http://xxx.lanl.gov/abs/hep-th/0001012}{{\tt hep-th/0001012}}.

\bibitem{TeschbL}
J.~Teschner, {\it Remarks on {Liouville} theory with boundary},
  \href{http://xxx.lanl.gov/abs/hep-th/0009138}{{\tt hep-th/0009138}}.

\bibitem{Hoso}
K.~Hosomichi, {\it Bulk boundary propagator in {Liouville} theory on a disc},
  {\em JHEP} {\bf 11} (2001) 044,
  \href{http://xxx.lanl.gov/abs/hep-th/0108093}{{\tt hep-th/0108093}}.

\bibitem{PonTes3}
B.~Ponsot and J.~Teschner, {\it Boundary {Liouville field theory: Boundary} three
  point function}, {\em Nucl. Phys.} {\bf B622} (2002) 309--327,
  \href{http://xxx.lanl.gov/abs/hep-th/0110244}{{\tt hep-th/0110244}}.

\bibitem{RasSta}
R.~C. Rashkov and M.~Stanishkov, {\it Three-point correlation functions in {$N=1$}
  {super-Liouville} theory}, {\em Phys. Lett.} {\bf B380} (1996) 49--58,
  \href{http://xxx.lanl.gov/abs/hep-th/9602148}{{\tt hep-th/9602148}}.

\bibitem{Pogh}
R.~H. Poghosian, {\it Structure constants in the {$N = 1$ super-Liouville} field
  theory}, {\em Nucl. Phys.} {\bf B496} (1997) 451--464,
  \href{http://xxx.lanl.gov/abs/hep-th/9607120}{{\tt hep-th/9607120}}.

\bibitem{FukHos}
T.~Fukuda and K.~Hosomichi, {\it Super {Liouville} theory with boundary}, {\em
  Nucl. Phys.} {\bf B635} (2002) 215--254,
  \href{http://xxx.lanl.gov/abs/hep-th/0202032}{{\tt hep-th/0202032}}.

\bibitem{AhRiSt}
C.~Ahn, C.~Rim, and M.~Stanishkov, {\it Exact one-point function of {$N = 1$
  super-Liouville} theory with boundary}, {\em Nucl. Phys.} {\bf B636} (2002)
  497--513, \href{http://xxx.lanl.gov/abs/hep-th/0202043}{{\tt
  hep-th/0202043}}.

\bibitem{TeschH31}
J.~Teschner, {\it On structure constants and fusion rules in the {SL(2,C)/SU(2)
  WZNW} model}, {\em Nucl. Phys.} {\bf B546} (1999) 390--422,
  \href{http://xxx.lanl.gov/abs/hep-th/9712256}{{\tt hep-th/9712256}}.

\bibitem{TeschH32}
J.~Teschner, {\it The mini-superspace limit of the {SL(2,C)/SU(2) WZNW} model},
  {\em Nucl. Phys.} {\bf B546} (1999) 369--389,
  \href{http://xxx.lanl.gov/abs/hep-th/9712258}{{\tt hep-th/9712258}}.

\bibitem{TeschH33}
J.~Teschner, {\it Operator product expansion and factorization in the {$H_3^+$
  WZNW} model}, {\em Nucl. Phys.} {\bf B571} (2000) 555--582,
  \href{http://xxx.lanl.gov/abs/hep-th/9906215}{{\tt hep-th/9906215}}.

\bibitem{TeschH34}
J.~Teschner, {\it Crossing symmetry in the {$H_3^+$ WZNW} model}, {\em Phys.
  Lett.} {\bf B521} (2001) 127--132,
  \href{http://xxx.lanl.gov/abs/hep-th/0108121}{{\tt hep-th/0108121}}.

\bibitem{GawNC}
K.~Gawedzki, {\it Noncompact {WZW} conformal field theories},
  \href{http://xxx.lanl.gov/abs/hep-th/9110076}{{\tt hep-th/9110076}}.

\bibitem{MaOo1}
J.~M. Maldacena and H.~Ooguri, {\it Strings in {$AdS_3$ and SL(2,R)} {WZW} model.
  {I}}, {\em J. Math. Phys.} {\bf 42} (2001) 2929--2960,
  \href{http://xxx.lanl.gov/abs/hep-th/0001053}{{\tt hep-th/0001053}}.

\bibitem{MaOo2}
J.~M. Maldacena, H.~Ooguri, and J.~Son, {\it Strings in {$AdS_3$ and the SL(2,R)
  WZW model. II: Euclidean} black hole}, {\em J. Math. Phys.} {\bf 42} (2001)
  2961--2977, \href{http://xxx.lanl.gov/abs/hep-th/0005183}{{\tt
  hep-th/0005183}}.

\bibitem{MaOo3}
J.~M. Maldacena and H.~Ooguri, {\it Strings in {$AdS_3$ and the SL(2,R) WZW model.
  III: Correlation} functions}, {\em Phys. Rev.} {\bf D65} (2002) 106006,
  \href{http://xxx.lanl.gov/abs/hep-th/0111180}{{\tt hep-th/0111180}}.

\bibitem{Wit1}
E.~Witten, {\it On string theory and black holes}, {\em Phys. Rev.} {\bf D44}
  (1991) 314--324.

\bibitem{HoSt}
G.~T. Horowitz and A.~Strominger, {\it Black strings and p-branes}, {\em Nucl.
  Phys.} {\bf B360} (1991) 197--209.

\bibitem{MaSt}
J.~M. Maldacena and A.~Strominger, {\it Semiclassical decay of near-extremal
  fivebranes}, {\em JHEP} {\bf 12} (1997) 008,
  \href{http://xxx.lanl.gov/abs/hep-th/9710014}{{\tt hep-th/9710014}}.

\bibitem{OoVa}
H.~Ooguri and C.~Vafa, {\it Two-dimensional black hole and singularities of {CY}
  manifolds}, {\em Nucl. Phys.} {\bf B463} (1996) 55--72,
  \href{http://xxx.lanl.gov/abs/hep-th/9511164}{{\tt hep-th/9511164}}.

\bibitem{HaPrTr}
A.~Hanany, N.~Prezas, and J.~Troost, {\it The partition function of the
  two-dimensional black hole conformal field theory}, {\em JHEP} {\bf 04}
  (2002) 014, \href{http://xxx.lanl.gov/abs/hep-th/0202129}{{\tt
  hep-th/0202129}}.

\bibitem{DVV}
R.~Dijkgraaf, H.~Verlinde, and E.~Verlinde, {\it String propagation in a black
  hole geometry}, {\em Nucl. Phys.} {\bf B371} (1992) 269--314.

\bibitem{KaKoKu}
V.~Kazakov, I.~K. Kostov, and D.~Kutasov, {\it A matrix model for the
  two-dimensional black hole}, {\em Nucl. Phys.} {\bf B622} (2002) 141--188,
  \href{http://xxx.lanl.gov/abs/hep-th/0101011}{{\tt hep-th/0101011}}.

\bibitem{FukHossL}
T.~Fukuda and K.~Hosomichi, {\it Three-point functions in {sine-Liouville}
  theory}, {\em JHEP} {\bf 09} (2001) 003,
  \href{http://xxx.lanl.gov/abs/hep-th/0105217}{{\tt hep-th/0105217}}.

\bibitem{GivKut}
A.~Giveon and D.~Kutasov, {\it Little string theory in a double scaling limit},
  {\em JHEP} {\bf 10} (1999) 034,
  \href{http://xxx.lanl.gov/abs/hep-th/9909110}{{\tt hep-th/9909110}}.

\bibitem{HorKap}
K.~Hori and A.~Kapustin, {\it Duality of the fermionic 2d black hole and {$N = 2$
  Liouville} theory as mirror symmetry}, {\em JHEP} {\bf 08} (2001) 045,
  \href{http://xxx.lanl.gov/abs/hep-th/0104202}{{\tt hep-th/0104202}}.

\bibitem{BORT}
J.~de~Boer, H.~Ooguri, H.~Robins, and J.~Tannenhauser, {\it String theory on
  {$AdS_3$}}, {\em JHEP} {\bf 12} (1998) 026,
  \href{http://xxx.lanl.gov/abs/hep-th/9812046}{{\tt hep-th/9812046}}.

\bibitem{StanAdS}
S.~Stanciu, {\it D-branes in an {$AdS_3$ background}}, {\em JHEP} {\bf 09} (1999)
  028, \href{http://xxx.lanl.gov/abs/hep-th/9901122}{{\tt hep-th/9901122}}.

\bibitem{BacPet}
C.~Bachas and M.~Petropoulos, {\it {Anti-de-Sitter D-branes}}, {\em JHEP} {\bf
  02} (2001) 025, \href{http://xxx.lanl.gov/abs/hep-th/0012234}{{\tt
  hep-th/0012234}}.

\bibitem{PoScTe}
B.~Ponsot, V.~Schomerus, and J.~Teschner, {\it Branes in the {Euclidean
  $AdS_3$}}, {\em JHEP} {\bf 02} (2002) 016,
  \href{http://xxx.lanl.gov/abs/hep-th/0112198}{{\tt hep-th/0112198}}.

\bibitem{LeOoPa}
P.~Lee, H.~Ooguri, and J.~Park, {\it Boundary states for {$AdS_2$ branes in
  $AdS_3$}}, {\em Nucl. Phys.} {\bf B632} (2002) 283--302,
  \href{http://xxx.lanl.gov/abs/hep-th/0112188}{{\tt hep-th/0112188}}.

\bibitem{Mess}
A.~Messiah, {\em {Quantum mechanics, Vol. 1/2}}.
\newblock North-Holland, 1975.

\bibitem{GiKuSc}
A.~Giveon, D.~Kutasov, and A.~Schwimmer, {\it Comments on {D-branes in $AdS_3$}},
  {\em Nucl. Phys.} {\bf B615} (2001) 133--168,
  \href{http://xxx.lanl.gov/abs/hep-th/0106005}{{\tt hep-th/0106005}}.

\bibitem{ParSah}
A.~Parnachev and D.~A. Sahakyan, {\it Some remarks on {D-branes in $AdS_3$}},
  {\em JHEP} {\bf 10} (2001) 022,
  \href{http://xxx.lanl.gov/abs/hep-th/0109150}{{\tt hep-th/0109150}}.

\bibitem{Ponsot}
B.~Ponsot, {\it Monodromy of solutions of the {Knizhnik-Zamolodchikov equation:
  SL(2,C)/SU(2) WZNW} model},
  \href{http://xxx.lanl.gov/abs/hep-th/0204085}{{\tt hep-th/0204085}}.

\end{thebibliography}
\end{document}